\documentclass{article}

\usepackage{PRIMEarxiv}

\usepackage{enumitem}
\usepackage{graphicx}
\usepackage{multirow}
\usepackage{makecell}
\usepackage{booktabs, colortbl}
\usepackage{orcidlink}
\hypersetup{colorlinks=true, allcolors=black, breaklinks=true}
\usepackage[framemethod=TikZ]{mdframed}
\usepackage{amsthm}
\usepackage{cleveref}
\usepackage{float} 
\usepackage{subfigure}
\usepackage{natbib}

\usepackage{amsfonts}
\usepackage{fancyhdr}
\usepackage{nicefrac}       
\usepackage{microtype}

\newcounter{rqres}[section] 
\newenvironment{rqres}[2][]{%
\refstepcounter{rqres}%
\ifstrempty{#1}%
{\mdfsetup{%
frametitle={%
\tikz[baseline=(current bounding box.east),outer sep=0pt]
\node[anchor=east,rectangle,fill=blue!15]
{\strut RQ\therqres~Findings};}}
}%
{\mdfsetup{%
frametitle={%
\tikz[baseline=(current bounding box.east),outer sep=0pt]
\node[anchor=east,rectangle,fill=blue!20]
{\strut RQ\therqres~Findings:~#1};}}%
}%
\mdfsetup{innertopmargin=2pt,linecolor=blue!15,%
linewidth=2pt,topline=true,%
frametitleaboveskip=\dimexpr-\ht\strutbox\relax
}
\begin{mdframed}[]\relax%
\label{#2}}{\end{mdframed}}

\usepackage{soul}
\sethlcolor{blue!10}

\makeatletter
\def\SOUL@hlpreamble{%
    \setul{}{3.5ex}
    \let\SOUL@stcolor\SOUL@hlcolor
    \dimen@\SOUL@ulthickness
    \dimen@i=-.75ex 
    \advance\dimen@i-.5\dimen@
    \edef\SOUL@uldepth{\the\dimen@i}%
    \let\SOUL@ulcolor\SOUL@stcolor
    \SOUL@ulpreamble
}
\makeatother
\newcommand*{\codebox}[1]{{\relax\hl{#1}}}

\newcommand*\rot{\rotatebox{70}}

\newcommand*\rotnd{\rotatebox{90}}

\title{REST API Testing in DevOps: A Study on an Evolving Healthcare IoT Application
}

\author{
  Hassan Sartaj, Shaukat Ali \\
  Simula Research Laboratory \\
  Oslo, Norway\\
  \texttt{\{hassan, shaukat\}@simula.no} \\
   \And
  Julie Marie Gjøby \\
  Welfare Technologies Section, Oslo Kommune Helseetaten \\
  Oslo, Norway\\
  \texttt{julie-marie.gjoby@hel.oslo.kommune.no} \\
}

\begin{document}
\maketitle

\newcommand{\RESTest}{{\mbox{RESTest}}}
\newcommand{\EvoMaster}{{\mbox{EvoMaster}}}
\newcommand{\Schemathesis}{{\mbox{Schemathesis}}}
\newcommand{\RESTler}{{\mbox{RESTler}}}
\newcommand{\RestTestGen}{{\mbox{RestTestGen}}}

\newcommand{\RESTestCBT}{{\fontfamily{cmtt}\selectfont \mbox{RESTest-CBT}}}
\newcommand{\RESTestART}{{\fontfamily{cmtt}\selectfont \mbox{RESTest-ART}}}
\newcommand{\RESTestRT}{{\fontfamily{cmtt}\selectfont \mbox{RESTest-RT}}}
\newcommand{\RESTestFT}{{\fontfamily{cmtt}\selectfont \mbox{RESTest-FT}}}

\newcommand{\EvoMasterBB}{{\fontfamily{cmtt}\selectfont \mbox{EvoMaster-BB}}}
\newcommand{\SchemathesisPT}{{\fontfamily{cmtt}\selectfont \mbox{Schemathesis-PT}}}

\newcommand{\RESTlerF}{{\fontfamily{cmtt}\selectfont \mbox{RESTler-F}}}
\newcommand{\RESTlerFL}{{\fontfamily{cmtt}\selectfont \mbox{RESTler-FL}}}
\newcommand{\RestTestGenNET}{{\fontfamily{cmtt}\selectfont \mbox{RestTestGen-NET}}}
\newcommand{\RestTestGenMAST}{{\fontfamily{cmtt}\selectfont \mbox{RestTestGen-MAST}}}

\newcommand{\RTCBT}{{\fontfamily{cmtt}\selectfont \mbox{RT-CBT}}}
\newcommand{\RTART}{{\fontfamily{cmtt}\selectfont \mbox{RT-ART}}}
\newcommand{\RTRT}{{\fontfamily{cmtt}\selectfont \mbox{RT-RT}}}
\newcommand{\RTFT}{{\fontfamily{cmtt}\selectfont \mbox{RT-FT}}}

\newcommand{\EMBB}{{\fontfamily{cmtt}\selectfont \mbox{EM-BB}}}
\newcommand{\STPT}{{\fontfamily{cmtt}\selectfont \mbox{ST-PT}}}

\newcommand{\RLFz}{{\fontfamily{cmtt}\selectfont \mbox{RL-Fz}}}
\newcommand{\RLFL}{{\fontfamily{cmtt}\selectfont \mbox{RL-FL}}}
\newcommand{\RTGNET}{{\fontfamily{cmtt}\selectfont \mbox{RTG-NET}}}
\newcommand{\RTGMAST}{{\fontfamily{cmtt}\selectfont \mbox{RTG-MAST}}}

\begin{abstract} 
Healthcare Internet of Things (IoT) applications often integrate various third-party healthcare applications and medical devices through REST APIs, resulting in complex and interdependent networks of REST APIs. 
Oslo City's healthcare department collaborates with various industry partners to develop these applications, enriched with diverse REST APIs that evolve during the DevOps process to accommodate evolving needs such as new features, services, and devices.
Oslo City's primary goal is to utilize automated solutions for continuous testing of REST APIs at each evolution stage to ensure dependability. 
Although the literature offers various automated REST API testing tools, their effectiveness in regression testing of the evolving REST APIs of healthcare IoT applications within a DevOps context remains undetermined. 
This paper evaluates state-of-the-art and well-established REST API testing tools---specifically, \RESTest{}, \EvoMaster{}, \Schemathesis{}, \RESTler{}, and \RestTestGen{}---for the regression testing of a real-world healthcare IoT application, considering failures, faults, coverage, regressions, and cost. 
We conducted experiments using all accessible REST APIs (17 APIs with 120 endpoints), and 14 releases evolved during DevOps. 
Overall, all tools generated tests leading to several failures, 18 potential faults, up to 84\% coverage, and 23 regressions.  
Over 70\% of tests generated by all tools fail to detect failures, resulting in significant overhead. 
\end{abstract}


\section{Introduction}
Healthcare Internet of Things (IoT) applications commonly use a cloud-based architecture as a hub for stakeholders such as healthcare professionals, caretakers, and patients~\cite{sartaj2023hita}. 
This involves integrating various healthcare applications serving different purposes (such as electronic health records (EHR) for managing patient data) and medical devices (like medicine dispensers) assigned to patients. 
These integrations into the IoT cloud are facilitated through Application Programming Interfaces (APIs) following the Representational State Transfer (REST) architecture---concisely REST APIs. 
A malfunction in REST APIs can significantly dismantle the healthcare IoT cloud, potentially making crucial healthcare services unavailable to patients. 
This could fail to provide timely healthcare services, which can have serious consequences. 
Given the critical nature of healthcare IoT applications, rigorous testing at various levels, such as the system level, is essential to ensure their dependability.

Oslo City's healthcare department~\cite{oslocity} collaborates with various healthcare industries to develop IoT applications, aiming to deliver efficient healthcare services to residents. 
These collaborations typically adhere to DevOps practices, where the Oslo City healthcare team continuously manages, analyzes, and reports problems within the IoT healthcare application under development. 
One of their activities is to test the REST APIs of such an application at the system level with integrated medical devices and third-party applications. 
However, manual testing of REST APIs at each evolution stage can be time-consuming and laborious due to the complex and interdependent nature of these APIs, thus necessitating an automated testing solution.

Over time, automated testing of REST APIs has attracted significant interest. 
Numerous approaches, such as \RESTest{}~\cite{martin2021restest}, \RESTler{}~\cite{atlidakis2019restler}, \Schemathesis{}~\cite{hatfield2022deriving}, RESTCT~\cite{wu2022combinatorial}, \RestTestGen{}~\cite{corradini2022resttestgen}, and \EvoMaster{}~\cite{arcuri2018evomaster}, have been proposed. 
The effectiveness of REST API testing approaches has been examined in different empirical studies (e.g.,~\cite{kim2022automated,martin2021black,arcuri2023building}) for various open-source and industrial web applications. 
However, REST APIs in IoT-based applications differ significantly due to their hybrid architecture (integrating cloud and edge computing), dependence on heterogeneous devices and third-party services, and frequent vendor-driven API evolution, factors that directly impact their dependability. 
In addition, healthcare applications, such as the one in our study, enforce stricter integrity, quality, and security requirements due to regulatory compliance, such as the EU's General Data Protection Regulation
(GDPR), requiring more stringent authentication and authorization mechanisms. 
Our analysis of the literature revealed two key observations. 
Firstly, the effectiveness of REST API testing approaches in testing IoT applications' REST APIs remains unexplored. 
Secondly, the use of REST API testing approaches in the regression testing of evolving REST APIs within the DevOps context is also unexplored. 
We found only one study by Godefroid et al.~\cite{godefroid2020differential} that employed \RESTler{} for regression testing of open-source APIs using historical release data. 
Thus, the evaluation of REST API testing approaches in the regression testing of continuously evolving REST APIs of real-world IoT applications remains a research gap. 
Considering this gap, we assess well-established approaches for testing the REST API of a real-world healthcare IoT application deployed in Oslo City and analyze their practicality in this context.

This paper builds upon our previous work~\cite{sartaj2023testing} in which we reported an empirical evaluation of \RESTest{} for testing six APIs with 41 endpoints of a healthcare IoT application's production release. 
This paper presents a large-scale empirical study that evaluates state-of-the-art and well-established tools---specifically \RESTest{}, \EvoMaster{}, \Schemathesis{}, \RESTler{}, and \RestTestGen{}---for the regression testing of evolving REST APIs associated with a real-world healthcare IoT application, which is progressively developed via the DevOps process. 
Our experiments encompass 14 different types of releases (daily, weekly, and monthly) and feature all accessible REST APIs of the application under test, i.e., 17 APIs with 120 endpoints. 
We evaluated REST API testing tools, considering their effectiveness in terms of failures, faults, coverage, regressions, and cost overhead. 
Results show that \EvoMaster{} generated tests led to the highest number of failures among all the tools.
The fault analysis revealed that the tests generated by all tools led to various potential faults, totaling 18 unique faults across 14 releases. 
The coverage results indicate that \RESTest{} achieved approximately 82\% coverage, while \EvoMaster{}, \Schemathesis{}, \RESTler{}, and \RestTestGen{} attained an overall coverage of around 84\%. 
Moreover, analysis of regressions across all 14 releases demonstrated that the tests generated by each tool resulted in the identification of several potential regressions, with \EvoMaster{} at the top. 
In total, 23 unique regressions were discovered through testing results from all tools across the 14 releases.
Through these regressions, we analyzed the stability of APIs during continuous releases, specifically focusing on the state of API implementation (i.e., feature upgrades) and the propagation of failures. 
The results of this regression analysis offer valuable insights for practitioners, assisting them in identifying potential areas of API vulnerability, determining recurring failures, and choosing a suitable testing tool. 
Finally, the cost analysis demonstrated that each tool generated a significant number of tests, often leading to numerous duplicate failures. 
For all tools, more than 70\% of the generated test cases did not detect any failures, thus incurring considerable overhead.

The specific contributions of this work are summarized below:
\begin{itemize}
    \item A large-scale empirical study for testing REST APIs of a real-world healthcare IoT application featuring:  
    \begin{itemize}
        \item[+] Five REST API testing tools, i.e., \RESTest{}, \EvoMaster{}, \Schemathesis{}, \RESTler{}, and \RestTestGen{} configured with different black-box testing techniques, making a total of 10 tools. 
        \item[+] A total of 17 REST APIs comprising 120 endpoints. 
        \item[+] 14 rapid releases (including daily, weekly, and monthly) of our industrial healthcare IoT application, continuously built in a staging environment following DevOps practices. 
    \end{itemize}
    \item A guide for practitioners on selecting REST API testing tools, derived from our empirical findings. 
    \item Experiences, insights, and challenges for researchers and practitioners.
\end{itemize}

The remainder of the paper is organized as follows. 
An overview of healthcare IoT applications and our industrial context are described in \Cref{sec:background}. 
Related works are discussed in \Cref{sec:relatedworks}. 
The experiments are reported in \Cref{sec:experiment}. 
Recommendations for selecting REST API testing tools are presented in \Cref{sec:recommendations}. 
Experiences, insights, and challenges are outlined in \Cref{sec:insights}. 
The paper concludes in \Cref{sec:conclusion}.

\section{Background}\label{sec:background}
We provide an overview of regression testing, healthcare IoT applications, our industrial context, and the associated challenges.

\subsection{Regression Testing}\label{ssec:regression}
Regressions are unintended defects or behavioral differences between two subsequent versions of the software, typically resulting from various modifications such as bug fixes, the addition of new features, or updates to existing functionality~\cite{beizer2003software}. 
These regressions often arise from integration issues, especially when modifications in one module impact others due to interdependencies. 
For instance, a database-related bug fix might inadvertently affect the functionality of the graphical user interface. 
Similarly, introducing new features or upgrading existing ones may disrupt software behaviors due to potential conflicts with current logic or workflows. 
Furthermore, regressions may result from the propagation of failures across versions, often triggered by external updates like third-party library or API changes, or the prioritization of critical bugs over lower-priority ones, especially common in rapid release environments. 
To identify regressions, practitioners use various criteria, including behavioral consistency checks that compare current functionality with expected outcomes from previous versions. 
Regression testing plays a key role in this regard by systematically running tests on both previous and updated software versions to ensure that recent changes have not adversely impacted existing functionalities~\cite{yoo2012regression}. 
A straightforward approach to regression testing is the \emph{retest-all} strategy, which involves rerunning all existing tests on the new version~\cite{yoo2012regression,greca2023state}. 
Although simple, this approach can be resource-intensive due to the execution of redundant tests. 
Therefore, many techniques are also available for selective regression testing, categorized as test case selection, test case minimization, and test case prioritization~\cite{yoo2012regression}. 
Test case selection techniques focus on choosing a subset of tests that target only the modified software components. 
Test case minimization techniques aim to remove redundant or obsolete tests from the test suite. 
Test case prioritization techniques determine the suitable ordering of tests to run with the aim of detecting faults as early as possible. 
These techniques are particularly useful in Continuous Integration and Continuous Deployment (CI/CD) or DevOps settings, where rapid and frequent releases demand efficient use of testing resources.

\subsection{Healthcare IoT Applications}\label{ssec:hiot}
\begin{figure}[htbp]
\centerline{\includegraphics[width=12.1cm, keepaspectratio]{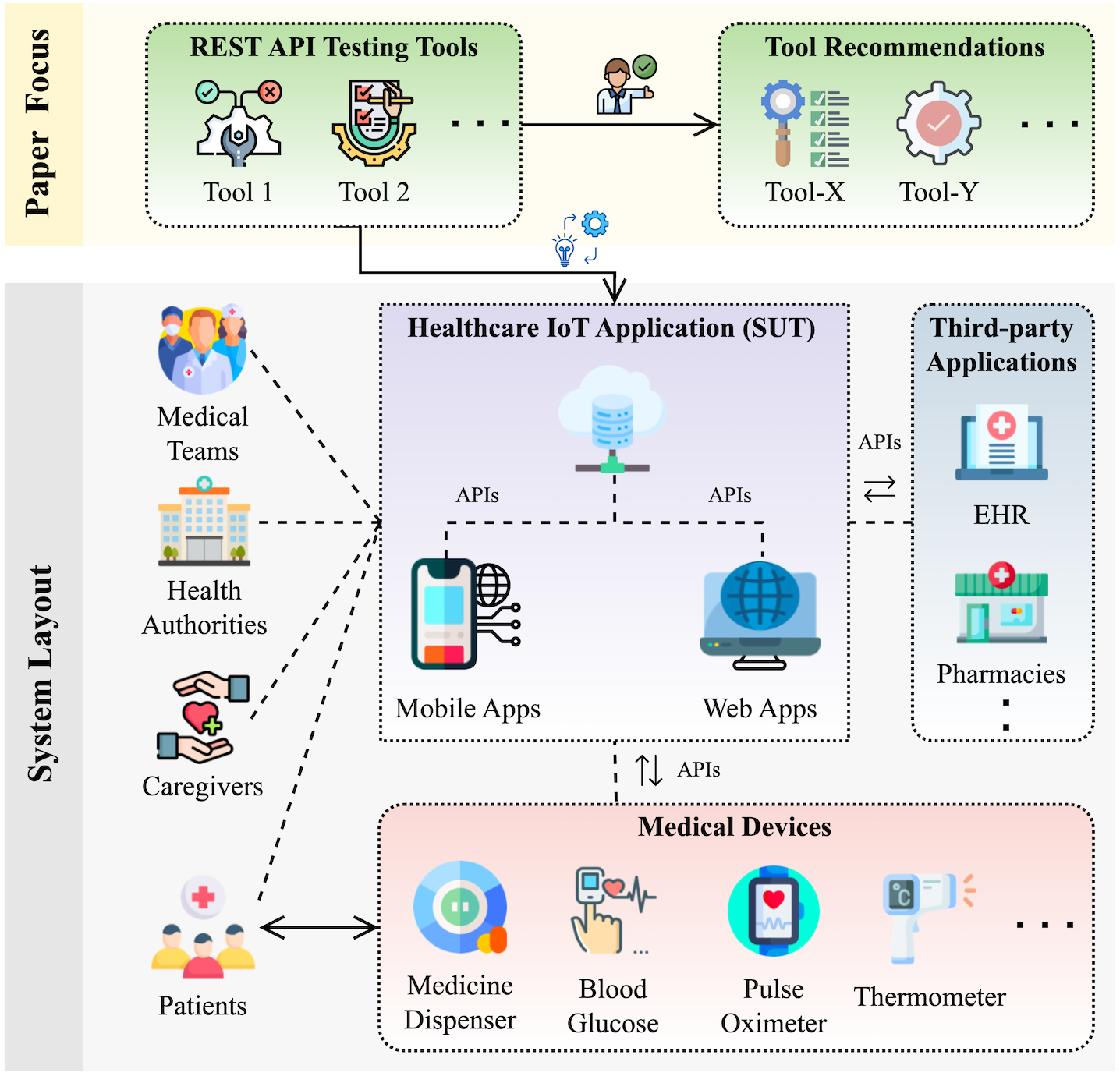}}
\caption{A typical layout of a healthcare IoT application and the focus of our work}
\label{fig:iot}
\end{figure}

A typical layout of a healthcare IoT application is depicted in \Cref{fig:iot}. 
A healthcare IoT application utilizes a cloud platform that offers web and mobile interfaces for various users, including medical professionals, healthcare authorities, caregivers, and patients. 
Within the IoT cloud, it links to numerous third-party applications that provide necessary healthcare services, such as EHR and pharmacies. 
It is also connected to various medical devices assigned to patients based on specific healthcare needs, such as medicine dispensers for timely medication and pulse oximeters for heart rate monitoring. 

All communications with third-party applications and medical devices are facilitated through REST APIs. 
Each service provider, whether for a device or an external application, has its own server and API implementation. 
The REST APIs of a healthcare IoT application communicate with the APIs of medical devices and external applications using the JavaScript Object Notation (JSON) format for data exchange. 
Consequently, the healthcare IoT application possesses a rich set of REST APIs with complicated structures due to their dependencies on the APIs of multiple types of applications. 
For example, in a patient medication scenario, the healthcare IoT application initially retrieves a medication plan via an API request to the pharmacy's API. 
It then interprets this medication plan and updates the web interface for the healthcare personnel. 
Finally, it sends the medication plan to the patient's medicine dispenser via another API call to the device's API.

\subsection{Industrial Context and Challenges}\label{ssec:context} 
\begin{figure}[htbp]
\centerline{\includegraphics[width=13.0cm, keepaspectratio]{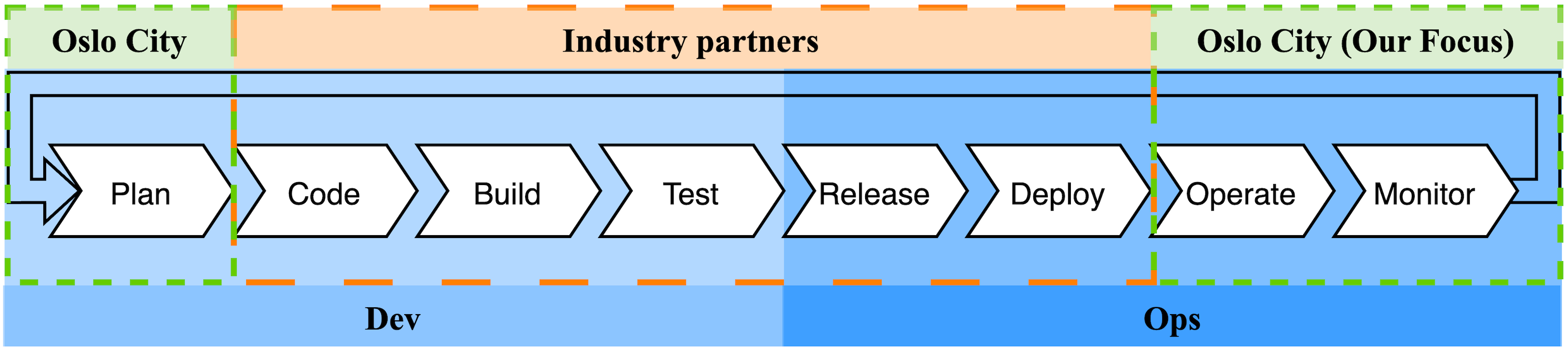}}
\caption{An overview of the DevOps process for the healthcare IoT application in our context, highlighting the phases involving industry partners, the involvement of Oslo City, and the specific focus of our paper. }
\label{fig:devops}
\end{figure}

The Directorate of Health in Norway launched the national welfare technology program~\cite{wtsoslo} to offer efficient digital home monitoring and healthcare services.
As part of this program, Oslo City's healthcare department~\cite{oslocity} worked with different industry partners to create a cloud-based platform with an integrated network of smart medical devices and healthcare service providers---a platform that resembles the layout of the IoT healthcare applications described in \Cref{ssec:hiot}. 
Oslo City follows the DevOps process to foster these collaborations with industry partners. 
In this process, Oslo City's healthcare team continuously manages, analyzes, and reports any issues that arise within the under-development healthcare IoT application. 
\Cref{fig:devops} illustrates the commonly followed DevOps process for healthcare IoT applications. 
In this process, Oslo City initially gets involved during the \emph{Plan} phase, where they define system requirements and plan various project aspects. 
The subsequent phases, including \emph{Code}, \emph{Build}, \emph{Test}, \emph{Release}, and \emph{Deploy}, involve contributions from industry partners. 
Upon deploying a specific release, Oslo City's healthcare team handles the \emph{Operate} and \emph{Monitor} phases, in which they analyze the delivered functionalities to ensure dependability and report any issues or necessary enhancements.

Throughout the DevOps process, healthcare IoT applications continuously evolve, adding new services or upgrading existing ones. 
REST APIs, which make up a significant part of these applications' backend, also undergo regular evolution. 
These stages of evolution typically correspond to daily builds for minor changes or bug fixes and weekly builds for major updates or new features.
Testing REST APIs manually at each stage of evolution is labor-intensive and time-consuming; therefore, automating this testing process is essential to ensure the dependability of healthcare IoT applications. 
To address this challenge, Oslo City's primary objective in conducting automated regression testing is to ensure that REST APIs function as intended despite constantly evolving requirements and regular feature upgrades.

Automated REST API testing of evolving REST APIs presents several challenges. 
The requirement for frequent testing at each stage of evolution, especially when connected to medical devices and third-party applications, may risk damaging the device or causing service blocking from external applications~\cite{sartaj2023hita,sartaj2024modelbased}. 
Moreover, identifying the root causes of faults or failures due to the dependencies of REST APIs on various third-party applications is a considerable challenge~\cite{sartaj2023testing}. 
For testing REST APIs of healthcare IoT applications, generating test data requires specific knowledge of the healthcare domain and an understanding of the application context.  
Lastly, the complexity and interdependent structure of REST APIs make automated test oracle generation a challenging task. 
These challenges were primarily addressed in our previous works~\cite{sartaj2024modelbased,sartaj2024uncertainty,sartaj2024medet}. 
In this study, we aim to evaluate REST API testing tools within the context of an evolving healthcare IoT application and to determine the most effective tools for integration into the overall test infrastructure (as indicated in \Cref{fig:iot}).

\section{Related Works}\label{sec:relatedworks}
We present related works on testing IoT applications, REST API testing, automated testing in DevOps, and empirical evaluations of REST API testing tools.

\subsection{IoT Applications Testing}
Numerous studies focus on various aspects of testing IoT applications~\cite{dias2018brief}. 
These include model-based conformance testing of the IoT~\cite{ahmad2016model}, user interface-driven acceptance testing of IoT systems~\cite{leotta2018acceptance}, testing the heterogeneity of IoT devices in the loop~\cite{amalfitano2017towards}, and identifying faults in the integration of devices with IoT applications~\cite{wang2022understanding}. 
Other research areas include combinatorial testing and coverage criteria for IoT systems~\cite{hu2022ct}, combinatorial testing for IoT-based smart home applications~\cite{garn2022combinatorial}, and simulating the health monitoring activities of healthcare IoT applications~\cite{sotiriadis2014towards}. 
In contrast to the studies mentioned above, our work primarily focuses on the regression testing of evolving REST APIs within a healthcare IoT application. 
Furthermore, our research involves the empirical evaluation of REST API testing tools, as opposed to proposing a new testing technique. 
The insights derived from our empirical study could potentially be employed to enhance existing techniques for testing IoT applications across various domains.

\subsection{REST API Testing}
Automated testing of REST APIs has gained significant interest, and many approaches have been proposed to address this~\cite{golmohammadi2022testing}. 
These comprise \RESTest{}~\cite{martin2021restest,martin2022online}, ARTE~\cite{alonso2022arte}---an extension to---\RESTest{}, \RESTler{}~\cite{atlidakis2019restler,godefroid2020intelligent}, \Schemathesis{}~\cite{hatfield2022deriving}, RESTCT~\cite{wu2022combinatorial}, \RestTestGen{}~\cite{corradini2022resttestgen,corradini2022automated}, \EvoMaster{}~\cite{arcuri2018evomaster,arcuri2019restful}, RapiTest~\cite{felicio2023rapitest}, QuickREST~\cite{karlsson2020quickrest}, NLP2REST~\cite{kim2023enhancing}, ARAT-RL~\cite{kim2023adaptive}, APIRL~\cite{foley2025apirl}, and LlamaRestTest~\cite{kim2025llamaresttest}. 
Some open-source tools like APIFuzzer~\cite{apifuzzer}, Tcases~\cite{tcases}, and Dredd~\cite{dredd} also support testing of REST APIs. 
However, as reported in~\cite{kim2022automated}, these tools did not outperform research-based tools. 
In addition, numerous studies target various forms of REST API testing, including model-based testing~\cite{liu2022morest}, specification-based testing~\cite{ed2018automatic}, robustness testing~\cite{laranjeiro2021black}, metamorphic testing~\cite{segura2018metamorphic}, search-based test case improvement~\cite{stallenberg2021improving}, security testing~\cite{atlidakis2020checking}, test input validation using deep learning~\cite{mirabella2021deep}, and identifying vulnerabilities in REST APIs~\cite{corradini2023automated}. 
Although these studies propose various approaches for testing REST APIs, our work explicitly evaluates selected tools for testing the REST APIs of a real-world healthcare IoT application within a DevOps environment. 
In another relevant work, Godefroid et al.~\cite{godefroid2020differential} proposed an approach for regression testing of REST APIs and evaluated their approach using the \RESTler{} tool on open-source APIs. 
Compared to this work, our work differs in two aspects. 
First, our analysis focuses on assessing REST API testing tools in a continuous regression testing environment, whereas their analysis is based on historical release data. 
Second, we utilized five REST API testing tools along with their supported testing techniques (amounting to a total of 10 tools), in contrast to their work, which only considered one tool, specifically \RESTler{}. 

\subsection{Automated Testing in DevOps}
There has been growing interest in developing automated testing approaches for web applications in DevOps across various levels, including unit, integration, and system testing~\cite{pando2022software,poth2022integration}. 
Several studies have also focused on specific aspects of testing web applications and services in DevOps, such as security testing~\cite{rangnau2020continuous}, performance-aware regression testing~\cite{chen2020performance}, and reliability testing~\cite{bertolino2023devopret}. 
However, there is a lack of research on applying automated REST API testing tools for regression testing in DevOps. 
This gap is particularly significant in the context of IoT-based applications, where REST APIs must operate across hybrid architectures (cloud and edge), accommodate evolving APIs from heterogeneous devices and third-party services, comply with strict data privacy regulations, and maintain consistency across environments such as development, staging, and production. 
In the context of existing recommended practices for automated testing in DevOps, Abdulla et al.~\cite{abdulla2023automated} provided guidelines for the GitHub development environment, while Patel et al.~\cite{patel2022state} presented general test automation practices within DevOps workflows. 
However, to the best of our knowledge, we could not find any guidelines specifically addressing the use of REST API testing tools in DevOps for IoT-based healthcare applications, which highlights a unique aspect of our work.

\subsection{Empirical Evaluations of REST API Testing Tools}
Several studies have conducted empirical research comparing various REST API testing tools.
In this regard, Martin-Lopez et al.~\cite{martin2021black} compared black-box and white-box testing tools, specifically \RESTest{} and \EvoMaster{}, by using them in four open-source APIs. 
Corradini et al.~\cite{corradini2021empirical} also delved into this subject, analyzing four tools: \RESTest{}, \RESTler{}, \RestTestGen{}, and bBOXRT. 
Their study involved testing the REST APIs of 14 open-source applications. 
Furthermore, Kim et al.~\cite{kim2022automated} evaluated 10 different tools designed for testing REST APIs, using 20 open-source applications in their study.
More recently, Zhang et al.~\cite{zhang2023open} performed an empirical comparison of seven REST API testing tools using 20 different case studies, which included an industrial case study. 
While the studies outlined above assess various REST API testing tools across different case studies, two key aspects remain unexplored. 
The first is the effectiveness of these testing tools when applied to REST APIs of large-scale IoT applications. 
The second is the applicability of these tools for regression testing of evolving REST APIs within a DevOps context. 
Our work specifically targets these aspects.

In our previous work~\cite{sartaj2023testing}, we evaluated the effectiveness of \RESTest{} in testing six REST APIs of an IoT application in industrial healthcare. 
In this paper, we extend this scope by evaluating five REST API testing tools for testing continuously evolving REST APIs of a real-world healthcare IoT application in a DevOps context. 
The specific additions to previous work are outlined below.
\begin{itemize}
    \item We incorporate an additional 11 REST APIs with 79 endpoints of a healthcare IoT application targeting different features, resulting in a total of 17 APIs with 120 endpoints. 
    \item In addition to \RESTest{}, we include four other REST API testing tools in our study, namely \EvoMaster{}, \Schemathesis{}, \RESTler{}, and \RestTestGen{}. Each tool is configured with its supported testing technique, resulting in 10 tools utilized in our experiment. 
    \item We evaluate REST API testing tools for continuous regression testing by utilizing 14 API releases (daily, weekly, and monthly) progressively built in a staging environment following DevOps practices. 
    \item We extended the evaluation scope by introducing two additional research questions: one focusing on regression analysis and the other on cost overhead analysis. 
    \item We present recommendations to assist practitioners in selecting appropriate tools. Furthermore, based on our large-scale empirical evaluation, we provide valuable insights and lessons for researchers and practitioners. 
\end{itemize}

\section{Experiments}\label{sec:experiment}
This section presents experiments to evaluate REST API testing tools in a real-world healthcare IoT application context. 
Our evaluation primarily focuses on regression testing of REST APIs that progressively evolved during the DevOps process. 
Specifically, we analyze REST API testing tools regarding failures, faults, coverage, regressions, and cost overhead. 
Given this focus, we formulate the following research questions (RQs).

\begin{itemize}[leftmargin=33pt]
    \item[\codebox{\textbf{RQ1:}}] \textit{How does the failure detection ability of REST API testing tools vary as APIs under test evolve?}\\
    As APIs undergo upgrades with new/enhanced features, it becomes essential to ensure that the REST API testing tools can detect any API failures arising from these changes. The findings from this RQ can assist practitioners in selecting appropriate tools for testing in the continuous integration and deployment phases of the DevOps process. 
    \item[\codebox{\textbf{RQ2:}}] \textit{What kinds of potential faults can be discovered using REST API testing tools during API evolution?}\\
    Faults in APIs, which can result in API failures, are important to investigate, especially in critical systems like healthcare.
    The potential faults identified in this RQ could assist practitioners in fixing API implementations. 
    Furthermore, they can guide practitioners during API development to avoid frequently occurring faults throughout the API's evolution. 
    \item[\codebox{\textbf{RQ3:}}] \textit{What level of API coverage does each REST API testing tool achieve across various API releases?}\\
    API coverage serves as an indicator of the effectiveness of the test cases generated by a testing tool. 
    Therefore, the coverage analysis in this RQ demonstrates the API coverage that each REST API testing tool achieves during test execution. 
    These findings could assist practitioners in selecting tools that consistently exhibit high and stable coverage across multiple evolving releases. 
    \item[\codebox{\textbf{RQ4:}}] \textit{What types of potential regressions can be uncovered from the outcomes of various REST API testing tools?}\\
    In this RQ, we investigate the test results generated by each REST API testing tool to find potential regressions in evolving API releases. 
    \item[\codebox{\textbf{RQ5:}}] \textit{What are the cost implications of using different REST API testing tools?}\\
    Testing frequent and rapid releases in DevOps could involve substantial costs, such as overheads associated with API calls. 
    From this RQ, we evaluate the relative cost-effectiveness of various REST API testing tools. 
    This can guide decision-making regarding which tool to adopt, based on a balance of cost and testing effectiveness. 
\end{itemize}

\subsection{REST APIs Under Test}\label{sec:apisinfo}
\Cref{tab:apis} presents the characteristics of all accessible APIs\footnote{APIs for which the access was provided by Oslo City as a part of an experimental apparatus for research purposes. Inaccessible APIs are for sensitive data with privacy concerns.} used in our experiments. 
These APIs are part of the staging environment set up by Oslo City’s industry partner specifically for testing purposes. 
Each API is associated with a specific functionality within the healthcare IoT application under test. 
These APIs consist of multiple endpoints, each utilizing different HTTP request methods and targeting various features. 
As observed in~\Cref{tab:apis}, most endpoints primarily use GET and POST methods for CRUD operations. 
However, requests such as PUT and DELETE, which involve creating and deleting resources, are inaccessible to us to ensure controlled access for testing and prevent unintended changes to the application’s functionality. 
The \textit{Alerts} API has five features for assigning alerts, closing alerts, all alerts, the total number of alerts, and dashboard alerts. 
The \textit{Authentication} API targets different methods for authenticating different users. 
This includes authentication through email, password, SMS, phone number, device's IMEI, token, medical device, and login credentials. 
In addition, it includes features such as password recovery for forgotten passwords, updating login credentials, and creating an emergency profile for a user. 
Three device APIs are related to three medicine dispensers (i.e., Karie~\cite{karie}, Medido~\cite{medido}, and Pilly~\cite{pilly}) that are integrated with the healthcare IoT application. 
Karie API provides features for modifying device settings, creating medication plans, and loading medical plans from pharmacies. 
Medido API enables modifying device settings and creating mock and real medication plans. 
Pilly API only allows for the configuration of device settings. 

The API for \textit{Patients} targets features are related to patients' health information, billings, time logs, health tracking, patients' actions to perform, and reimbursements. 
The \textit{Users} API focuses on user actions, that is, adding notes, creating events, threshold, searching users based on name or birth date, validating a user, updating user information, counting or listing down users, logging user interactions, and matching with existing patients. 
The \textit{Courses} API is concerned with e-learning courses with features for course category, courses list, course modules, linking a course, uploading a new course, and finishing a course. 
The API for \textit{Mobile App} targets features related to app version, app information, patient's timeline, password information, and checking old and new passwords.
The \textit{Reimbursements} API includes features for creating billing summaries, reimbursement time logs, patient reimbursement status, measurements related reimbursements, task identifiers, and tracking reimbursements. 

The API for \textit{User Tasks} provides features for various user tasks, i.e., assigning tasks to the concerned users, patient health surveys, task lists, task reminders, general tasks for administrative purposes, and global surveys for different users. 
The \textit{Invoicing} API provides features for adding, canceling, replacing, terminating, and subscribing to invoices. 
The \textit{Catalog} API supplies catalogs with various types, models, units, features, and lists of catalogs. 
The API for \textit{Reports} provides reports in different formats, patients' reports, patients' medication reports, zone-wise reports, and event-specific reports. 
The \textit{Care} API is related to patient care messages, events, and reminder status.  
The \textit{Development} API is concerned with creating smart alert rules, user profiles, and temporary interface generation.

\begin{table}
  \caption{Characteristics of APIs selected for the experiments, i.e., API label, number of endpoints (EPs), EPs corresponding to HTTP request methods, and features}
  \label{tab:apis}
    \begin{tabular}{p{.14\textwidth}p{.04\textwidth}p{.18\textwidth}p{.55\textwidth}}
    \toprule
    \textbf{API Label}&\textbf{EPs}&\textbf{EPs w.r.t. Requests}&\textbf{Features}\\
    \midrule
    
    Alerts & 5 &2 POST, 3 GET& Assign, Close, All, Total, Dashboard\\ 
    \makecell[l]{Authentication} & 13 &13 POST& Email, Password, SMS, Phone, IMEI, Token, Device, Log-in, Forgot password (Send/Resend), Credentials, Emergency profile\\ 
    Device-Karie & 3 &1 GET+POST, 2 POST& Settings, Plan, Pharmacy \\
    Device-Medido & 3 &2 GET+POST, 1 GET& Settings, Plan, Mock Plan\\
    Device-Pilly & 1 &1 GET+POST& Settings \\
    Patients & 6 &6 POST& Health data, Billings, Time logs, Tracking, Actions, Reimbursements  \\ 
    \makecell[l]{Measurements} & 5 &5 POST& Single \& Multiple inputs (Manual/Auto), Time series \\
    Users & 14 &1 GET, 13 POST, 1 GET+POST& Notes, Events, Threshold, Search methods, Validate, Update, Count/List, Logs, Match patient \\
    Courses & 6 &2 GET, 1 POST, 3 GET+POST& Category, List, Module, Link, Upload, Finish \\
    Mobile App & 6 &2 POST, 4 GET& Version, Information, Timeline, Password info, Check old \& new password \\
    Reimbursements & 11 &3 POST, 8 GET& Billing summary, Time logs, Patients, Measurements, Task identifiers, Track \\
    User Tasks & 9 &9 GET+POST& Assignments, Survey, Tasks, Reminder, General tasks, Global survey \\
    Invoicing & 9 &5 POST, 4 GET& Add, Cancel, Replace, Terminate, Subscriptions (Check, Fix, Aggregate) \\
    Catalog & 8 &8 GET& Types, Models, Units, Features, List \\
    Reports & 10 &1 POST, 9 GET& Formats (HTML, JSON, PDF), Patients, Medication, Zone, Events \\
    Care & 4 &4 GET& Messages, Events, Reminder status  \\
    Development & 7 &7 GET& Smart alert rules, Profiles, Auto UI generation \\
    \midrule
    Total (APIs: 17) & 120 &-& - \\
  \bottomrule
\end{tabular}
\end{table}

\subsection{Testing Tools Selection}
\Cref{tab:tools} presents information for REST API testing tools selected for our experiments, including details such as tool name, version, release date, and applied testing techniques. 
The most recent version of each tool was picked as of May 10, 2024. 
To select tools for our experiments, we first analyzed empirical studies focusing on tool comparison. 
Drawing from the recommendations of effective and widely-used tools in previous studies~\cite{golmohammadi2022testing,zhang2023open,kim2022automated}, we initially shortlisted eight tools for our experiments, including RESTest, EvoMaster, RESTler, RESTCT, RestTestGen, bBOXRT, Morest, and Schemathesis. 
We later expanded our list to include TnT-Fuzzer, Got-Swag, RapiTest, QuickREST, and three recent tools: NLPtoREST, ARAT-RL, and foREST. 
Initially, we configured all tools to test the REST APIs of our application under test. 
Upon running these tools, we encountered various compatibility issues and crashes.
After diagnosing these problems, we identified their common causes. 
Based on our investigation and insights from previous empirical studies~\cite{zhang2023open, kim2022automated}, we formulated a set of criteria required for a tool to qualify for testing our application under test. 
Note that resolving issues related to the tools could potentially introduce an internal validity threat, which also falls outside the scope of our paper.
Next, we outline the criteria and elaborate on our systematic approach to the tool selection process.

\begin{itemize}
    \item[\textbf{C1.}] Tools that support OpenAPI specification version 3.0, ensuring compatibility with the REST APIs of our healthcare IoT application under test. 
    \item[\textbf{C2.}] Well-established tools that have undergone continuous development phases, showcasing a commitment to improvement through the addition of new features, bug fixes, and the release of stable versions. 
    \item[\textbf{C3.}] Tools applied to numerous cross-domain industrial systems and support execution on multiple platforms.
    \item[\textbf{C4.}] Research-based tools derived from various research works with comprehensive and well-founded techniques (similar to earlier works~\cite{zhang2023open, kim2022automated}). 
\end{itemize}

Among all available REST API testing tools, RESTCT only supports the OpenAPI specification version 2.0 (violating \textbf{C1}). 
Tools like TnT-Fuzzer and Got-Swag, as referenced in~\cite{hatfield2022deriving}, also support OpenAPI version 2.0, thereby violating \textbf{C1}. 
The public version of Morest only provides a replication package, and its repository was last updated on September 13, 2021 (violating \textbf{C2}). 
bBOXRT provides an initial version as a replication package, last updated on January 30, 2020 (violating \textbf{C2}). 
In addition, bBOXRT requires an input API file in Java, which is not available in our case. 
In a similar line of work, RapiTest and QuickREST only provide replication packages, violating \textbf{C2}.  
Recent tools NLPtoREST~\cite{kim2023enhancing}, ARAT-RL~\cite{kim2023adaptive}, APIRL~\cite{foley2025apirl}, and LlamaRestTest~\cite{kim2025llamaresttest} provide replication packages, are still in initial development stages, with NLPtoREST, ARAT-RL, and LlamaRestTest necessitating execution on Ubuntu (violating \textbf{C2} \& \textbf{C3}). 
The tool link for another recent black-box testing tool, foREST~\cite{lin2023forest}, is also unavailable. 
Moreover, foREST is in early development phases, violating \textbf{C2}.  
Tools like Tcases, APIFuzzer, and Dredd are open-source tools without a research background, violating criterion \textbf{C4}. 
After careful inspection, we selected \RESTest{}, \EvoMaster{}, \Schemathesis{}, \RESTler{}, and \RestTestGen{} for our experiments. 
Below is a brief overview of these selected REST API testing tools. 

\textbf{\RESTest{}}~\cite{martin2021restest} is an automated tool designed to test REST APIs of web applications, which work mainly in a black-box testing mode. 
This tool utilizes the API specification and test configurations as input to generate test cases. 
For test generation, \RESTest{} supports various testing techniques, including constraint-based testing (CBT), adaptive random testing (ART), random testing (RT), and fuzz testing (FT).
In addition, \RESTest{} offers advanced features such as online testing in a production environment, handling inter-parameter dependencies, and generating realistic test data.
These features are enabled through various techniques such as the CBT approach for handling inter-parameter dependencies~\cite{martin2020restest,martin2021specification}. 

\textbf{\EvoMaster{}}~\cite{arcuri2018evomaster} is a Web API testing tool that employs evolutionary techniques to generate test cases. 
It supports black-box and white-box testing modes and is compatible with various API types, including REST and GraphQL. 
In black-box testing, \EvoMaster{} takes the API specification as input. White-box testing requires additional inputs, including source code and a driver.
For black-box mode, \EvoMaster{} employs fuzzing techniques for test generation, while in white-box mode, it utilizes many objective search algorithms for test generation.

\textbf{\Schemathesis{}}~\cite{hatfield2022deriving} is a web API fuzz testing tool that utilizes property-based testing for generating test cases. 
It operates in black-box and white-box testing modes and is compatible with APIs specified in OpenAPI or GraphQL formats. 
This tool is designed to test cross-platform web APIs, irrespective of the programming language used for development. 
In addition, \Schemathesis{} has been used in various industrial projects, thereby streamlining the process of automated REST API testing.

\textbf{\RESTler{}}~\cite{atlidakis2019restler} is a web services fuzz testing tool mainly focusing on identifying security and reliability faults in REST APIs. 
It takes an API specification in OpenAPI format, analyzes the specification to identify dependencies among requests, generates fuzz tests, and executes tests on the API to explore web services. 
Based on responses originating from various test executions, it learns distinct bug patterns. 
It supports two types of fuzzing, i.e., fuzz and fuzz-lean.

\textbf{\RestTestGen{}}~\cite{corradini2022resttestgen} is a REST API testing tool with a primary focus on black-box testing.
It supports two testing techniques: nominal and error testing, and mass assignment security testing. 
\RestTestGen{} offers an extensible interface that facilitates adding new features and implementing novel testing techniques. 
In this context, a recent tool, NLP2REST~\cite{kim2023enhancing}, is integrated with \RestTestGen{}, leveraging natural language processing techniques to enhance functionality in API testing scenarios.

\begin{table}
    \centering
    \caption{REST API testing tools used in the experiments, comprising details like the tool's name, version, release date, and applied testing techniques} 
    \label{tab:tools}
    \begin{tabular}{p{.15\textwidth}p{.1\textwidth}p{.15\textwidth}p{.5\textwidth}}
    \toprule
    \textbf{Tool}&\textbf{Version}&\textbf{Release Date}&\textbf{Testing Techniques}\\
    \midrule
       \RESTest{}  & 1.4.0 &Nov. 12, 2023& CBT, ART, RT, FT\\ 
       \EvoMaster{}  & 3.0.0 &Apr. 05, 2024& Fuzzing\\
       \Schemathesis{}  & 3.27.1 &Apr. 29, 2024& Property-based testing\\
       \RESTler{}  & 9.2.4 &Mar. 06, 2024& Fuzz, Fuzz-lean\\
       \RestTestGen{}  & 24.09 &Mar. 30, 2024& Nominal and error testing, Mass assignment security testing\\
    \bottomrule
    \end{tabular}
\end{table}

\begin{figure}[htbp]
\centerline{\includegraphics[width=\linewidth, keepaspectratio]{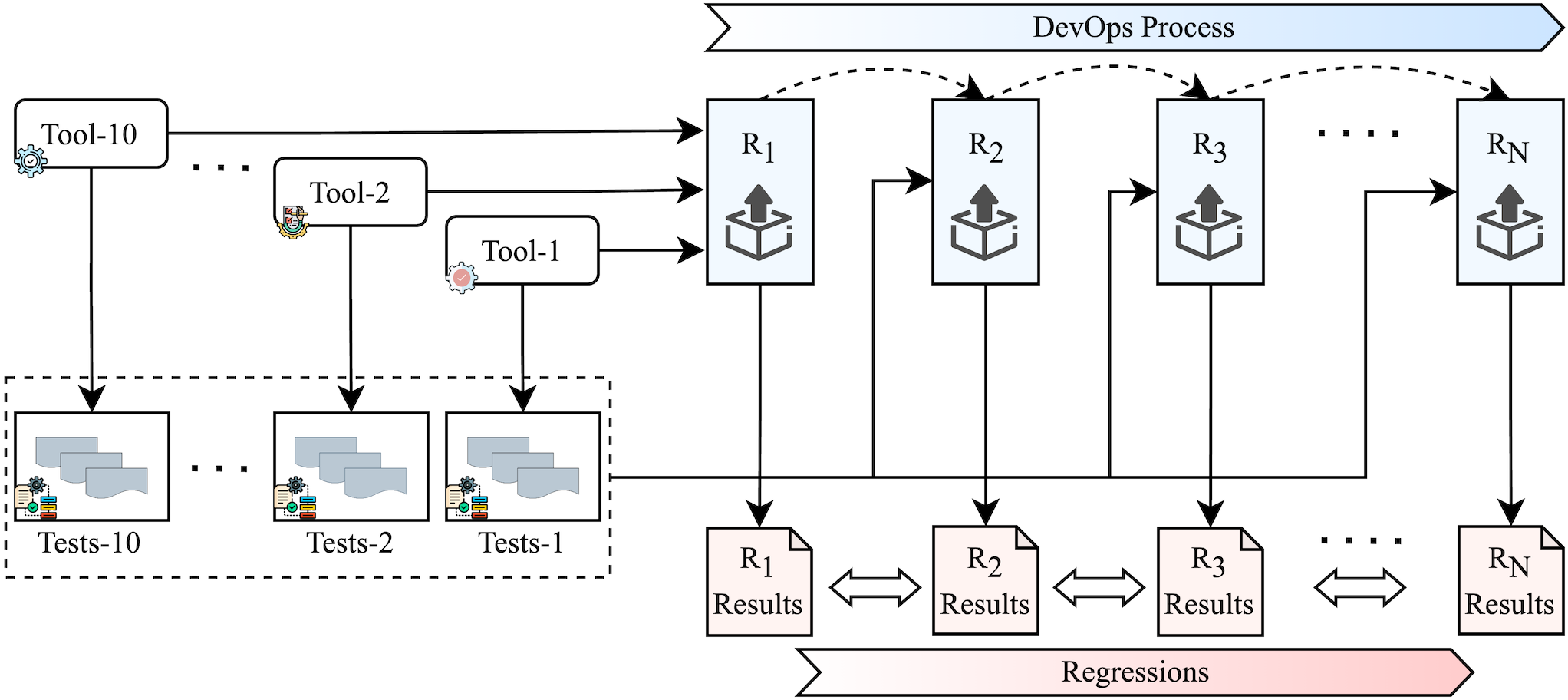}}
\caption{Workflow of experiments design, setup, and execution for testing multiple releases of REST APIs, a healthcare IoT application continuously developed during the DevOps process.}
\label{fig:expdesign}
\end{figure}

\subsection{Experiments Design}
We designed our experiments to test the REST APIs of a real-world healthcare IoT application that continually evolved in DevOps. 
In the following, we elaborate on the details of the preparation and execution of the experiment.

\subsubsection{Experiment Preparation}
Our experiments involve five REST API testing tools, each supporting one or more testing techniques. 
For each testing mode, we set up different configurations of these tools. 
\RESTest{} configured with the CBT, ART, RT, and FT testing techniques is termed \RESTestCBT{}, \RESTestART{}, \RESTestRT{}, and \RESTestFT{}, abbreviated as \RTCBT{}, \RTART{}, \RTRT{}, and \RTFT{}, respectively. 
\EvoMaster{} configured with back-box testing mode is termed \EvoMasterBB{} (in short \EMBB{}). 
\Schemathesis{} configured with property-based testing mode is termed \SchemathesisPT{} (in short \STPT{}).  
\RESTler{} configured with fuzz and fuzz-lean testing modes is termed \RESTlerF{} and \RESTlerFL{}, abbreviated as \RLFz{} and \RLFL{}, respectively. 
\RestTestGen{} configured with nominal and error testing and mass assignment security testing is termed \RestTestGenNET{} and \RestTestGenMAST{}, abbreviated as \RTGNET{} and \RTGMAST{} respectively. 
These configurations result in a total of 10 tools for our experiments.

We set up our experiments based on information from Oslo City's industry partner about planned releases, i.e., daily, weekly, and monthly. 
Our experiments were set up to use 14 distinct releases of our application under test, denoted as R1 through R14.
This includes eight daily and five weekly builds in the staging environment and one monthly production release. 
The releases were structured as follows: R1, R4, R7, R10, and R13 were weekly releases; R2, R3, R5, R6, R8, R9, R11, and R12 were daily releases; finally, R14 was a monthly release.
Note that the specific dates corresponding to the releases on which we conducted our experiments are not disclosed to maintain the confidentiality of Oslo City's industry partner.

\Cref{fig:expdesign} shows the design of the experiment. 
We started with the first weekly release (R1) and ran all testing tools on that release. 
For each tool's execution on R1, we set up a proxy server to collect HTTP requests generated by each tool and responses obtained from the APIs. 
Note that instead of relying on third-party proxy solutions, which could introduce execution overhead or raise data privacy concerns for the industry partner, we developed a lightweight proxy to ensure minimal interference and precise data capture. 
To enable consistent test replay across subsequent releases, we also implemented a test runner that sequentially picks up each HTTP request initially generated by a tool, sends it to the appropriate API endpoint, and records the resulting response. 
\textcolor{black}{
Since the APIs under test only use GET and POST methods, as shown in \Cref{tab:apis} and discussed in \Cref{sec:apisinfo}, managing dynamic data dependencies like resource IDs---typically required for methods such as CREATE and DELETE---was not applicable in our study. 
Therefore, we did not implement adaptations to handle such dependencies in the replay test runner. 
However, in scenarios involving methods like CREATE and DELETE, a potential approach could be to maintain a dynamic map to track resource creation and deletion, ensuring consistent reuse of resource IDs across operations, regardless of the request execution order. 
}
This allowed us to reuse the same set of test cases across all subsequent releases (R2--R14), ensuring consistency in regression testing across evolving versions. 
Reusing the same test set is essential to the inherent randomness factor involved in each testing tool, leading to potential variability in the tests generated at different times.
For example, the tests generated during the initial run may differ from those generated during subsequent runs. This variability could hinder the detection of regressions in evolving releases. 
Therefore, to effectively identify regressions across different releases, we used the set of tests generated for R1 to test releases R2--R14. 
This approach to regression testing is commonly referred to as the \emph{retest-all} strategy in literature~\cite{yoo2012regression,greca2023state}. 
We chose this strategy because it does not necessitate a test selection, minimization, or prioritization approach to determine which specific test cases should be rerun. 
To the best of our knowledge, no such approach exists for REST API test cases. 
Furthermore, adapting existing test selection, minimization, or prioritization approaches for test cases created for REST APIs could potentially introduce significant bias, such as implementation bias~\cite{sartaj2024uncertainty}. 
This also deviates from the primary focus of our study.

One of the potential implications of using the \emph{retest-all} strategy in our experiments is simplicity in test generation and consistency in test execution across releases, which aligns with our experimental goals. 
However, other potential implications include increased resource consumption and redundancy due to rerunning all tests. 
It is important to note that for the APIs under test, we lacked prior knowledge of planned feature changes (e.g., additions, removals, or updates) across releases. 
Without such information and the availability of automated test selection, minimization, or prioritization approaches, we relied on the \emph{retest-all} strategy for regression testing. 
This involved generating automated test cases once for R1 using each tool and reusing them for subsequent releases. 
If some tools supported these functions, our evaluation could have been designed accordingly. 
Instead of rerunning all tests from R1, we could have used test selection or minimization techniques to execute a targeted or reduced set of tests for each subsequent release, reducing resource consumption and avoiding redundancy. 
Another important consideration is the maintenance of the test suite, which requires regular review and evolution to align with API changes during DevOps. 
Although developing a dedicated approach for automated test suite maintenance is essential, it falls beyond the scope of our study and presents a promising direction for future research.

\subsubsection{Experiment Execution}
We scheduled our experiments for each daily build to run overnight. 
For the weekly release, experiments were set to run over the weekend. 
The experiment for the monthly release was also arranged to run overnight, considering they were scheduled during the workweek.
The overall duration of our experiments was more than a month, specifically lasting 4 weeks and 5 days. 
We executed all experiments on one machine with specifications of a four-core 3.6 GHz CPU, 32 GB RAM, and Windows 10 operating system.

To account for the inherent randomness of REST API testing tools, we executed each tool 10 times in parallel, with each run limited to one hour. 
Our parallel execution setup was inspired by the existing work on tool comparison~\cite{zhang2023open}. 
Specifically, we configured eight threads to run eight parallel jobs using eight logical processors, ensuring that each run adhered to the one-hour limit and was forcibly terminated if it exceeded this duration. 
During execution, we observed that all tools consistently generated tests and utilized the full time budget, except \RestTestGenMAST{}, which did not generate any tests (as shown in~\Cref{tab-tests}) and terminated early without errors. 
Since no tests were produced (as detailed in~\Cref{sec:resultsdicussion}), rerunning \RestTestGenMAST{} to fully utilize the allocated time---similar to the setup in~\cite{zhang2023open}---would have unnecessarily consumed resources. 
Therefore, the freed-up resources became available for other tools.

Running tools in parallel continuously can potentially lead to APIs becoming overloaded. 
To avoid this problem, we coordinated with Oslo City’s team and their industry partner while planning our experiments. 
They provided access to a dedicated staging environment, likely similar to the one used internally for testing. 
We communicated our testing schedule, estimated execution durations, and expected API call overload. 
During planned testing periods, they configured the API servers to efficiently manage the traffic generated from our designated test account to ensure smooth execution without affecting other system functionalities or users. 
Furthermore, our experiments involve three real medical devices in the loop. 
Given the constraints on the ability to handle requests from these devices and the potential damage to the device~\cite{sartaj2023hita,sartaj2024modelbased}, we introduced a delay of three seconds for APIs that require interaction with medical devices. 
This execution setup for device APIs reduced API overload and aligns with the setup used in our previous work~\cite{sartaj2023testing}. 

Throughout the execution of our experiments, we continuously compiled tests generated by each REST API testing tool. 
This data comprised comprehensive details of the requests generated by each testing tool and the corresponding responses returned from the APIs under test. 
At the end of each experiment run, we collected and reviewed the data, and then stored it in an organized manner corresponding to each release, tool, and API.
The data collection process was designed to facilitate data analysis and interpretation for each RQ. 

\begin{table}[!t]
	\centering
    \small
	\noindent
	\caption{Evaluation aspect and analyses metrics, statistical tests, and measures corresponding to each RQ}
    \begin{tabular}{p{.1\textwidth}p{.1\textwidth}p{.36\textwidth}p{.15\textwidth}p{.17\textwidth}}
        \toprule
		\multicolumn{1}{ l }{\textbf{RQ}} & \textbf{Aspect} & \textbf{Analysis Metrics} & \textbf{Statistical Tests}& \textbf{Measures}\\
		\cmidrule(lr){1-1}\cmidrule(lr){2-2}\cmidrule(ll){3-3} \cmidrule(ll){4-4} \cmidrule(ll){5-5} 
		\multicolumn{1}{l}{RQ1} &Failure& Gestalt similarity, Total failures, Ratio & Wilcoxon test& Vargha-Delaney \^{A}${}_{12}$\\
		\multicolumn{1}{l}{RQ2} &Fault& Gestalt similarity, Unique faults, Ratio, APFD&--&--\\
        \multicolumn{1}{l}{RQ3} &Coverage& Percentage coverage& Wilcoxon test& Vargha-Delaney \^{A}${}_{12}$\\
        \multicolumn{1}{l}{RQ4} &Regression& Gestalt similarity, Unique regressions&--&--\\
        \multicolumn{1}{l}{RQ5} &Cost& Percentage cost&--&--\\
		\bottomrule
	\end{tabular}
	\label{tab:rqmapping}
\end{table}

\begin{figure}[htbp]
\centerline{\includegraphics[width=12cm, keepaspectratio]{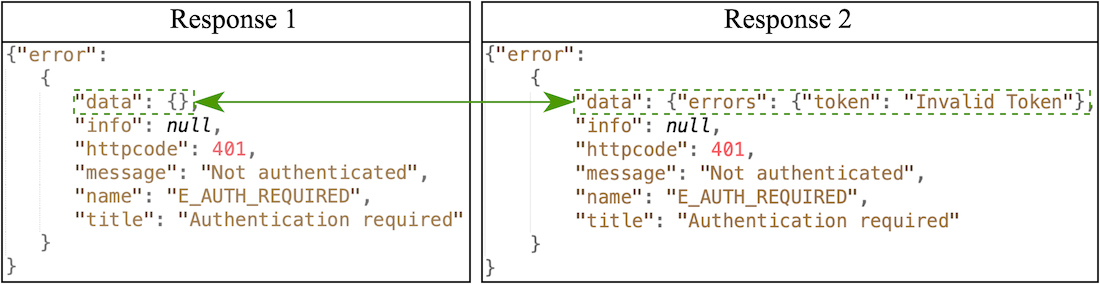}}
\caption{Example comparison of error responses in JSON format, highlighting one property with different values while other property values are the same.} 
\label{fig:responsecomp}
\end{figure}

\subsection{Data Analysis and Interpretation}
\Cref{tab:rqmapping} presents analysis metrics, measures, and statistical tests for each RQ, along with its corresponding evaluation aspect. 
To analyze RQ1 results related to failures, we computed total API failures representing server-side (5XX) errors. 
We also calculate failure ratios using~\Cref{eq1}, where $N_{failures}$ represents the total number of failures, $T$ denotes the total number of tests generated by a tool.

To analyze RQ2 results concerning faults, we followed a semi-automatic approach inspired by Martin-Lopez et al.~\cite{martin2022online} to discover potential faults. 
Initially, we grouped all failure and success responses (string data) based on their similarity score, calculated using the Gestalt algorithm with the same threshold (i.e., 0.5) set for RQ1. 
We chose to include the success responses in this process, as our previous work has shown that even successful responses can potentially lead to faults~\cite{sartaj2021testing}.
To calculate the similarity between responses (string data), we used the Gestalt approach for string matching proposed by Ratcliff and Metzener~\cite{ratcliff1988pattern}, implemented in a Python library named \emph{difflib}~\cite{python3difflib}. 
Given that the responses obtained in our study contain data in JSON and HTML formats, we opted for the Gestalt string matching algorithm, considering its application for code comparison~\cite{tsikerdekis2018persistent,gabel2010scalable}. 
The Gestalt string matching algorithm calculates the similarity between two strings on a range from 0 to 1, where 0 indicates no similarity and 1 denotes a 100\% similarity~\cite{tsikerdekis2018persistent,python3difflib}. 
To identify unique responses in our study, we set a threshold of 0.5 (or 50\%) as a balanced midpoint on the similarity scale. 
Responses with a similarity score below this threshold are considered to exhibit low similarity, while those above are deemed highly similar. 
This threshold is particularly relevant in contexts where structural similarity is important, such as pattern matching and code or data comparisons~\cite{ratcliff1988pattern,tsikerdekis2018persistent,gabel2010scalable}. 
Our choice was inspired by threshold values previously determined through experiments in code similarity research~\cite{gabel2010scalable}. 
The potential implications of using this threshold include tolerance for minor variations, handling structural similarities, and flexibility in determining what constitutes a unique response. 
We illustrate the usage of Gestalt string matching in our context with an example shown in~\Cref{fig:responsecomp}, where two error responses in JSON format differ only in a single property value. 
Although the \emph{data} property in Response 2 contains a slightly longer string, the similarity score computed using the Gestalt algorithm is 0.89, which exceeds the defined threshold and indicates a high similarity. 
This is primarily due to the Gestalt algorithm's ability to intuitively compare text, manage minor variations, and handle structural similarities of JSON responses. 
In another case, when a property such as \emph{data} in Response 2 contains a significantly longer and differing string compared to Response 1, Gestalt may return a similarity score close to or below the threshold. 
Therefore, as a follow-up step, we conducted manual inspections on all Gestalt similarity results to ensure the accuracy of the responses' uniqueness.

After calculating similarity scores, we created clusters for different groups by parsing the JSON and HTML responses to obtain information about failures, such as error messages. 
Lastly, we carefully inspected the resulting list of potential faults and assigned each a unique identifier (ID) to facilitate the precise interpretation and reporting of these faults. 
In addition, we calculate fault ratios using~\Cref{eq1}, where $N_{faults}$ represents the total number of unique faults and $T$ denotes the total number of tests generated by a tool. 
To further analyze fault detection effectiveness of testing tools over the test suite lifecycle, we calculate the Average Percentage of Faults Detected (APFD), originally introduced by Elbaum et al.~\cite{elbaum2002test} and later applied in a similar context by Martin-Lopez et al.~\cite{martin2022online}.

To analyze coverage results for RQ3, we used Restats~\cite{corradini2021restats} to calculate the coverage (numerical data) of test cases generated by each REST API testing tool. 
The coverage reported by Restats encompasses different aspects, including paths, operations, status classes, statuses, response types, request types, parameters, and parameter values. 
For our analysis, we evaluated the overall coverage achieved by each testing tool for all releases.  
Since we lack access to the source code, we rely on the black-box coverage criteria proposed by Martin-Lopez et al.~\cite{martin2019test} and implemented in Restats.

To analyze RQ4 results related to regressions (string data), we first assembled responses into pairs of two consecutive releases, such as R1-R2, R2-R3, etc. 
Subsequently, we compared the response status codes of each pair (for example, R1 and R2) to analyze any discrepancy between the status codes of two consecutive releases. 
If two status codes differed, we further compared the detailed responses of each pair by calculating their similarity using the Gestalt algorithm with the same threshold set for RQ1. 
Finally, we thoroughly inspected the set of potential regressions and marked each with a unique identifier (ID) to differentiate regressions while reporting results.
It is important to note that comparing differences between releases was impossible for us due to the lack of access to API implementations in various releases. 
Consequently, our regression analysis is based on API responses obtained during test executions across these releases.

In the cost analysis for RQ5, we initially computed the failure and fault detection rates (numerical data) and subsequently calculated the percentages of cost overhead. 
We utilized \Cref{eq1} to calculate the failure detection rates, and \Cref{eq2} for computing fault detection rates. 
In these equations, $N_{failures}$ represents the total number of failures, $T$ denotes the total number of tests generated by a tool, and $M_{faults}$ symbolizes the total number of unique faults identified from failure results. 
To compute the cost overhead for the rates corresponding to failures or faults, we use \Cref{eq3}. 
In this equation, we subtract the rates associated with failures or faults from 100. 
The primary objective is to determine the percentage of tests that failed to identify failures or faults, thus representing the overhead. 
In DevOps environments with rapid and frequent releases, it becomes important for practitioners to understand how effectively their testing budget is used, specifically in terms of whether tests contribute to detecting failures or faults. 
For example, consider a scenario where the entire testing budget (100\%) is allocated to execute 300 tests for a specific release. 
If 12 failures are detected from 300 test cases, a failure detection ratio of 4\% is obtained with~\Cref{eq1}. 
Subsequently, applying~\Cref{eq3}, the cost overhead is calculated as 100–4=96\%, indicating that 96\% of the tests did not reveal any failures. 
This measurement allows practitioners to estimate testing cost overheads and make better decisions about carefully selecting or prioritizing tests, which is especially useful during continuous delivery cycles in DevOps.

\begin{equation}\label{eq1}
Ratio_{failure} = \left( \frac{N_{failures}}{T} \right) \times 100
\end{equation}

\begin{equation}\label{eq2}
Ratio_{fault} = \left( \frac{M_{faults}}{T} \right) \times 100
\end{equation}

\begin{equation}\label{eq3}
Cost = 100 - Ratio_{failure/fault}
\end{equation}

To statistically compare REST API testing tools for RQ1 and RQ3, we first assessed (numerical) data normality using the Shapiro-Wilk test~\cite{shapiro1965analysis}, which indicated non-normal distributions with p-values close to 0. 
As a result, we used non-parametric tests, specifically the Wilcoxon signed-rank test to identify statistically significant differences, and the Vargha-Delaney \^{A}${}_{12}$ effect size measure to analyze which tool performed better, following the recommended guidelines by Arcuri and Briand~\cite{arcuri2011practical}. 
In addition, we applied Holm’s method~\cite{holm1979simple} to adjust p-values for multiple comparisons, using a significance level of $\alpha = 0.05$. 
The Holm method is a stepwise correction technique that is known for a higher statistical power compared to methods such as Bonferroni correction~\cite{wagner2021code}, which makes it well-suited for our experimental analysis. 
For RQ2 and RQ4, we present the results descriptively, focusing on discussions related to potential faults and regressions, similar to existing studies~\cite{martin2022online,godefroid2020differential}.

\begin{table}[htbp]
    \centering
    \noindent
    \caption{Number of test cases generated by different REST API testing tools for each API under test}
    \begin{tabular}{@{} cl*{10}l @{}}\toprule
        & & \rot{\RESTestCBT{}} & \rot{\RESTestART{}} & \rot{\RESTestRT{}} & \rot{\RESTestFT{}} 
        & \rot{\EvoMasterBB{}} & \rot{\SchemathesisPT{}} & \rot{\RESTlerF{}} 
        & \rot{\RESTlerFL{}} & \rot{\RestTestGenNET{}} & \rot{\RestTestGenMAST{}} \\

        \cmidrule{2-12}
        & \textbf{Alerts}             &400&400&400&401&7000&2101&731&392&800&0\\
        & \textbf{Authentication}     &1308&1308&1308&1308&5511&2534&1892&1769&2440&0\\
        & \textbf{Device-Karie}       &210&210&210&210&903&101&649&649&800&0\\
        & \textbf{Device-Medido}      &0&455&0&0&9377&1837&860&860&1000&0\\
        & \textbf{Device-Pilly}       &238&238&238&22&8802&102&395&398&400&0\\
        & \textbf{Patients}           &545&545&545&471&5191&168&868&868&964&0\\
        & \textbf{Measurements}       &510&510&510&510&5803&2136&4217&4734&1000&0\\
        & \textbf{Users}              &574&0&574&0&8765&1363&476&476&3771&0\\
        & \textbf{Courses}             &729&729&729&0&9740&78&423&418&0&0\\
        & \textbf{Mobile App}          &528&528&528&528&9746&43&937&608&1020&0\\
        & \textbf{Reimbursements}      &1182&1182&1182&1182&10393&370&1861&1861&1924&0\\
        & \textbf{User Tasks}         &308&0&308&0&4719&733&255&255&2029&0\\
        & \textbf{Invoicing}           &729&729&729&0&8448&493&1409&1405&1620&0\\
        & \textbf{Catalog}             &592&592&592&592&11704&16&188&13&1060&0\\
        & \textbf{Reports}             &689&689&689&301&8089&205&1129&1129&1512&0\\
        & \textbf{Care}                &565&565&565&429&3160&6&20&2&600&0\\
        & \textbf{Development}         &0&0&0&0&8592&9&1434&4&340&0\\
        \cmidrule{2-12}
        & \textbf{Total}               &9107&8680&9107&5954&125943&12295&17744&15841&21280&0\\

        \bottomrule
    \end{tabular}
    \label{tab-tests}
\end{table}

\subsection{Results Discussion}\label{sec:resultsdicussion}
\Cref{tab-tests} presents the number of tests each tool generated for every API. 
It can be observed that \EvoMasterBB{} consistently generated tests for every API and produced the greatest number of tests compared to all other tools. 
In addition, \RestTestGenNET{}, \RESTlerF{}, \RESTlerFL{}, and \SchemathesisPT{} were ranked second, and four variants of \RESTest{} were ranked third in terms of the total number of tests generated. 
From the results, it is noticeable that \RestTestGenMAST{} did not generate any tests for all APIs, even after successful runs. 
Given that \RestTestGenMAST{} is primarily designed to test for security vulnerabilities, it only generates tests upon identifying potential vulnerabilities (as mentioned in Github issue\footnote{https://github.com/SeUniVr/RestTestGen/issues/17}). 
This could potentially be attributed to the presence of read-only parameters in the API schema. 
However, we found that the APIs accessible to us do not contain such parameters.  
Alternatively, this could be related to the API implementation, which remains a black box to us due to our lack of access to the source code.

In subsequent subsections, we discuss results associated with failures, faults, coverage, regressions, and cost overhead corresponding to each RQ. 
For a discussion on the results, we refer to \Cref{tab-rq1results,tab-rq1results-4xx,tab-rq1results-agg,tab-stcomp-rq1} and \Cref{fig:rq1failures} for RQ1 concerning failures, \Cref{tab-rq2results,tab-rq2results-agg,tab-rq2faultsinfo} and \Cref{fig:rq2faults,fig:apfd} for RQ2 pertaining to faults, \Cref{tab-rq3results,tab-rq3results-agg,tab-stcomp-rq3} for RQ3 associated with coverage, \Cref{tab-rq4results,tab-rq4results-agg,tab-rq4regsinfo} and \Cref{fig:rq4regressions} for RQ4 related to regressions, and \Cref{tab-rq5cost2,tab-rq5cost3} for RQ5 results concerning cost overhead. 
Note that while discussing the experimental results for each RQ, we maintain a balance between conveying essential information and omitting specific details to prevent disclosing sensitive information that could potentially affect the confidentiality and privacy of Oslo City's industry partner.

\subsubsection{RQ1 Results -- REST API Failures}\label{RQ1Results}

\begin{figure}[H]
\centerline{\includegraphics[width=13cm, keepaspectratio]{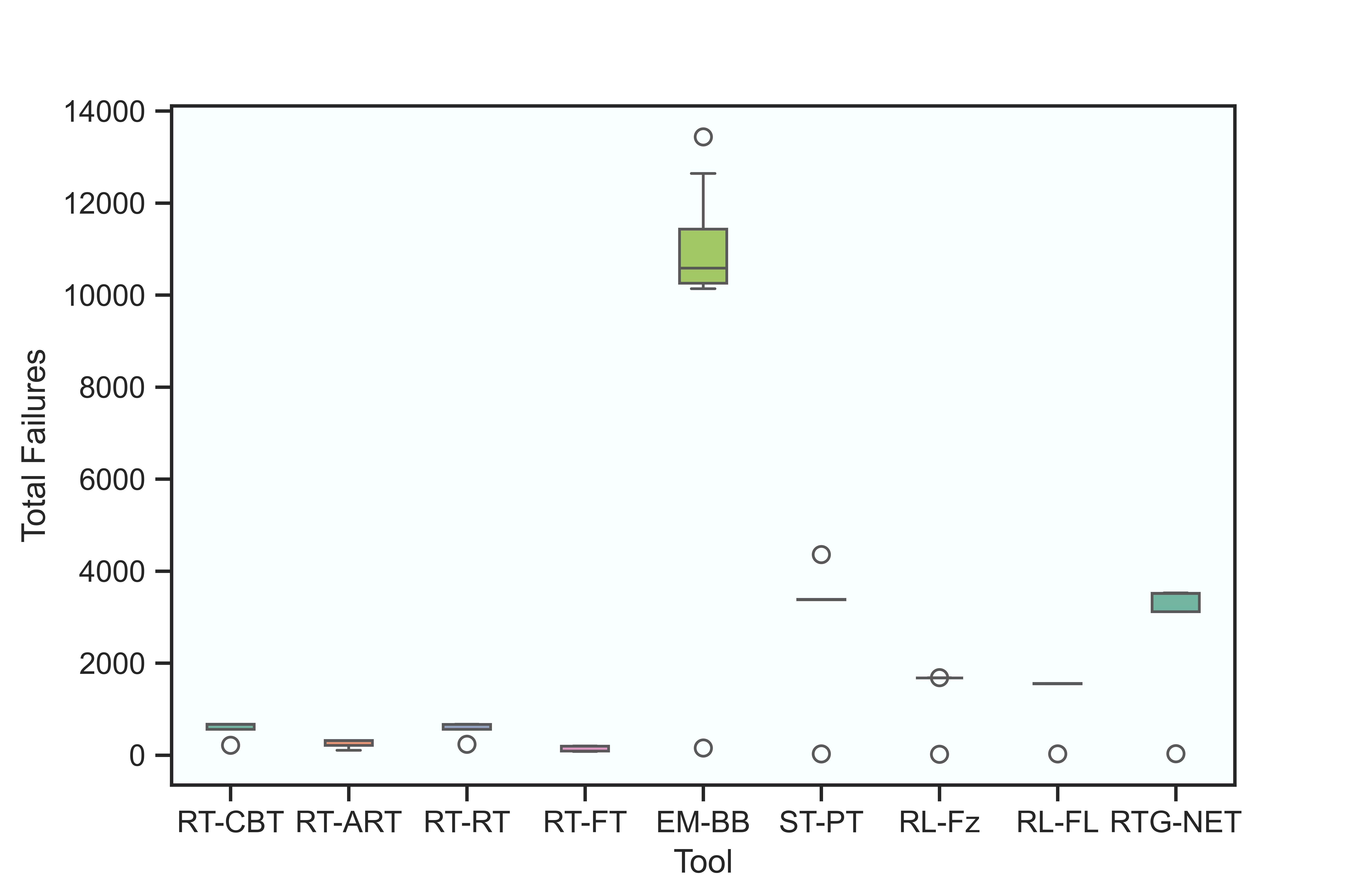}}
\caption{RQ1 results: Boxplot representing the median number of API 5XX failures detected by each tool across all releases and all runs. }
\label{fig:rq1failures}
\end{figure}

From the results in \Cref{tab-rq1results}, it can be observed that \EvoMasterBB{}, which generated the most tests, also led to the highest number of API failures compared to other tools. 
The second highest number of failures was detected in the case of \SchemathesisPT{} and \RestTestGenNET{}. 
This is also evident from \Cref{fig:rq1failures} that \EvoMasterBB{} has the highest median number of failures, while \SchemathesisPT{} and \RestTestGenNET{} are in the second position. 
Moreover, the median number of detected failures remains comparable across all variants of \RESTest{}, with \RESTlerF{} and \RESTlerFL{} slightly exceeding the \RESTest{} variants in failure detection. 
When examining the failure detection ratios, \EvoMasterBB{} shows lower failure detection ratios compared to \SchemathesisPT{} and \RestTestGenNET{}. 
Among these, \SchemathesisPT{} demonstrates the highest failure detection ratios, outperforming all other tools. 
The analysis of the overall results presented in \Cref{tab-rq1results-agg} shows that \EvoMaster{} maintains the highest total failures, but lower failure detection ratios. 
This is mainly because \EvoMaster{} generated the highest number of tests, with only a small fraction resulting in failures. 
\textcolor{black}{
Similarly, while \RESTler{} and \RestTestGen{} show comparable total failures, failure detection ratios for \RESTler{} are lower due to the high number of tests generated by its two variants \RESTlerF{} and \RESTlerFL{}. 
In the case of \RESTest{}, the failure detection ratios are lower compared to all other tools, except for the release R1. 
}

When comparing releases from R1 to R14, it is evident that all tools detected fewer failures in R1 than in subsequent releases. 
Notably, only the tests generated by \RESTestRT{} and \RESTestCBT{} led to identifying a high number of failures in R1, while \RESTlerF{} demonstrated the lowest failure detection. 
As shown in the overall tool comparison in \Cref{tab-rq1results-agg}, \RESTest{} detected the highest number of failures in R1. 
Another observation is that tests generated by some tools resulted in a higher number of client-side (4XX) errors in R1. 
This is especially noticeable with \SchemathesisPT{}, \RESTlerF{}, \RESTlerFL{}, and \RestTestGenNET{}, where the ratio of 4XX errors exceeds 90\%. 
From release R2 onward, failure detection by each tool remained relatively consistent with minor variations. 
Our observations revealed that a significant change in the number of failures from R1 to R2 was primarily due to a shift in the test environment. 
Release R1, in the staging environment, was transitioned from the prior release in the production environment. This environmental shift induced instability in R1, resulting in a lower detection of failures and higher counts of 4XX errors in the releases from R2 onward.

The results of the statistical comparison of all REST API testing tools, based on their ability to detect failures, are presented in \Cref{tab-stcomp-rq1}.
The comparative analysis reveals that \RESTestCBT{} outperformed both \RESTestART{} and \RESTestFT{}, 
while its failure detection performance was comparable to \RESTestRT{} because no statistically significant difference was observed between them. 
While \RESTestART{} managed to outperform \RESTestFT{}, it falls short when compared to \RESTestRT{}.   
\RESTestFT{}, on the other hand, was unable to surpass any of the tools. 
Notably, \EvoMasterBB{} demonstrated superior performance over all REST API testing tools.
In the case of \SchemathesisPT{}, it outperformed all variants of \RESTest{}, including \RESTestCBT{}, \RESTestART{}, \RESTestRT{}, and \RESTestFT{}, as well as \RESTlerF{} and \RESTlerFL{}. 
However, no statistically significant difference was observed between \SchemathesisPT{} and \RestTestGenNET{}. 
\RESTlerF{} performed well compared to five tools, i.e., \RESTestCBT{}, \RESTestART{}, \RESTestRT{}, \RESTestFT{}, and \RESTlerFL{}, whereas \RESTlerFL{} only outperformed all variants of \RESTest{}. 
Lastly, \RestTestGenNET{} performed better than all tools except \EvoMasterBB{}.

\vspace{5pt}
\begin{rqres}{rq1res}
The tests generated using \EvoMasterBB{} lead to the highest number of failures compared to the other REST API testing tools. 
The overall statistical comparison also shows that \EvoMasterBB{} outperformed all other tools, indicating its higher failure detection effectiveness. 
Furthermore, tests obtained from each tool mostly led to client-side errors, suggesting that these tools often produced requests with invalid data, format, or headers. 
This observation highlights the need to improve data generation capabilities within tools and ensure robust input validation mechanisms on the API side. 
\end{rqres}

\begin{table}[H]
    \noindent
    \tiny
    \centering
    \caption{RQ1 results: Total number of API 5XX failures detected by each tool and their ratios across all APIs from releases R1 to R14, derived from all runs. Releases tagged with D, W, and M, denote daily, weekly, and monthly releases, respectively. }
    
    \begin{tabular}{@{}l l l l l l l l l l l l l l l l l l l@{}}
    \toprule
      \multicolumn{1}{l }{\textbf{}} & \multicolumn{2}{c }{\textbf{\RTCBT{}}} & \multicolumn{2}{c }{\textbf{\RTART{}}}& \multicolumn{2}{c }{\textbf{\RTRT{}}} & \multicolumn{2}{c }{\textbf{\RTFT{}}} & \multicolumn{2}{c }{\textbf{\EMBB{}}} & \multicolumn{2}{c }{\textbf{\STPT{}}}& \multicolumn{2}{c }{\textbf{\RLFz{}{}}} & \multicolumn{2}{c }{\textbf{\RLFL{}}} & \multicolumn{2}{c }{\textbf{\RTGNET{}}} \\ 
	\cmidrule(lr){2-3} \cmidrule(ll){4-5} \cmidrule(ll){6-7}\cmidrule(ll){8-9}\cmidrule(ll){10-11}\cmidrule(ll){12-13}\cmidrule(ll){14-15}\cmidrule(ll){16-17}\cmidrule(ll){18-19}
	\multicolumn{1}{ l }{\textbf{Release}} & 5XX& Ratio & 5XX& Ratio& 5XX& Ratio& 5XX& Ratio& 5XX& Ratio& 5XX& Ratio& 5XX& Ratio& 5XX& Ratio& 5XX& Ratio\\ 
	\cmidrule(lr){1-1}\cmidrule(lr){2-3}\cmidrule(ll){4-5} \cmidrule(ll){6-7}\cmidrule(ll){8-9}\cmidrule(ll){10-11}\cmidrule(ll){12-13}\cmidrule(ll){14-15}\cmidrule(ll){16-17}\cmidrule(ll){18-19}
    
	\multicolumn{1}{ l }{\textbf{[W]R1}} 
        &218&2.4\%&106&1.2\%&237&2.6\%&80&1.3\%&159&0.1\%&31&0.3\%&23&0.1\%&29&0.2\%&32&0.2\%\\
        \multicolumn{1}{ l }{\textbf{[D]R2}} &564&6.2\%&212&2.4\%&562&6.2\%&91&1.5\%&10143&8.1\%&3385&27.5\%&1678&9.5\%&1557&9.8\%&3122&14.7\%\\
        \multicolumn{1}{ l }{\textbf{[D]R3}} &568&6.2\%&213&2.5\%&566&6.2\%&89&1.5\%&10139&8.1\%&3379&27.5\%&1678&9.5\%&1549&9.8\%&3120&14.7\%\\
        \multicolumn{1}{ l }{\textbf{[W]R4}} &669&7.3\%&317&3.7\%&666&7.3\%&194&3.3\%&10580&8.4\%&3386&27.5\%&1678&9.5\%&1550&9.8\%&3514&16.5\%\\
        \multicolumn{1}{ l }{\textbf{[D]R5}} &673&7.4\%&316&3.6\%&671&7.4\%&197&3.3\%&10585&8.4\%&4363&35.5\%&1680&9.5\%&1558&9.8\%&3522&16.6\%\\
        \multicolumn{1}{ l }{\textbf{[D]R6}} &671&7.4\%&318&3.7\%&670&7.4\%&195&3.3\%&11465&9.1\%&3379&27.5\%&1680&9.5\%&1553&9.8\%&3517&16.5\%\\
        \multicolumn{1}{ l }{\textbf{[W]R7}} &671&7.4\%&316&3.6\%&667&7.3\%&193&3.2\%&10583&8.4\%&3379&27.5\%&1678&9.5\%&1557&9.8\%&3519&16.5\%\\
        \multicolumn{1}{ l }{\textbf{[D]R8}} &666&7.3\%&317&3.7\%&670&7.4\%&198&3.3\%&10589&8.4\%&3383&27.5\%&1679&9.5\%&1551&9.8\%&3517&16.5\%\\
        \multicolumn{1}{ l }{\textbf{[D]R9}} &670&7.4\%&320&3.7\%&671&7.4\%&198&3.3\%&12642&10.0\%&3378&27.5\%&1681&9.5\%&1556&9.8\%&3520&16.5\%\\
        \multicolumn{1}{ l }{\textbf{[W]R10}} &670&7.4\%&317&3.7\%&667&7.3\%&194&3.3\%&10989&8.7\%&3381&27.5\%&1680&9.5\%&1555&9.8\%&3521&16.5\%\\
        \multicolumn{1}{ l }{\textbf{[D]R11}} &671&7.4\%&319&3.7\%&669&7.3\%&199&3.3\%&11921&9.5\%&3388&27.6\%&1678&9.5\%&1554&9.8\%&3525&16.6\%\\
        \multicolumn{1}{ l }{\textbf{[D]R12}} &562&6.2\%&212&2.4\%&560&6.1\%&91&1.5\%&13441&10.7\%&3379&27.5\%&1685&9.5\%&1558&9.8\%&3118&14.7\%\\
        \multicolumn{1}{ l }{\textbf{[W]R13}} &560&6.1\%&210&2.4\%&564&6.2\%&88&1.5\%&10148&8.1\%&3383&27.5\%&1677&9.5\%&1558&9.8\%&3119&14.7\%\\
        \multicolumn{1}{ l }{\textbf{[M]R14}} &563&6.2\%&212&2.4\%&561&6.2\%&92&1.5\%&11338&9.0\%&3381&27.5\%&1680&9.5\%&1557&9.8\%&3117&14.6\%\\
        
	\bottomrule
	\end{tabular}
    
 \label{tab-rq1results}

\end{table}

\begin{table}[H]
    \noindent
    \footnotesize
    \centering
    \caption{RQ1 results: Overall comparison of tools in terms of total number of API 5XX failures detected by each tool and their ratios across all APIs from releases R1 to R14, derived from all runs. Releases tagged with D, W, and M, denote daily, weekly, and monthly releases, respectively.} 
    
    \begin{tabular}{@{}l l l l l l l l l l l@{}}
    \toprule 
    \multicolumn{1}{l }{\textbf{}} & \multicolumn{2}{c }{\textbf{\RESTest{}}} & \multicolumn{2}{c }{\textbf{\EvoMaster{}}} & \multicolumn{2}{c }{\textbf{\Schemathesis{}}} & \multicolumn{2}{c }{\textbf{\RESTler{}}} & \multicolumn{2}{c }{\textbf{\RestTestGen{}}} \\ 
    \cmidrule(lr){2-3} \cmidrule(ll){4-5} \cmidrule(ll){6-7} \cmidrule(ll){8-9} \cmidrule(ll){10-11}
    \multicolumn{1}{ l }{\textbf{Release}} & 5XX & Ratio & 5XX & Ratio & 5XX & Ratio & 5XX & Ratio & 5XX & Ratio \\ 
    \cmidrule(lr){1-1}\cmidrule(lr){2-3}\cmidrule(ll){4-5} \cmidrule(ll){6-7} \cmidrule(ll){8-9} \cmidrule(ll){10-11}
    \multicolumn{1}{ l }{\textbf{[W]R1}} 
        & 641 & \textcolor{black}{2.0\%} & 159 & 0.1\% & 31 & 0.3\% & 52 & \textcolor{black}{0.2\%} & 32 & 0.2\% \\
    \multicolumn{1}{ l }{\textbf{[D]R2}} 
        & 1629 & \textcolor{black}{4.0\%} & 10143 & 8.1\% & 3385 & 27.5\% & 3235 & \textcolor{black}{9.6\%} & 3122 & 14.7\% \\
    \multicolumn{1}{ l }{\textbf{[D]R3}} 
        & 1636 & \textcolor{black}{5.0\%} & 10139 & 8.1\% & 3379 & 27.5\% & 3227 & \textcolor{black}{9.6\%} & 3120 & 14.7\% \\
    \multicolumn{1}{ l }{\textbf{[W]R4}} 
        & 1846 & \textcolor{black}{5.6\%} & 10580 & 8.4\% & 3386 & 27.5\% & 3230 & \textcolor{black}{9.6\%} & 3514 & 16.5\% \\
    \multicolumn{1}{ l }{\textbf{[D]R5}} 
        & 1857 & \textcolor{black}{5.7\%} & 10585 & 8.4\% & 4363 & 35.5\% & 3238 & \textcolor{black}{9.6\%} & 3522 & 16.6\% \\
    \multicolumn{1}{ l }{\textbf{[D]R6}} 
        & 1854 &\textcolor{black}{5.6\%} & 11465 & 9.1\% & 3379 & 27.5\% & 3206 & \textcolor{black}{9.5\%} & 3517 & 16.5\% \\
    \multicolumn{1}{ l }{\textbf{[W]R7}} 
        & 1847 & \textcolor{black}{5.6\%} & 10583 & 8.4\% & 3379 & 27.5\% & 3234 & \textcolor{black}{9.6\%} & 3519 & 16.5\% \\
    \multicolumn{1}{ l }{\textbf{[D]R8}} 
        & 1851 & \textcolor{black}{5.6\%} & 10589 & 8.4\% & 3383 & 27.5\% & 3230 & \textcolor{black}{9.6\%} & 3517 & 16.5\% \\
    \multicolumn{1}{ l }{\textbf{[D]R9}} 
        & 1859 & \textcolor{black}{5.7\%} & 12642 & 10.0\% & 3378 & 27.5\% & 3237 & \textcolor{black}{9.6\%} & 3520 & 16.5\% \\
    \multicolumn{1}{ l }{\textbf{[W]R10}} 
        & 1848 & \textcolor{black}{5.6\%} & 10989 & 8.7\% & 3381 & 27.5\% & 3235 & \textcolor{black}{9.6\%} & 3521 & 16.5\% \\
    \multicolumn{1}{ l }{\textbf{[D]R11}} 
        & 1858 & \textcolor{black}{5.7\%} & 11921 & 9.5\% & 3388 & 27.6\% & 3232 & \textcolor{black}{9.6\%} & 3525 & 16.6\% \\
    \multicolumn{1}{ l }{\textbf{[D]R12}} 
        & 1425 & \textcolor{black}{4.3\%} & 13441 & 10.7\% & 3379 & 27.5\% & 3243 & \textcolor{black}{9.7\%} & 3118 & 14.7\% \\
    \multicolumn{1}{ l }{\textbf{[W]R13}} 
        & 1422 & \textcolor{black}{4.3\%} & 10148 & 8.1\% & 3383 & 27.5\% & 3235 & \textcolor{black}{9.6\%} & 3119 & 14.7\% \\
    \multicolumn{1}{ l }{\textbf{[M]R14}} 
        & 1428 & \textcolor{black}{4.3\%} & 11338 & 9.0\% & 3381 & 27.5\% & 3237 & \textcolor{black}{9.6\%} & 3117 & 14.6\% \\
    \bottomrule
    
    \end{tabular}
    
 \label{tab-rq1results-agg}
\end{table}

\begin{table}[H]
    \noindent
    \tiny
    \centering
    \caption{Total number of client-side 4XX errors and their ratios across all APIs from releases R1 to R14, derived from all runs. Releases tagged with D, W, and M, denote daily, weekly, and monthly releases, respectively. }
    
    \begin{tabular}{@{}l l l l l l l l l l l l l l l l l l l@{}}
    \toprule
      \multicolumn{1}{l }{\textbf{}} & \multicolumn{2}{c }{\textbf{\RTCBT{}}} & \multicolumn{2}{c }{\textbf{\RTART{}}}& \multicolumn{2}{c }{\textbf{\RTRT{}}} & \multicolumn{2}{c }{\textbf{\RTFT{}}} & \multicolumn{2}{c }{\textbf{\EMBB{}}} & \multicolumn{2}{c }{\textbf{\STPT{}}}& \multicolumn{2}{c }{\textbf{\RLFz{}{}}} & \multicolumn{2}{c }{\textbf{\RLFL{}}} & \multicolumn{2}{c }{\textbf{\RTGNET{}}} \\ 
	\cmidrule(lr){2-3} \cmidrule(ll){4-5} \cmidrule(ll){6-7}\cmidrule(ll){8-9}\cmidrule(ll){10-11}\cmidrule(ll){12-13}\cmidrule(ll){14-15}\cmidrule(ll){16-17}\cmidrule(ll){18-19}
	\multicolumn{1}{ l }{\textbf{Release}} & 4XX& Ratio & 4XX& Ratio& 4XX& Ratio& 4XX& Ratio& 4XX& Ratio& 4XX& Ratio& 4XX& Ratio& 4XX& Ratio& 4XX& Ratio\\ 
	\cmidrule(lr){1-1}\cmidrule(lr){2-3}\cmidrule(ll){4-5} \cmidrule(ll){6-7}\cmidrule(ll){8-9}\cmidrule(ll){10-11}\cmidrule(ll){12-13}\cmidrule(ll){14-15}\cmidrule(ll){16-17}\cmidrule(ll){18-19}
    
	\multicolumn{1}{ l }{\textbf{[W]R1}} &5773&63.4\%&5842&67.3\%&5772&63.4\%&4270&71.7\%&109500&86.9\%&12175&99.0\%&16127&90.9\%&15728&99.3\%&20932&98.4\%\\
        \multicolumn{1}{ l }{\textbf{[D]R2}} &6062&66.6\%&6120&70.5\%&6064&66.6\%&4462&74.9\%&60586&48.1\%&8837&71.9\%&14346&80.8\%&14242&89.9\%&12918&60.7\%\\
        \multicolumn{1}{ l }{\textbf{[D]R3}} &6058&66.5\%&6119&70.5\%&6060&66.5\%&4464&75.0\%&60590&48.1\%&8843&71.9\%&14346&80.8\%&14250&90.0\%&12920&60.7\%\\
        \multicolumn{1}{ l }{\textbf{[W]R4}} &5957&65.4\%&6015&69.3\%&5960&65.4\%&4359&73.2\%&60149&47.8\%&8836&71.9\%&14346&80.8\%&14249&90.0\%&12526&58.9\%\\
        \multicolumn{1}{ l }{\textbf{[D]R5}} &5953&65.4\%&6016&69.3\%&5955&65.4\%&4356&73.2\%&60144&47.8\%&7859&63.9\%&14344&80.8\%&14241&89.9\%&12518&58.8\%\\
        \multicolumn{1}{ l }{\textbf{[D]R6}} &5955&65.4\%&6014&69.3\%&5956&65.4\%&4358&73.2\%&59264&47.1\%&8843&71.9\%&14344&80.8\%&14246&89.9\%&12523&58.8\%\\
        \multicolumn{1}{ l }{\textbf{[W]R7}} &5955&65.4\%&6016&69.3\%&5959&65.4\%&4360&73.2\%&60146&47.8\%&8843&71.9\%&14346&80.8\%&14242&89.9\%&12521&58.8\%\\
        \multicolumn{1}{ l }{\textbf{[D]R8}} &5960&65.4\%&6015&69.3\%&5956&65.4\%&4355&73.1\%&60140&47.8\%&8839&71.9\%&14345&80.8\%&14248&89.9\%&12523&58.8\%\\
        \multicolumn{1}{ l }{\textbf{[D]R9}} &5956&65.4\%&6012&69.3\%&5955&65.4\%&4355&73.1\%&58456&46.4\%&8844&71.9\%&14343&80.8\%&14243&89.9\%&12520&58.8\%\\
        \multicolumn{1}{ l }{\textbf{[W]R10}} &5956&65.4\%&6015&69.3\%&5959&65.4\%&4359&73.2\%&60150&47.8\%&8841&71.9\%&14344&80.8\%&14244&89.9\%&12519&58.8\%\\
        \multicolumn{1}{ l }{\textbf{[D]R11}} &5955&65.4\%&6013&69.3\%&5957&65.4\%&4354&73.1\%&60001&47.6\%&8834&71.9\%&14346&80.8\%&14245&89.9\%&12515&58.8\%\\
        \multicolumn{1}{ l }{\textbf{[D]R12}} &6064&66.6\%&6120&70.5\%&6066&66.6\%&4462&74.9\%&59953&47.6\%&8843&71.9\%&14339&80.8\%&14241&89.9\%&12922&60.7\%\\
        \multicolumn{1}{ l }{\textbf{[W]R13}} &6066&66.6\%&6122&70.5\%&6062&66.6\%&4465&75.0\%&60581&48.1\%&8839&71.9\%&14347&80.9\%&14241&89.9\%&12921&60.7\%\\
        \multicolumn{1}{ l }{\textbf{[M]R14}} &6063&66.6\%&6120&70.5\%&6065&66.6\%&4461&74.9\%&60603&48.1\%&8841&71.9\%&14344&80.8\%&14242&89.9\%&12923&60.7\%\\
	\bottomrule
	\end{tabular}
    
 \label{tab-rq1results-4xx}
\end{table}

\begin{table}[H]
    \noindent
    \small
    \rotatebox{90}{
    \begin{minipage}{1\textheight}
    \centering
    \caption{Statistical comparison of various testing tools based on the number of failures detected across releases R1--R14 and all runs. It includes the \textit{p}-values derived from the Wilcoxon test and the $\hat{A}_{12}$ values obtained through Vargha-Delaney's effect size measure. }
    
	\begin{tabular}{@{}l l l l l l l l l l l l l l l l l l l@{}}\toprule
        \multicolumn{1}{l }{\textbf{}} & \multicolumn{2}{c }{\textbf{\RTCBT{}}} & \multicolumn{2}{c }{\textbf{\RTART{}}}& \multicolumn{2}{c }{\textbf{\RTRT{}}} & \multicolumn{2}{c }{\textbf{\RTFT{}}} & \multicolumn{2}{c }{\textbf{\EMBB{}}} & \multicolumn{2}{c }{\textbf{\STPT{}}}& \multicolumn{2}{c }{\textbf{\RLFz{}{}}} & \multicolumn{2}{c }{\textbf{\RLFL{}}} & \multicolumn{2}{c }{\textbf{\RTGNET{}}} \\ 
	\cmidrule(lr){2-3}
	\cmidrule(ll){4-5} \cmidrule(ll){6-7}\cmidrule(ll){8-9}\cmidrule(ll){10-11}\cmidrule(ll){12-13}\cmidrule(ll){14-15}\cmidrule(ll){16-17}\cmidrule(ll){18-19}
	\multicolumn{1}{ l }{\textbf{}}  & \textit{p}-val& $\hat{A}_{12}$& \textit{p}-val& $\hat{A}_{12}$& \textit{p}-val& $\hat{A}_{12}$& \textit{p}-val& $\hat{A}_{12}$& \textit{p}-val& $\hat{A}_{12}$& \textit{p}-val& $\hat{A}_{12}$& \textit{p}-val& $\hat{A}_{12}$& \textit{p}-val& $\hat{A}_{12}$& \textit{p}-val& $\hat{A}_{12}$\\ 
	\cmidrule(lr){1-1}\cmidrule(lr){2-3}\cmidrule(ll){4-5} \cmidrule(ll){6-7}\cmidrule(ll){8-9}\cmidrule(ll){10-11}\cmidrule(ll){12-13}\cmidrule(ll){14-15}\cmidrule(ll){16-17}\cmidrule(ll){18-19} 
    
	\multicolumn{1}{ l }{\RTCBT{}} &-&-&0.009&\textbf{0.959}&\textcolor{gray}{1.0}&\textbf{0.564}&0.009&\textbf{1.0}&0.01&0.071&0.01&0.071&0.01&0.071&0.01&0.071&0.01&0.071\\
        \multicolumn{1}{ l }{\RTART{}} &0.009&0.041&-&-&0.009&0.041&0.009&\textbf{0.959}&0.009&0.066&0.01&0.071&0.01&0.071&0.01&0.071&0.01&0.071\\
        \multicolumn{1}{ l }{\RTRT{}} &\textcolor{gray}{1.0}&0.436&0.009&\textbf{0.959}&-&-&0.009&\textbf{1.0}&0.01&0.071&0.01&0.071&0.01&0.071&0.01&0.071&0.01&0.071\\
        \multicolumn{1}{ l }{\RTFT{}} &0.009&0.0&0.009&0.041&0.009&0.0&-&-&0.009&0.041&0.01&0.071&0.01&0.071&0.01&0.071&0.01&0.071\\
        \multicolumn{1}{ l }{\EMBB{}} &0.01&\textbf{0.929}&0.009&\textbf{0.934}&0.01&\textbf{0.929}&0.009&\textbf{0.959}&-&-&0.009&\textbf{0.934}&0.009&\textbf{0.934}&0.009&\textbf{0.934}&0.009&\textbf{0.934}\\
        \multicolumn{1}{ l }{\STPT{}} &0.01&\textbf{0.929}&0.01&\textbf{0.929}&0.01&\textbf{0.929}&0.01&\textbf{0.929}&0.009&0.066&-&-&0.009&\textbf{0.934}&0.009&\textbf{0.934}&\textcolor{gray}{1.0}&0.439\\
        \multicolumn{1}{ l }{\RLFz{}} &0.01&\textbf{0.929}&0.01&\textbf{0.929}&0.01&\textbf{0.929}&0.01&\textbf{0.929}&0.009&0.066&0.009&0.066&-&-&0.01&\textbf{0.929}&0.009&0.066\\
        \multicolumn{1}{ l }{\RLFL{}} &0.01&\textbf{0.929}&0.01&\textbf{0.929}&0.01&\textbf{0.929}&0.01&\textbf{0.929}&0.009&0.066&0.009&0.066&0.01&0.071&-&-&0.009&0.066\\
        \multicolumn{1}{ l }{\RTGNET{}} &0.01&\textbf{0.929}&0.01&\textbf{0.929}&0.01&\textbf{0.929}&0.01&\textbf{0.929}&0.009&0.066&\textcolor{gray}{1.0}&\textbf{0.561}&0.009&\textbf{0.934}&0.009&\textbf{0.934}&-&-\\

        \bottomrule
        
        \multicolumn{19}{l}{* The values shown in gray color indicate that $p-value>\alpha$, i.e., there is no statistically significant difference. The values presented in bold indicate that the}\\
        \multicolumn{19}{l}{tool listed on the left side outperformed the tool on the right side. Instances of self-comparison of tools are left blank in the table.}
	\end{tabular}
    
 \label{tab-stcomp-rq1}
 \end{minipage}}
\end{table}

\subsubsection{RQ2 Results -- REST APIs Faults}\label{RQ2Results}

The results in \Cref{tab-rq2results} present the number of unique potential faults discovered from the results generated by each testing tool, along with their respective fault detection ratios across all releases. 
In release R1, tests generated by \RESTestCBT{} and \RESTestRT{} led to finding the highest number of potential faults with the highest fault detection ratios compared to other releases and tools. 
From R2 onward, their performance remained relatively consistent. 
Notably, from R4 to R11, tests generated by \EvoMasterBB{} and \RestTestGenNET{} led to detecting the most potential faults, followed by \SchemathesisPT{}. 
However, in terms of fault detection ratios, \RESTestFT{} exhibited the highest, while \EvoMasterBB{} demonstrated the lowest. 
Analyzing the overall comparison results in~\Cref{tab-rq2results-agg}, it is evident that test cases generated by \RESTest{} led to the highest number of potential faults in most releases. 
The performance of \EvoMaster{} and \RestTestGen{} was second highest and comparable to each other, followed by \Schemathesis{} and \RESTler{}. 
When considering fault detection ratios, \EvoMaster{} demonstrated the lowest ratios, while \Schemathesis{} achieved the highest in most releases.

\begin{figure}[H]
\centerline{\includegraphics[width=\linewidth, keepaspectratio]{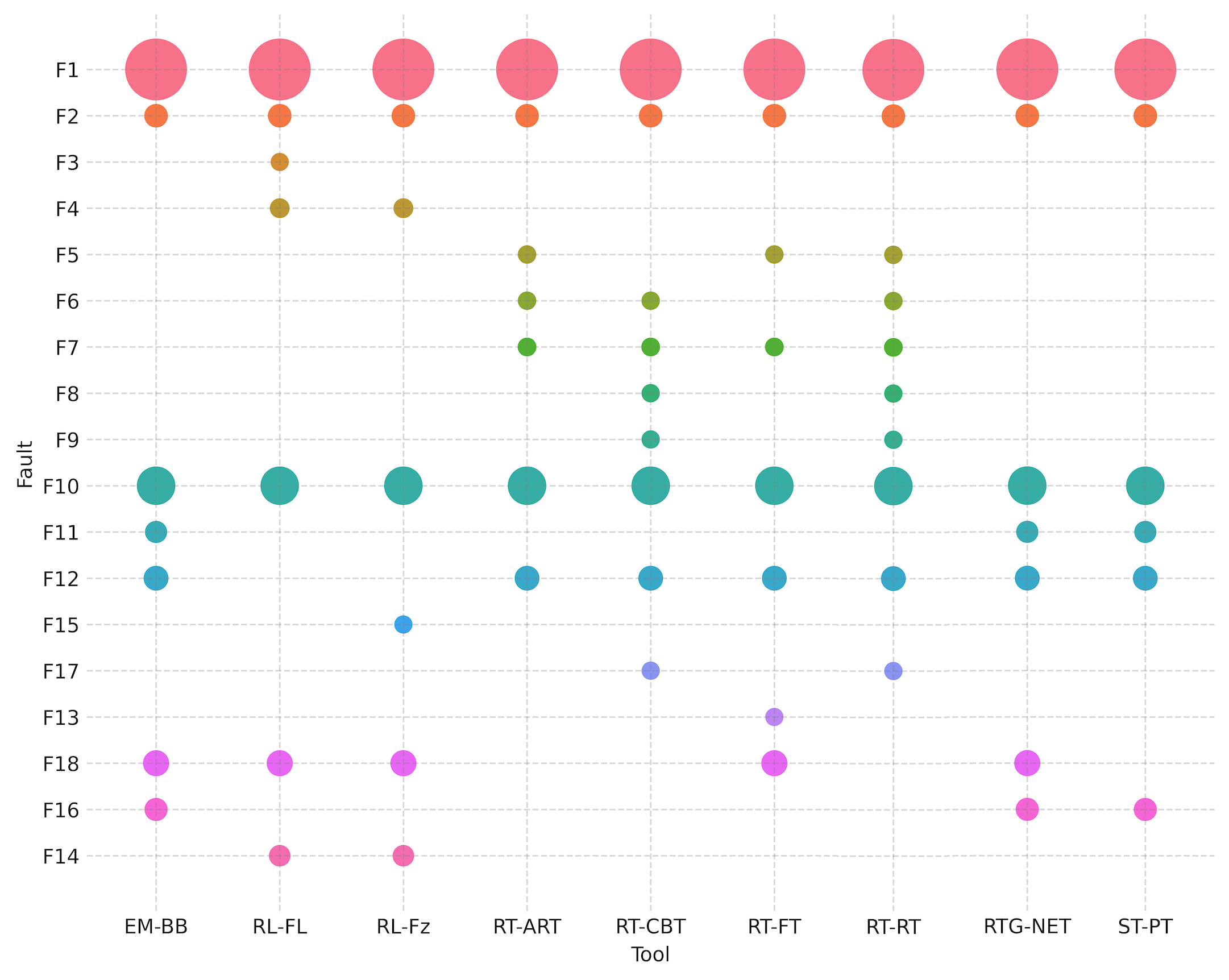}}
\caption{RQ2 results: Faults and their frequency of occurrence corresponding to each testing tool across releases R1--R14 and all runs. 
In this plot, each color-coded bubble represents a specific fault (shown on the y-axis) identified from the test cases generated by all tools (listed on the x-axis). The size of the bubble shows the frequency of the fault. 
Key findings indicate that faults F1, F2, and F10 were uncovered from tests generated by all tools, with F1 occurring most frequently. 
The majority of faults were found through tests generated by \RESTestCBT{} and \RESTestRT{}. 
Notably, fault F3 was identified from \RESTlerFL{}-generated tests, and F13 was detected by tests from \RESTestFT{}. 
Detailed descriptions of all faults (F1--F18) are provided in~\Cref{tab-rq2faultsinfo}. 
}
\label{fig:rq2faults}
\end{figure}

Among all identified faults, some were common across tools, while others were unique to a specific tool due to the distinct testing techniques employed. 
Moreover, some faults were discovered from successful testing results (i.e., those with a 200 status code), while the rest were found due to client-side (4XX) and server-side (5XX) errors. 
In total, across all releases, 18 unique potential faults (F1-F18) were identified that are common across all tools, as detailed in \Cref{tab-rq2faultsinfo}. 
It was observed that certain faults reoccurred multiple times across evolving releases (i.e., from R1 to R14). 
The frequency of each fault's occurrence w.r.t. each testing tool is illustrated in \Cref{fig:rq2faults}.

The fault F1 is related to a server error, which was triggered when the server could not process requests containing invalid (e.g., missing required fields) or large amounts of data (e.g., string data exceeding 125KB).  
This fault was also noted in responses from third-party applications.
Consequently, it is associated with two HTTP status codes: \emph{502 Bad Gateway} and \emph{413 Payload Too Large}. 
As depicted in the results (\Cref{fig:rq2faults}), fault F1 was identified multiple times from the results of all testing tools.
This fault was found in multiple APIs including \emph{Alerts}, \emph{Authentication}, \emph{Device-Karie}, \emph{Device-Medido}, \emph{Device-Pilly}, \emph{Measurements}, \emph{Users}, \emph{Courses}, \emph{Mobile App}, \emph{Reimbursements}, \emph{Invoicing}, and \emph{Reports}. 

Fault F2 was detected from the successful results that returned a 200 status code. 
This fault was observed in the \emph{Authentication} API, where requests containing invalid login information resulted in \emph{200 Ok} responses. 
This implies that the system's authentication process may not be functioning properly, permitting unauthorized access and thereby posing a considerable security risk. 
Therefore, F2 could potentially be classified as a security fault.

Fault F3 was also discovered in \emph{Authentication} API, resulting in \emph{500 Internal Server Error} with a message stating that the server could not understand the request. 
This fault was discovered only through the results of \RESTlerFL{}. 
It was noted that the API could not interpret the tokens generated by \RESTlerFL{} for device authentication. 
Similarly, fault F4 was identified from errors related to \emph{400 Bad Request}, accompanied by a message stating that the server could not parse the headers.  
This fault was discovered from tests generated by \RESTlerF{} and \RESTlerFL{}. 
This fault was found in APIs including \emph{Alerts}, \emph{Authentication}, \emph{Device-Karie}, \emph{Device-Medido}, \emph{Measurements}, \emph{Users}, \emph{Invoicing}, and \emph{Reports}. 

Fault F5 was discovered from the error \emph{429 Too Many Requests} originating from \emph{Authentication}. 
This fault was detected in the tests generated by \RESTestART{}, \RESTestRT{}, and \RESTestFT{}. 
In this case, it was observed that even after reaching the maximum limit for login attempts, the waiting time for the next attempts continued to increase from seconds to minutes to hours. 
This pattern suggests a potential fault in the API's implementation, which is intended to control or limit the incremental wait time.

Fault F6 was found in \emph{Measurements} API using the tests generated by \RESTestART{}, \RESTestCBT{}, and \RESTestRT{}. 
This fault was identified when incorrect user IDs resulted in successful responses with \emph{200 Ok} status. 
We observed this during manual inspection, where some randomly generated IDs by the tools appeared valid within the system but differed from the user ID assigned to us for testing, indicating a potential weakness in the API’s access control mechanism. 
Similarly, fault F7 was found in two device-related APIs, i.e., \textit{Medido} and \textit{Pilly}, and using the tests generated by the four variants of \RESTest{}. 
In this case, empty/invalid JSON objects for these devices led to \emph{200 Ok} responses that contained complete information about the corresponding devices.
Furthermore, a similar case was observed for the fault F9 that was identified in \emph{Users} API using tests generated by \RESTestCBT{} and \RESTestRT{}. 
It was noticed that JSON objects with incorrect property values led to \emph{200 Ok} responses, providing all user-relevant records such as tasks/surveys. 
These faults (i.e., F6, F7, and F9) could be due to missing input validations in the API implementation.

The faults F8, F11, F17, and F18 are associated with errors that originated from the database. 
Fault F8 was discovered in \emph{Measurements} API from the tests generated by \RESTestCBT{} and \RESTestRT{}. 
This fault, indicated by \emph{409 Conflict} status code, was identified when the API implementation tried to insert measurement records for users with IDs that do not exist in the database. 
The fault F11 was found in \emph{Users} API from the status code \emph{500 Internal Server Error} and from tests generated by \EvoMasterBB{}, \SchemathesisPT{}, and \RestTestGenNET{}.
This fault was identified from a database error when the API implementation tried to insert records with missing fields.
Similarly to F8, fault F17 was also found in \emph{Measurements} API due to \emph{409 Conflict} error from tests generated by \RESTestCBT{} and \RESTestRT{}. 
While the status code for both faults is the same, the detailed response for F17 indicated a violation of foreign key constraints in the database. 
This violation occurred when trying to retrieve measurement records using invalid foreign keys. 
We have marked this fault as unique and have left it to the API development team to determine if they are the same.
For fault F18, identified in the \emph{Mobile App} API, a \emph{500 Internal Server Error} occurred due to long values in the request object. 
In addition, the error response was accompanied by a complete database query. 
This fault was detected using the tests generated by \EvoMasterBB{}, \RESTlerFL{}, \RESTlerF{}, \RESTestFT{}, and \RestTestGenNET{}.

The faults F10, F13, F14, F15, and F16 are associated with programming errors or problems related to business logic, categories which align with the predefined classification~\cite{marculescu2022faults}. 
Fault F10 was discovered due to the \emph{500 Internal Server Error} response originating from APIs such as \emph{Device-Karie}, \emph{Device-Medido}, \emph{Device-Pilly}, and \emph{Users}. 
This fault occurred when the API implementation attempted to iterate over an object of the String type.
Fault F13 was identified due to a \emph{400 Bad Request} error in \emph{Measurements} API. 
This fault pertains to a type mismatch issue encountered during arithmetic operations.
The fault F14 was found due to \emph{500 Internal Server Error} response from \emph{Reimbursements} API. 
The primary cause of this fault was the conversion of a large value from Python's Int type to C's Long type. 
The fault F15 was located due to a \emph{400 Bad Request} error in the \emph{Authentication} API. 
This fault occurred during the execution of an array operation on a Boolean-type variable.
Fault F16 was discovered due to \emph{500 Internal Server Error} response from the \emph{Users} API. 
This fault occurred while attempting to access Dict items from an object of the String type.
Among these faults, F10 was the most frequently occurring, and tests from all tools led to its detection. 
In the case of F13, it was only identified from tests generated by \RESTestCBT{}. 
F14 was discovered from tests produced by both \RESTlerFL{} and \RESTlerF{}, while F15 was only located from the tests generated by \RESTlerFL{}. 
In addition, F16 was detected using tests obtained from \EvoMasterBB{}, \RestTestGenNET{}, and \SchemathesisPT{}.

Finally, fault F12 was discovered in the \emph{Karie} device API using tests generated by \EvoMasterBB{}, \RESTestART{}, \RESTestCBT{}, \RESTestRT{}, \RESTestFT{}, \SchemathesisPT{}, and \RestTestGenNET{}. 
It was noted that a \emph{500 Internal Server Error} was triggered whenever a request was made to the device APIs while the device was in the middle of a scheduled update.   
While the device software was undergoing an update, this server error originated from the application's API under test.
Note that medical devices come from different vendors, and their software update schedules are unknown to us. 

\begin{figure*}
    \subfigure[Alerts]{\includegraphics[width=3.7cm,height=2.5cm]{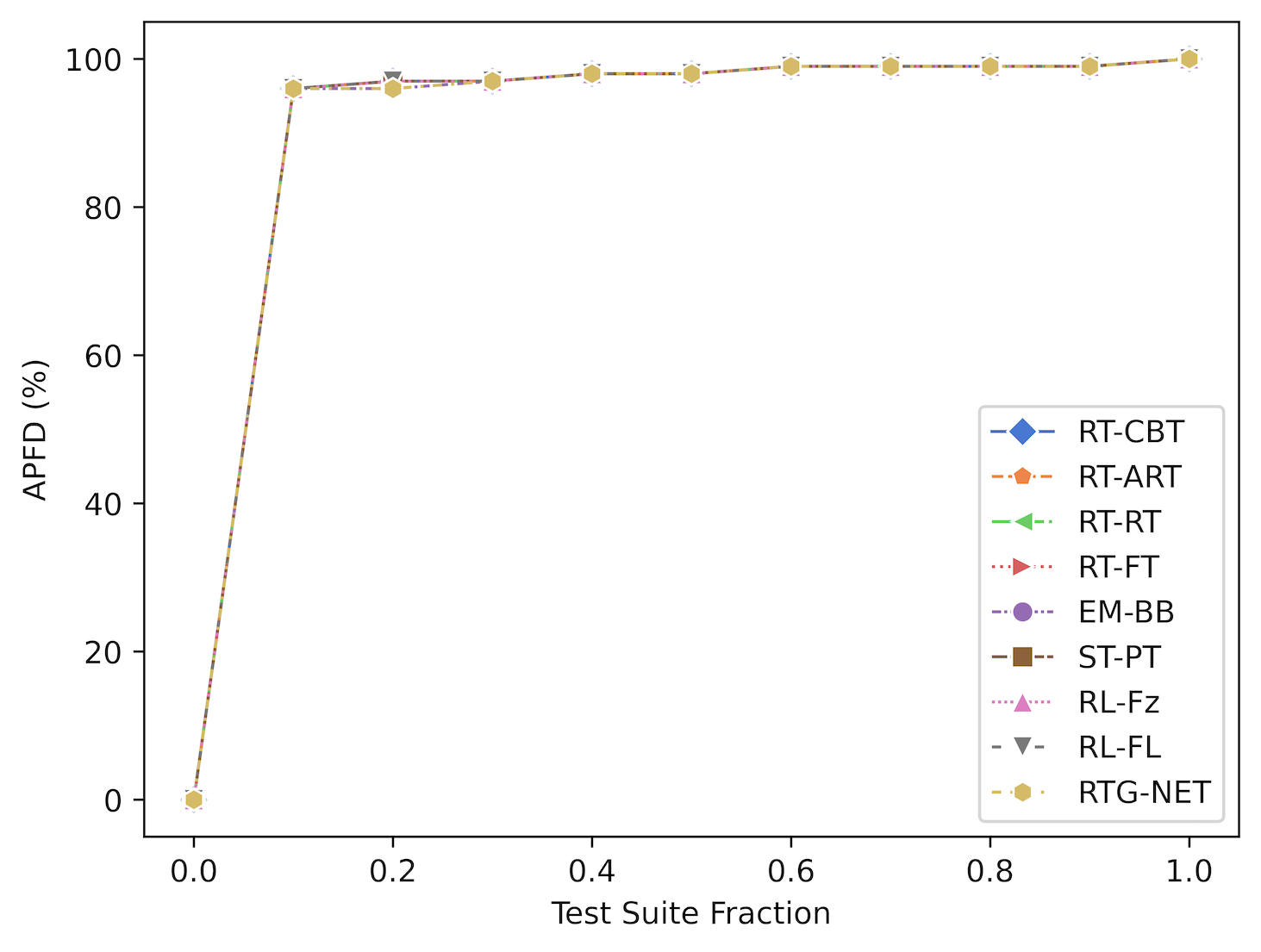}\label{fig:apfd1}}
    \subfigure[Authentication]{\includegraphics[width=3.7cm,height=2.5cm]{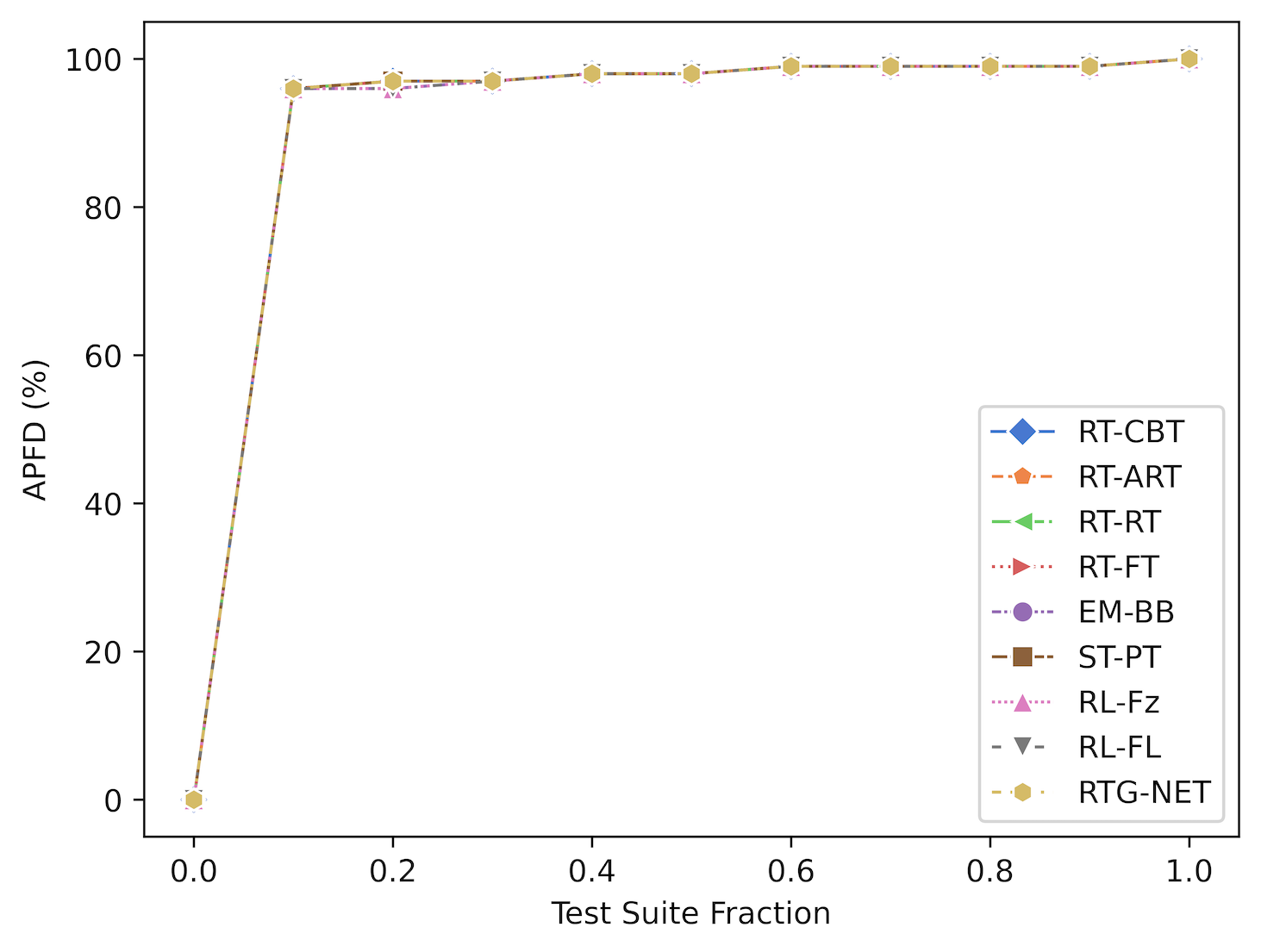}\label{fig:apfd2}}
    \subfigure[Device-Karie]{\includegraphics[width=3.7cm,height=2.5cm]{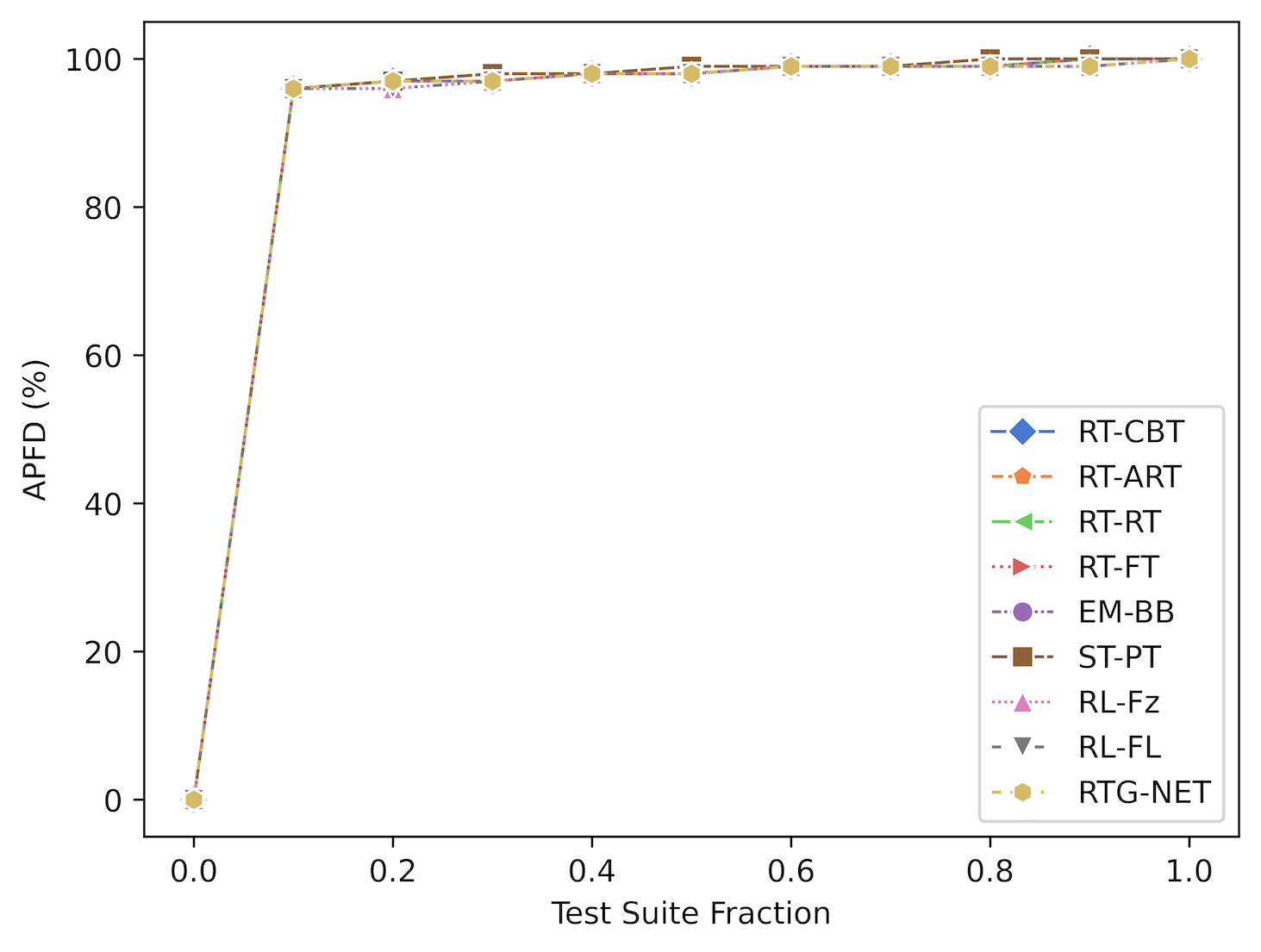}\label{fig:apfd3}}
    \subfigure[Device-Medido]{\includegraphics[width=3.7cm,height=2.5cm]{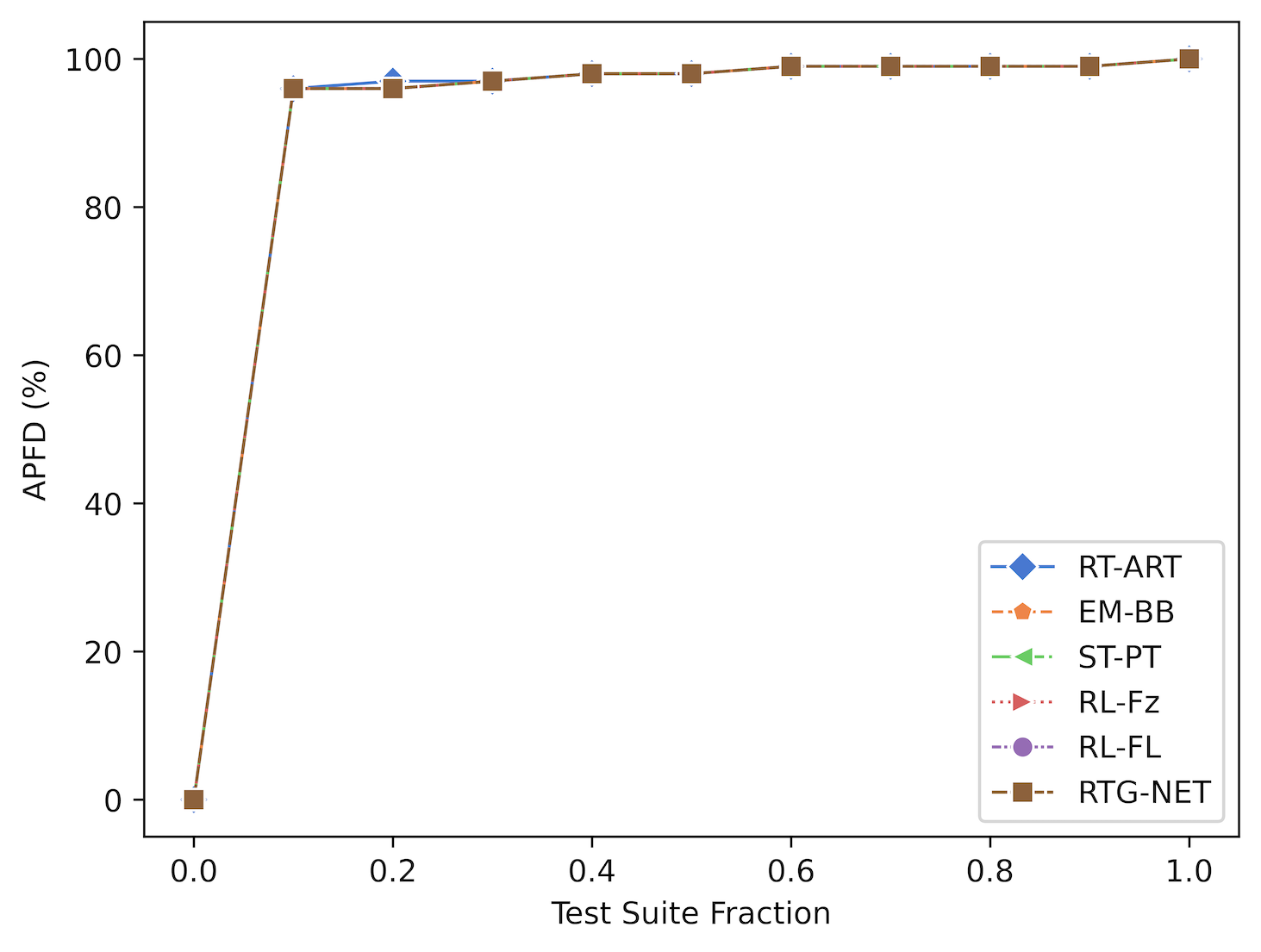}\label{fig:apfd4}}

    \subfigure[Device-Pilly]{\includegraphics[width=3.7cm,height=2.5cm]{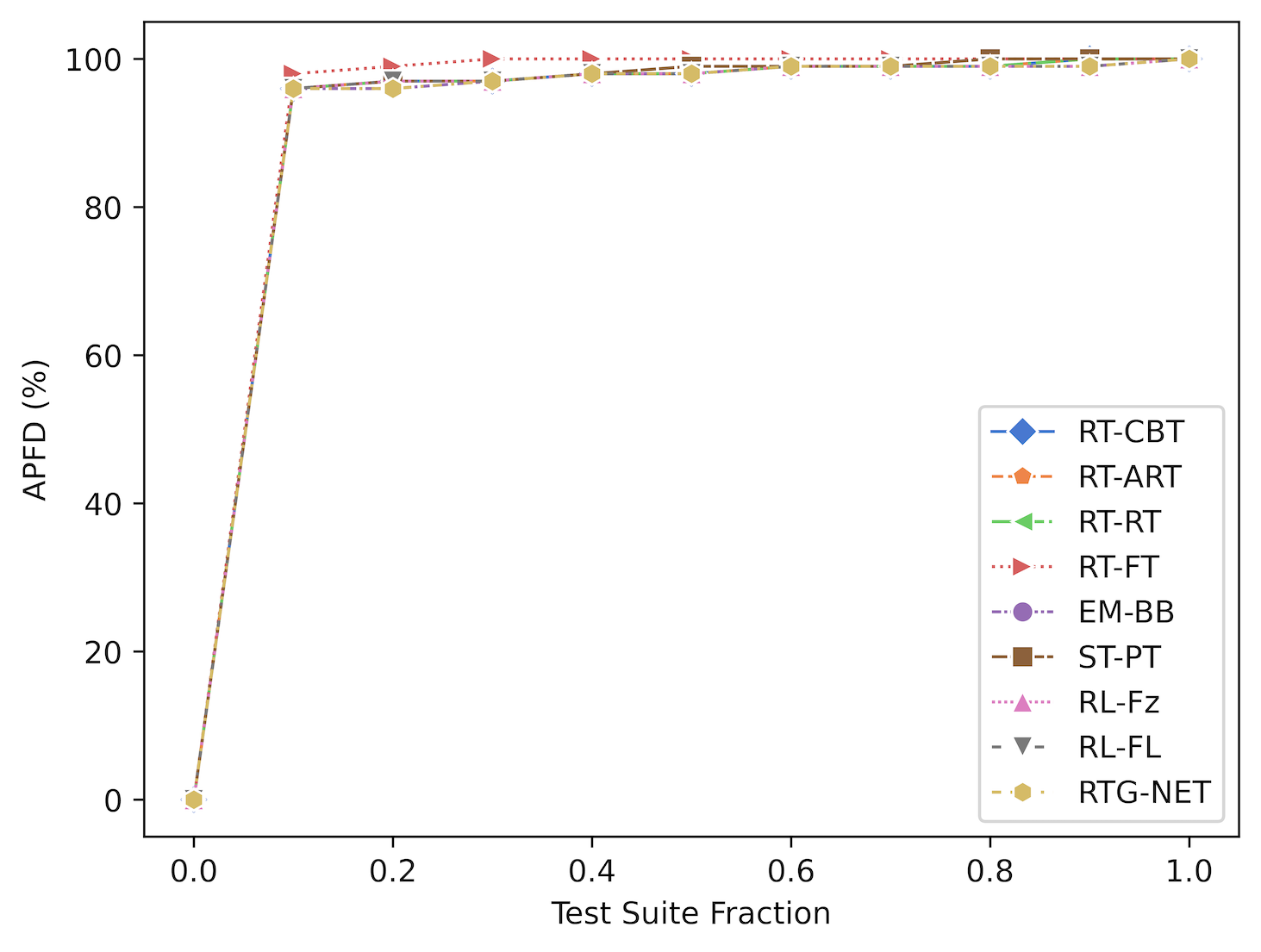}\label{fig:apfd5}}
    \subfigure[Patients]{\includegraphics[width=3.7cm,height=2.5cm]{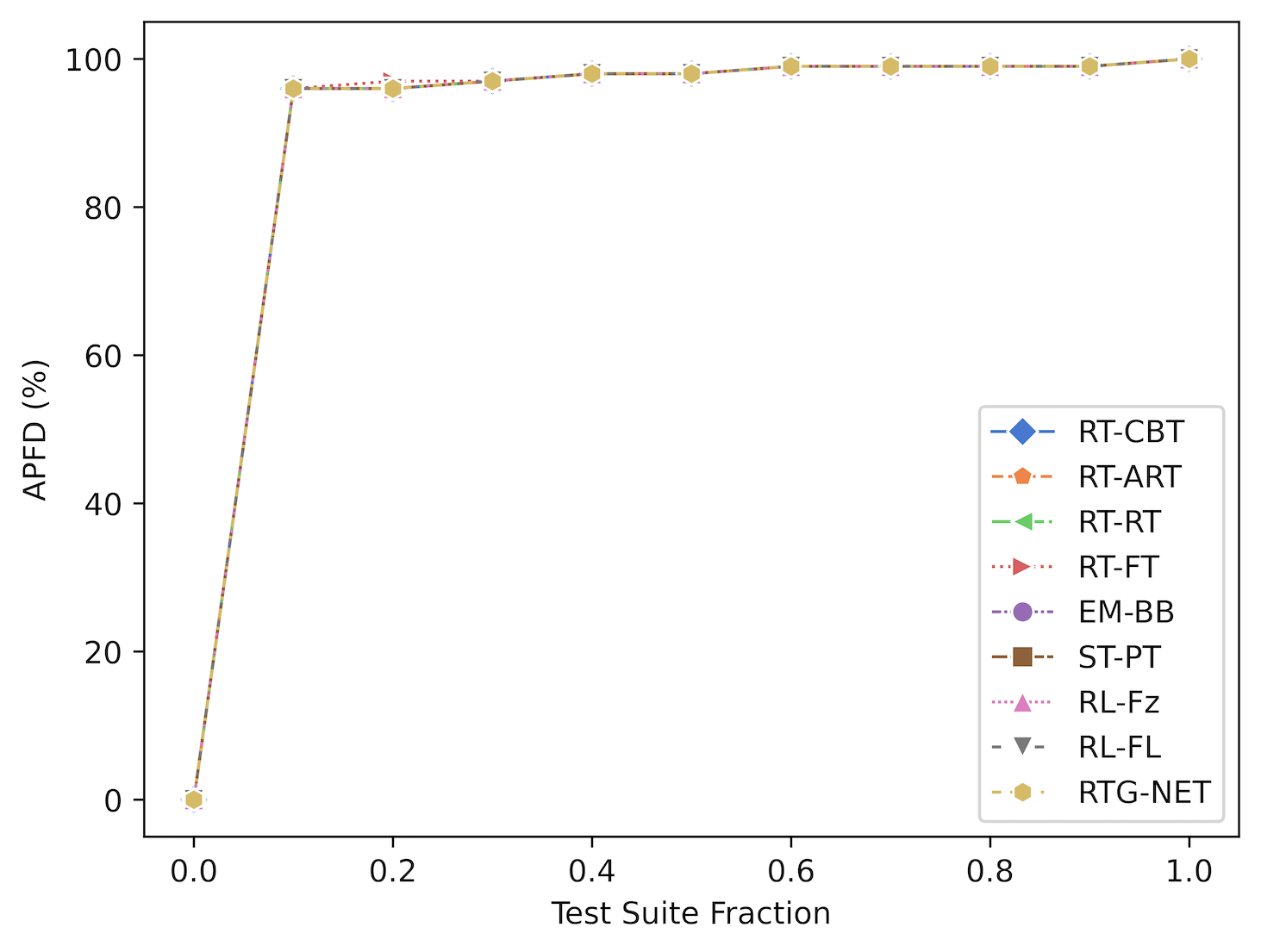}\label{fig:apfd6}}
    \subfigure[Measurements]{\includegraphics[width=3.7cm,height=2.5cm]{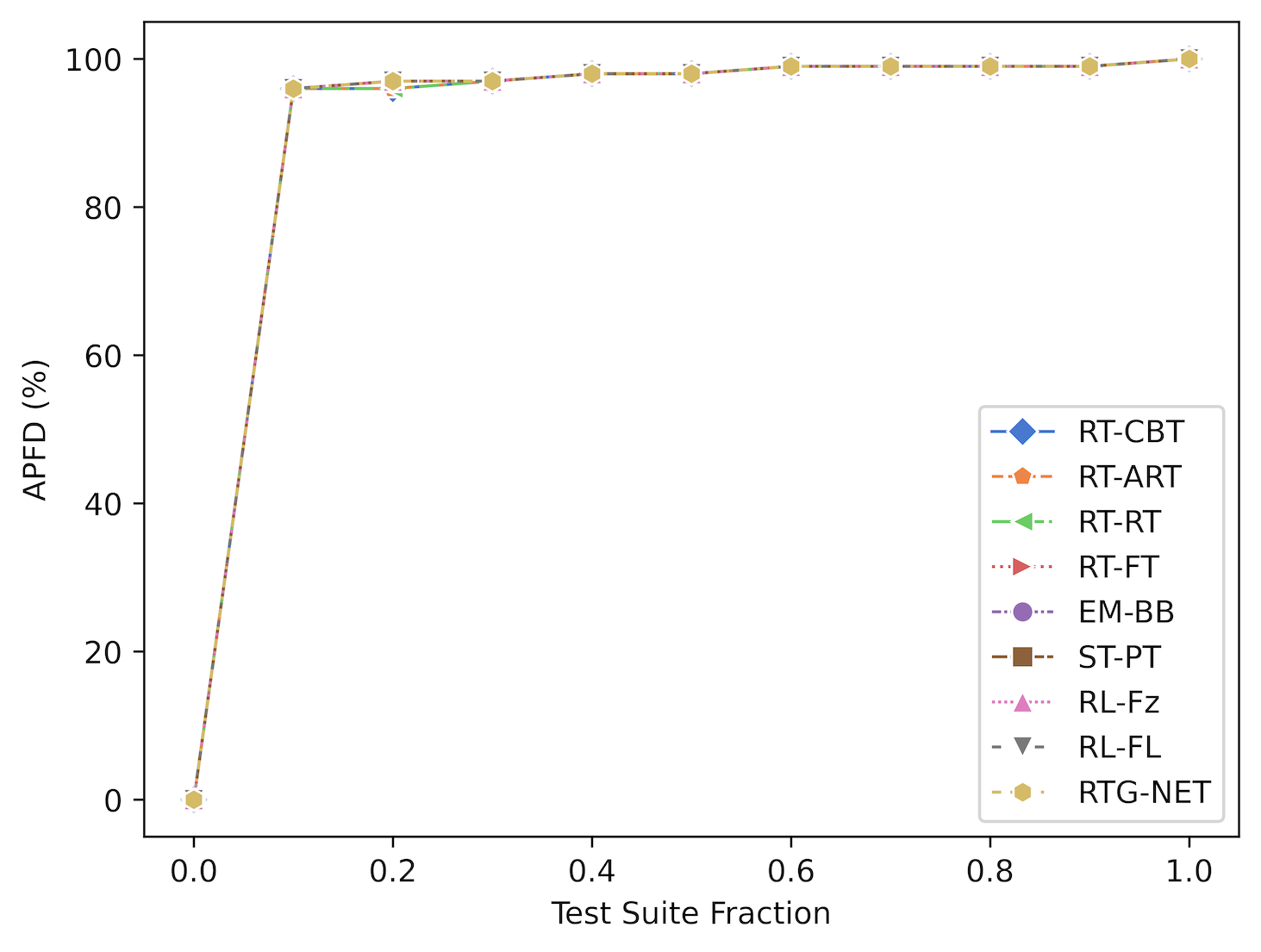}\label{fig:apfd7}}
    \subfigure[Users]{\includegraphics[width=3.7cm,height=2.5cm]{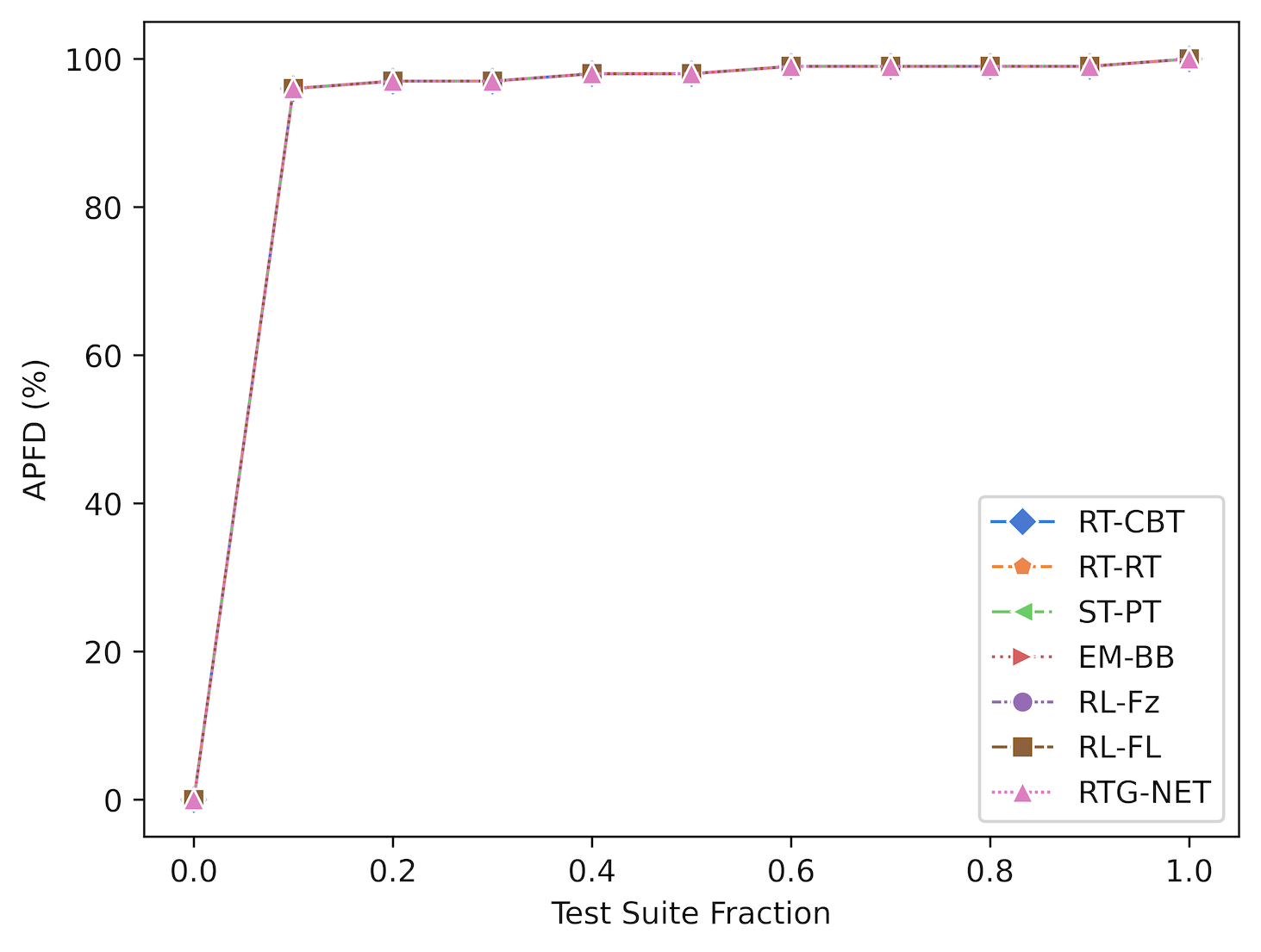}\label{fig:apfd8}}

    \subfigure[Courses]{\includegraphics[width=3.7cm,height=2.5cm]{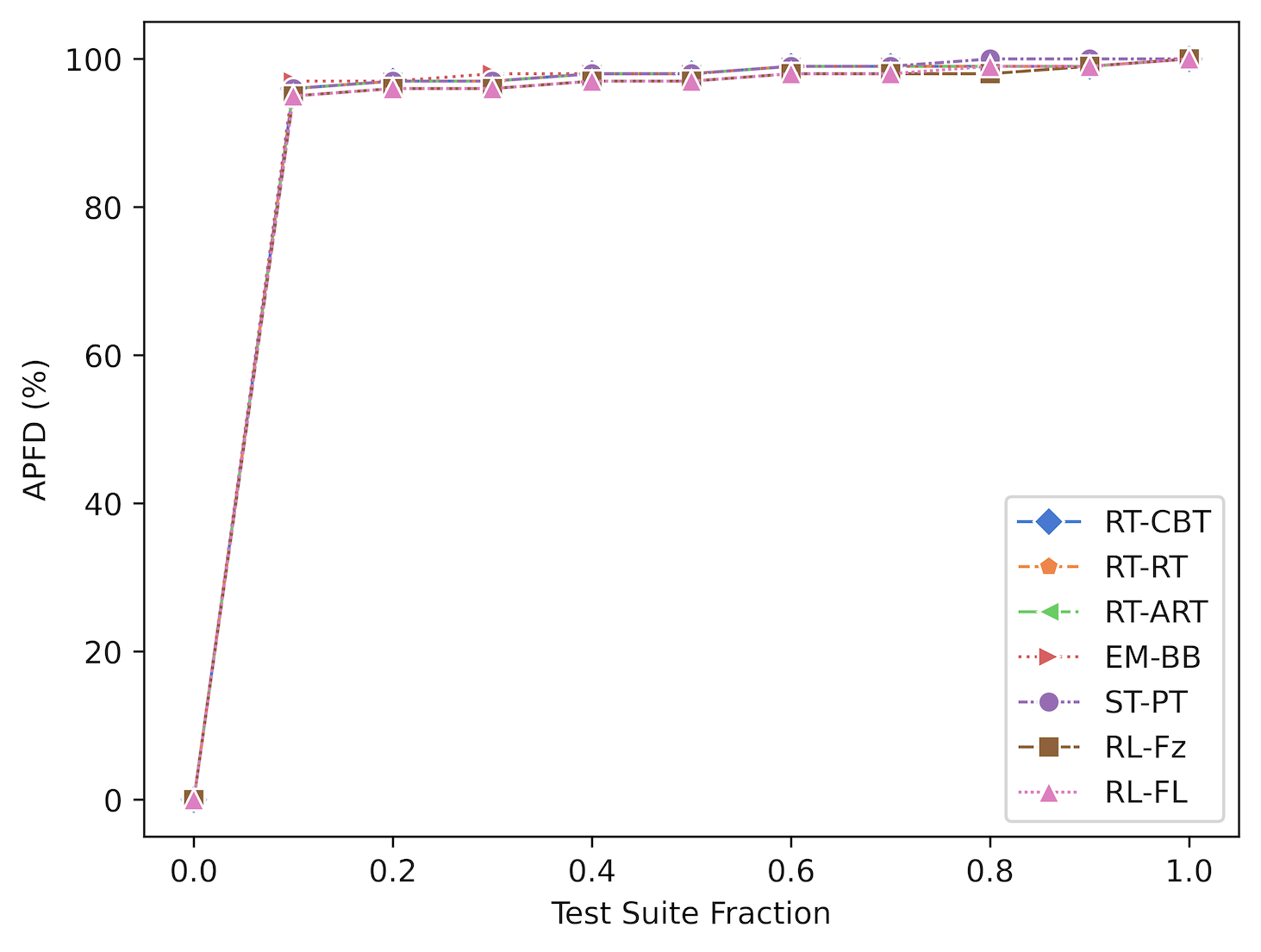}\label{fig:apfd9}}
    \subfigure[Mobile App]{\includegraphics[width=3.7cm,height=2.5cm]{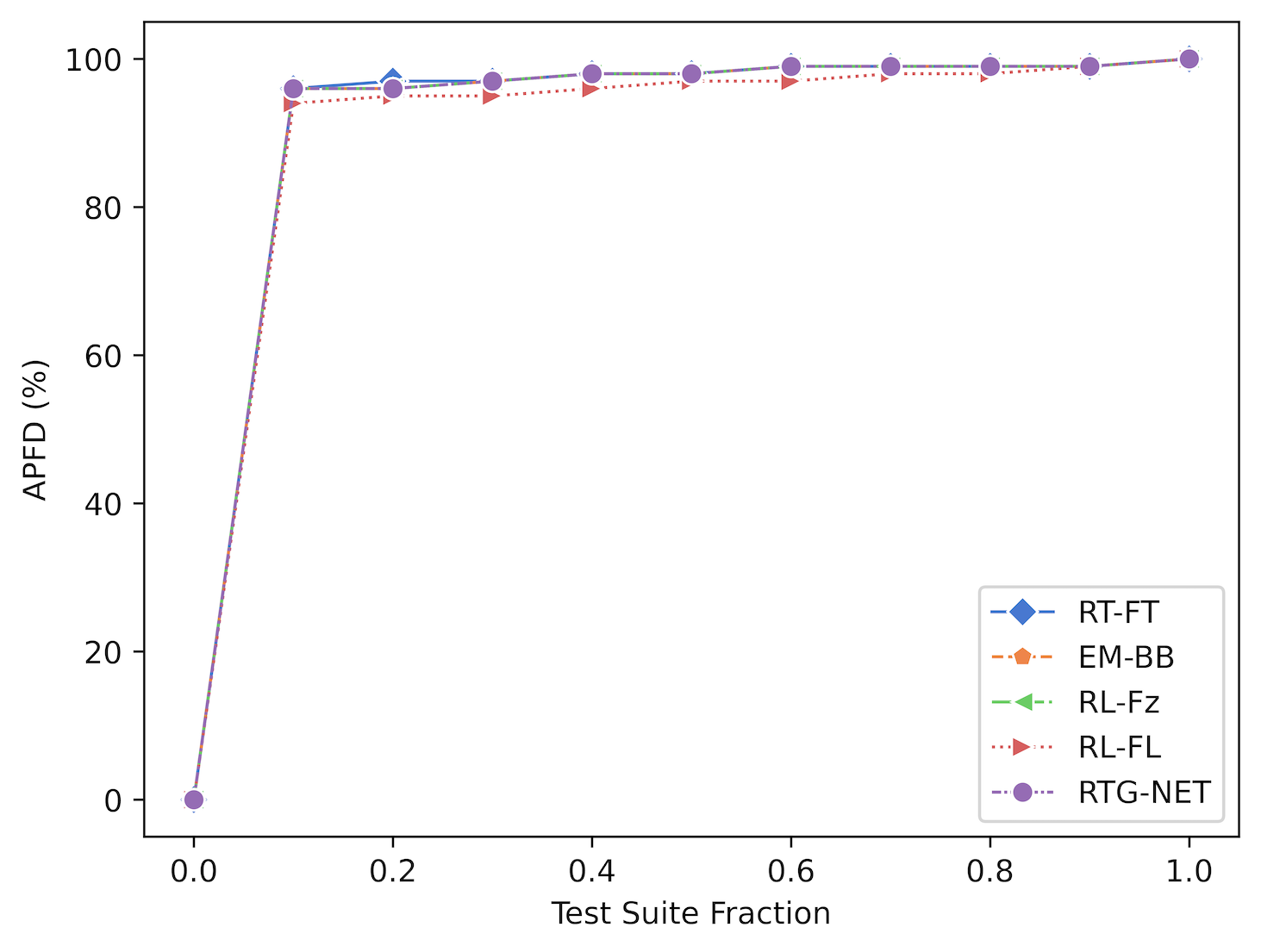}\label{fig:apfd10}}
    \subfigure[Reimbursements]{\includegraphics[width=3.7cm,height=2.5cm]{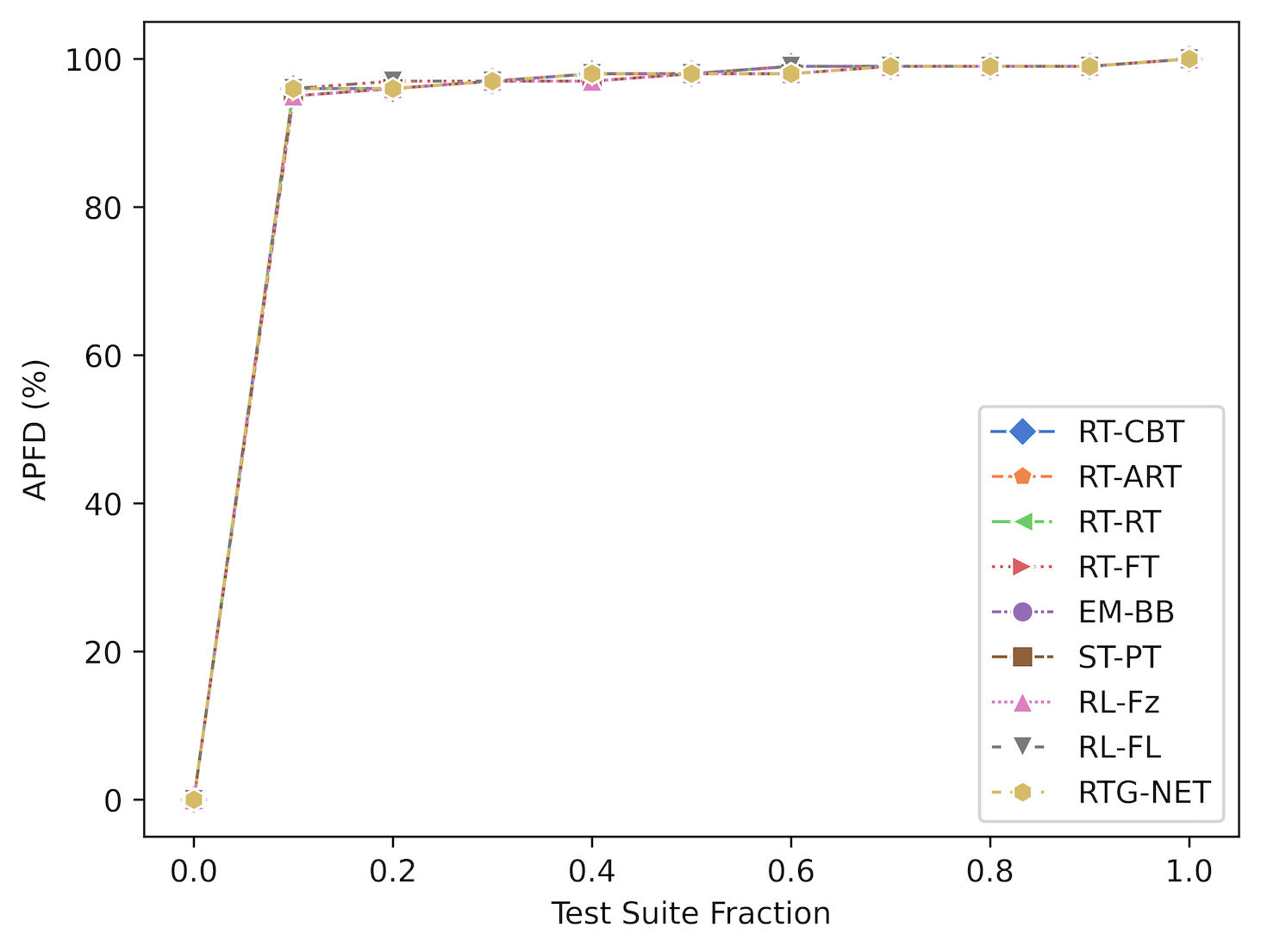}\label{fig:apfd11}}
    \subfigure[User Tasks]{\includegraphics[width=3.7cm,height=2.5cm]{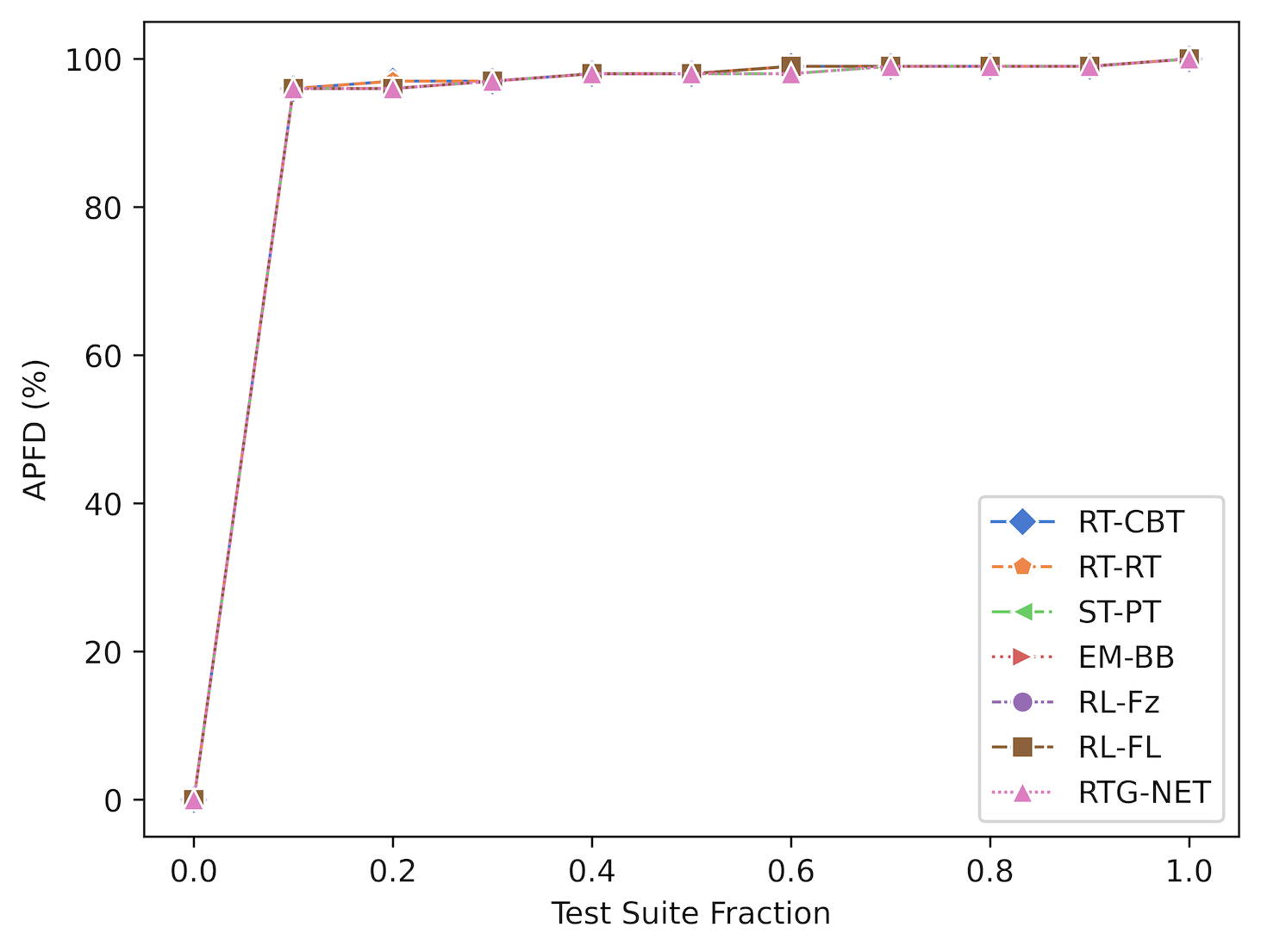}\label{fig:apfd12}} 

    \subfigure[Invoicing]{\includegraphics[width=3.7cm,height=2.5cm]{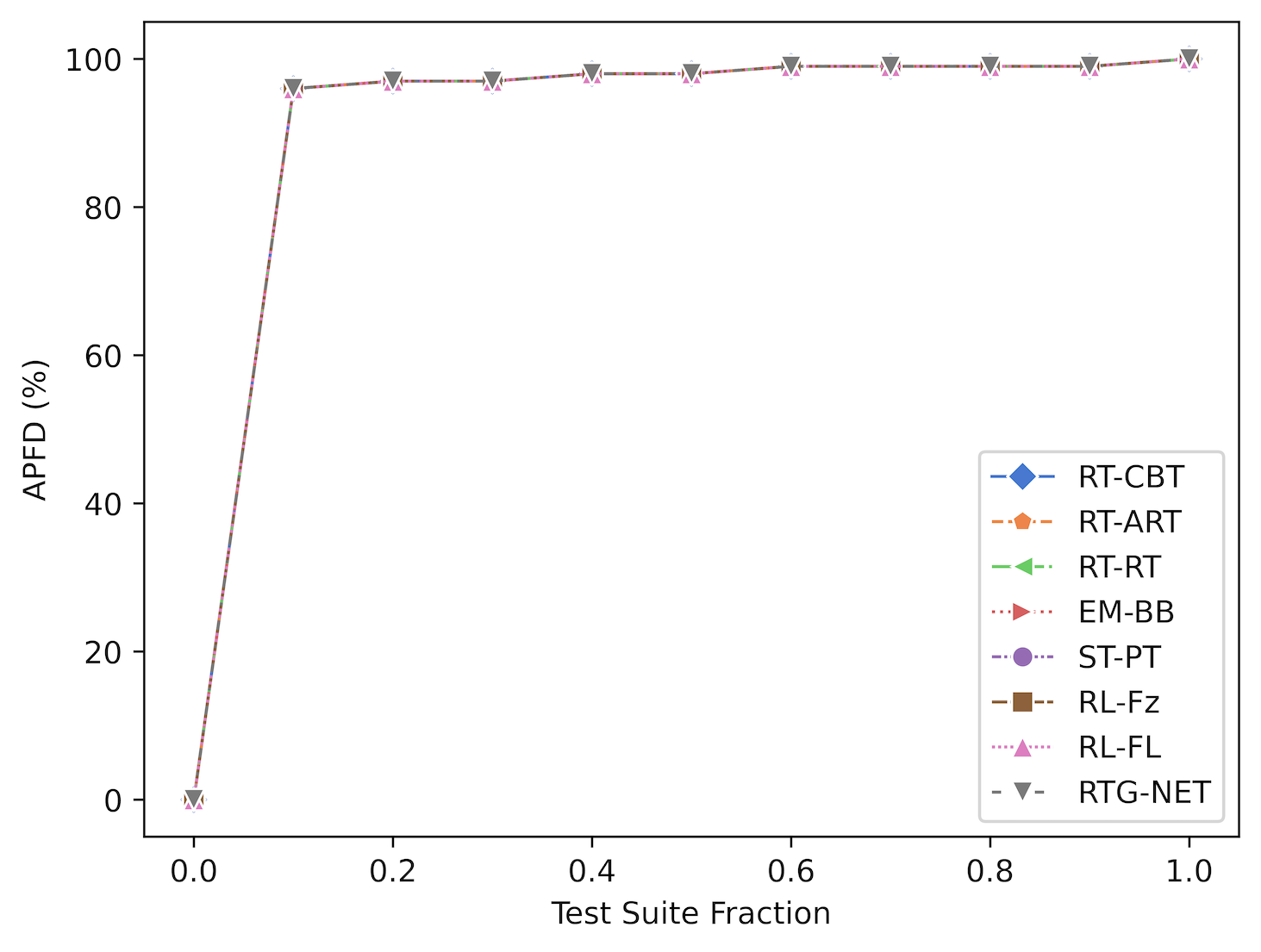}\label{fig:apfd13}}
    \subfigure[Catalog]{\includegraphics[width=3.7cm,height=2.5cm]{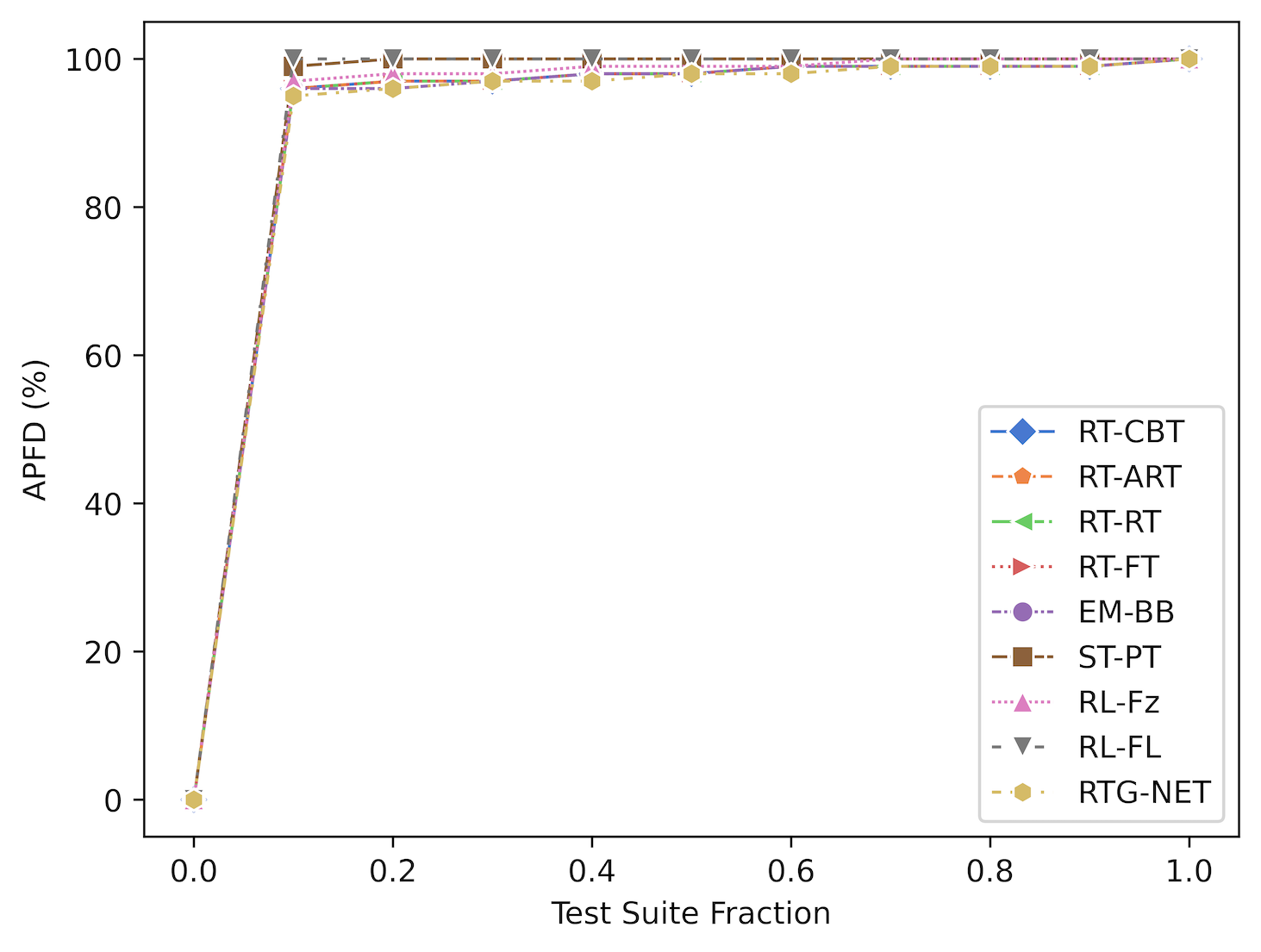}\label{fig:apfd14}}
    \subfigure[Report]{\includegraphics[width=3.7cm,height=2.5cm]{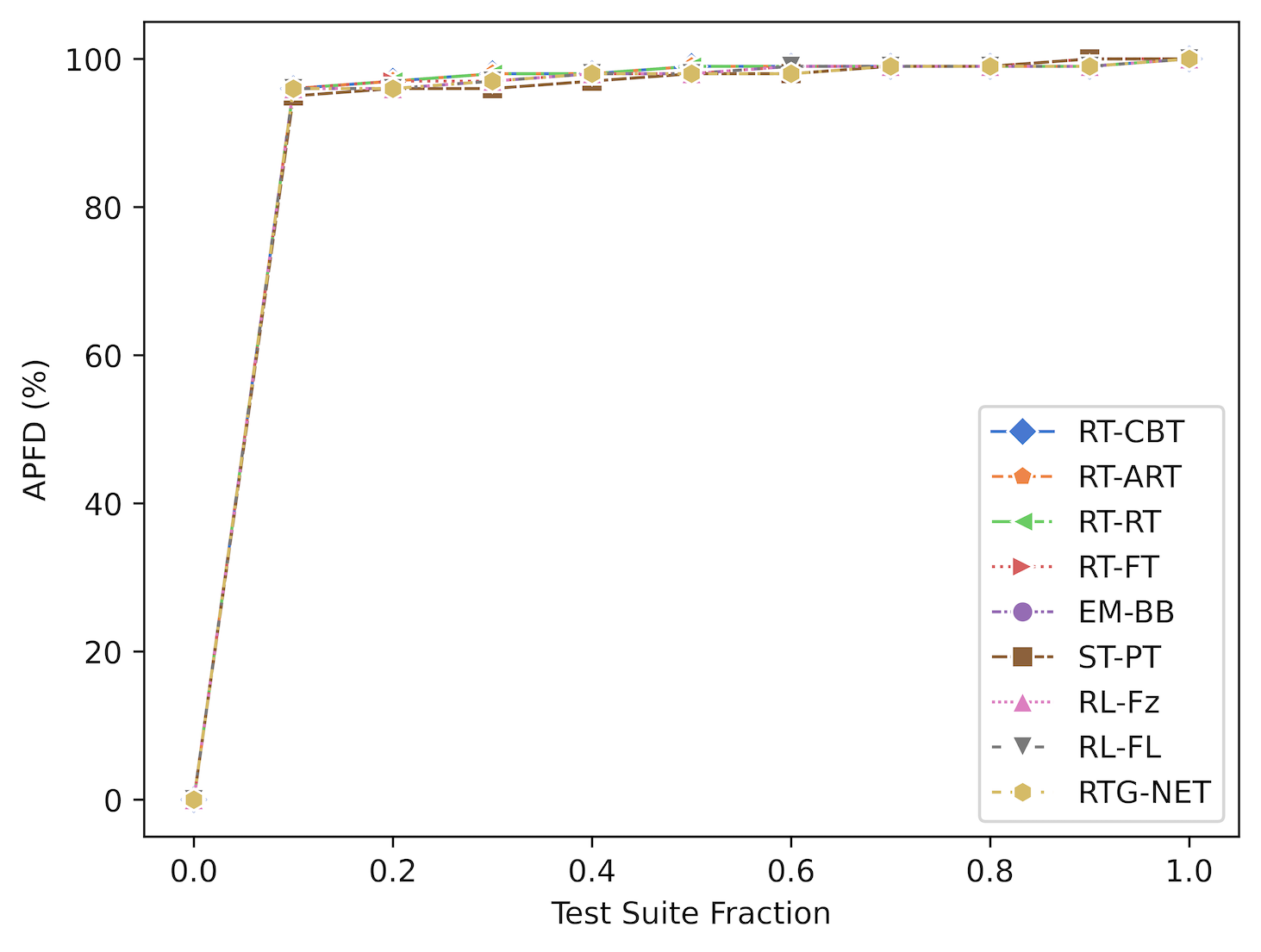}\label{fig:apfd15}}
    \subfigure[Care]{\includegraphics[width=3.7cm,height=2.5cm]{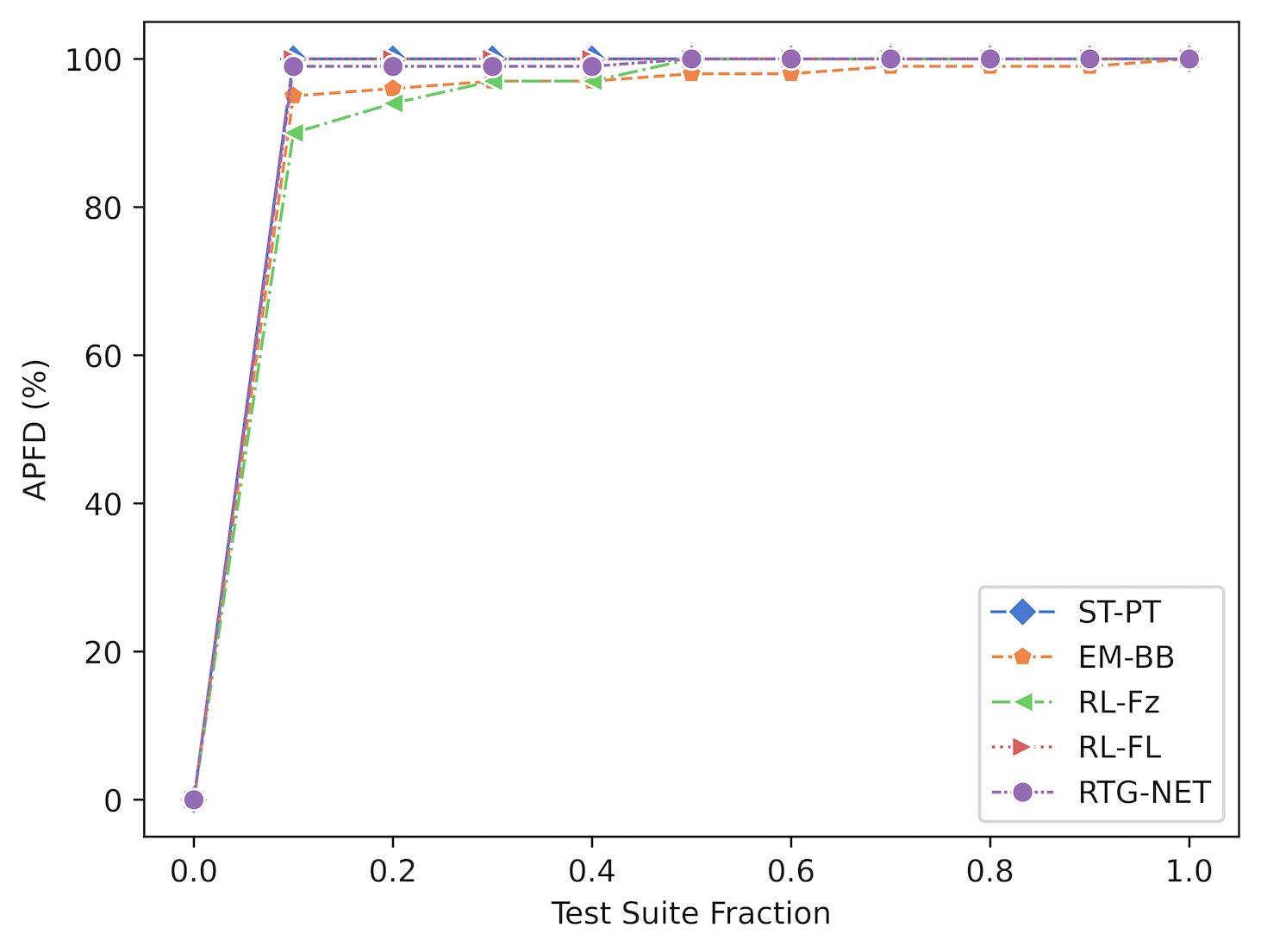}\label{fig:apfd16}}
    \caption{APFD over the test suite fraction for all individual tool variants across the APIs under test, illustrating the comparison of tools configured with specific testing techniques. } 
    \label{fig:apfd}
\end{figure*}

\textcolor{black}{
Regarding fault detection over time, \Cref{fig:apfd} presents APFD results for each tool across all tested APIs, excluding the \textit{Development} API, for which no faults were detected. 
Analyzing the APFDs indicates that all tools exhibited comparable performance across the entire test suite. 
For most APIs, almost all tools achieved close to 100\% APFD within the first 20\% of test executions. 
In some cases, some tools continued to detect faults up to 60\% of the test executions, such as \RestTestGenNET{} for the \textit{Catalog} API (\Cref{fig:apfd14}) and \RESTlerF{} for the \textit{Care} API (\Cref{fig:apfd16}). 
These findings suggest that allocating a testing budget of up to 60\% is generally sufficient for REST API testing tools to detect most faults. 
}

\vspace{5pt}
\begin{rqres}{rq2res}
Tests from \RESTestCBT{} and \RESTestRT{} led to finding the highest number of unique potential faults in different APIs. 
Overall, 18 potential faults were discovered across all releases from the tests generated by all REST API testing tools. 
These 18 faults belong to various categories, including database failures, business logic flaws, missing data validation, and security-related issues, fitting within the predefined classification~\cite{marculescu2022faults}. 
Furthermore, for most tools, the highest average percentage of faults was detected within the first 50\% of tests, whereas for some tools, additional faults were detected throughout the entire test suite. 
\end{rqres}

\begin{table}[H]
    \noindent
    \tiny
    \centering
    \caption{RQ2 results: Unique potential faults (PF) and their corresponding ratios for each tool across releases R1--R14 and all runs. Releases tagged with D, W, and M, denote daily, weekly, and monthly releases, respectively.}
    
    \begin{tabular}{@{}l l l l l l l l l l l l l l l l l l l@{}}
    \toprule
      \multicolumn{1}{l }{\textbf{}} & \multicolumn{2}{c }{\textbf{\RTCBT{}}} & \multicolumn{2}{c }{\textbf{\RTART{}}}& \multicolumn{2}{c }{\textbf{\RTRT{}}} & \multicolumn{2}{c }{\textbf{\RTFT{}}} & \multicolumn{2}{c }{\textbf{\EMBB{}}} & \multicolumn{2}{c }{\textbf{\STPT{}}}& \multicolumn{2}{c }{\textbf{\RLFz{}{}}} & \multicolumn{2}{c }{\textbf{\RLFL{}}} & \multicolumn{2}{c }{\textbf{\RTGNET{}}} \\ 
	\cmidrule(lr){2-3} \cmidrule(ll){4-5} \cmidrule(ll){6-7}\cmidrule(ll){8-9}\cmidrule(ll){10-11}\cmidrule(ll){12-13}\cmidrule(ll){14-15}\cmidrule(ll){16-17}\cmidrule(ll){18-19}
	\multicolumn{1}{ l }{\textbf{Release}} & PF& Ratio & PF& Ratio& PF& Ratio& PF& Ratio& PF& Ratio& PF& Ratio& PF& Ratio& PF& Ratio& PF& Ratio\\ 
	\cmidrule(lr){1-1}\cmidrule(lr){2-3}\cmidrule(ll){4-5} \cmidrule(ll){6-7}\cmidrule(ll){8-9}\cmidrule(ll){10-11}\cmidrule(ll){12-13}\cmidrule(ll){14-15}\cmidrule(ll){16-17}\cmidrule(ll){18-19}
    
	\multicolumn{1}{ l }{\textbf{[W]R1}} &8&0.088\%&6&0.069\%&9&0.099\%&7&0.118\%&3&0.002\%&3&0.024\%&5&0.028\%&5&0.032\%&3&0.014\%\\
        \multicolumn{1}{ l }{\textbf{[D]R2}} &2&0.022\%&3&0.035\%&2&0.022\%&3&0.05\%&5&0.004\%&4&0.033\%&5&0.028\%&5&0.032\%&4&0.019\%\\
        \multicolumn{1}{ l }{\textbf{[D]R3}} &2&0.022\%&3&0.035\%&2&0.022\%&4&0.067\%&5&0.004\%&4&0.033\%&5&0.028\%&5&0.032\%&6&0.028\%\\
        \multicolumn{1}{ l }{\textbf{[W]R4}} &4&0.044\%&4&0.046\%&4&0.044\%&5&0.084\%&7&0.006\%&6&0.049\%&5&0.028\%&5&0.032\%&6&0.028\%\\
        \multicolumn{1}{ l }{\textbf{[D]R5}} &4&0.044\%&4&0.046\%&4&0.044\%&5&0.084\%&7&0.006\%&6&0.049\%&5&0.028\%&5&0.032\%&7&0.033\%\\
        \multicolumn{1}{ l }{\textbf{[D]R6}} &4&0.044\%&4&0.046\%&4&0.044\%&4&0.067\%&7&0.006\%&6&0.049\%&5&0.028\%&5&0.032\%&7&0.033\%\\
        \multicolumn{1}{ l }{\textbf{[W]R7}} &4&0.044\%&4&0.046\%&4&0.044\%&4&0.067\%&7&0.006\%&6&0.049\%&5&0.028\%&5&0.032\%&7&0.033\%\\
        \multicolumn{1}{ l }{\textbf{[D]R8}} &4&0.044\%&4&0.046\%&4&0.044\%&5&0.084\%&7&0.006\%&6&0.049\%&5&0.028\%&5&0.032\%&6&0.028\%\\
        \multicolumn{1}{ l }{\textbf{[D]R9}} &4&0.044\%&4&0.046\%&4&0.044\%&5&0.084\%&6&0.005\%&6&0.049\%&5&0.028\%&5&0.032\%&6&0.028\%\\
        \multicolumn{1}{ l }{\textbf{[W]R10}} &4&0.044\%&4&0.046\%&4&0.044\%&5&0.084\%&6&0.005\%&6&0.049\%&5&0.028\%&5&0.032\%&7&0.033\%\\
        \multicolumn{1}{ l }{\textbf{[D]R11}} &4&0.044\%&4&0.046\%&4&0.044\%&5&0.084\%&7&0.006\%&6&0.049\%&5&0.028\%&5&0.032\%&7&0.033\%\\
        \multicolumn{1}{ l }{\textbf{[D]R12}} &3&0.033\%&3&0.035\%&2&0.022\%&4&0.067\%&6&0.005\%&5&0.041\%&5&0.028\%&5&0.032\%&6&0.028\%\\
        \multicolumn{1}{ l }{\textbf{[W]R13}} &2&0.022\%&2&0.023\%&3&0.033\%&3&0.05\%&5&0.004\%&5&0.041\%&5&0.028\%&5&0.032\%&6&0.028\%\\
        \multicolumn{1}{ l }{\textbf{[M]R14}} &3&0.033\%&3&0.035\%&2&0.022\%&4&0.067\%&5&0.004\%&5&0.041\%&5&0.028\%&5&0.032\%&6&0.028\%\\

	\bottomrule
	\end{tabular}
    
 \label{tab-rq2results}
\end{table}

\begin{table}[H]
    \noindent
    \footnotesize
    \centering
    \caption{RQ2 results: Overall comparison of tools in terms of unique potential faults (PF) and their corresponding ratios for each tool across releases R1--R14 and all runs. Releases tagged with D, W, and M, denote daily, weekly, and monthly releases, respectively. }
    
    \begin{tabular}{@{}l l l l l l l l l l l@{}}
    \toprule 
    \multicolumn{1}{l }{\textbf{}} & \multicolumn{2}{c }{\textbf{\RESTest{}}} & \multicolumn{2}{c }{\textbf{\EvoMaster{}}} & \multicolumn{2}{c }{\textbf{\Schemathesis{}}} & \multicolumn{2}{c }{\textbf{\RESTler{}}} & \multicolumn{2}{c }{\textbf{\RestTestGen{}}} \\ 
    \cmidrule(lr){2-3} \cmidrule(ll){4-5} \cmidrule(ll){6-7} \cmidrule(ll){8-9} \cmidrule(ll){10-11}
    \multicolumn{1}{ l }{\textbf{Release}} & PF & Ratio & PF & Ratio & PF & Ratio & PF & Ratio & PF & Ratio \\ 
    \cmidrule(lr){1-1}\cmidrule(lr){2-3}\cmidrule(ll){4-5} \cmidrule(ll){6-7} \cmidrule(ll){8-9} \cmidrule(ll){10-11}
    
        \multicolumn{1}{ l }{\textbf{[W]R1}} & 12 & 0.037\% & 3 & 0.002\% & 3 & 0.024\% & 5 & 0.015\% & 3 & 0.014\% \\
        \multicolumn{1}{ l }{\textbf{[D]R2}} & 5 & 0.015\% & 5 & 0.004\% & 4 & 0.033\% & 5 & 0.015\% & 4 & 0.019\% \\
        \multicolumn{1}{ l }{\textbf{[D]R3}} & 5 & 0.015\% & 5 & 0.004\% & 4 & 0.033\% & 5 & 0.015\% & 6 & 0.028\% \\
        \multicolumn{1}{ l }{\textbf{[W]R4}} & 8 & 0.024\% & 7 & 0.006\% & 6 & 0.049\% & 5 & 0.015\% & 6 & 0.028\% \\
        \multicolumn{1}{ l }{\textbf{[D]R5}} & 8 & 0.024\% & 7 & 0.006\% & 6 & 0.049\% & 5 & 0.015\% & 7 & 0.033\% \\
        \multicolumn{1}{ l }{\textbf{[D]R6}} & 7 & 0.021\% & 7 & 0.006\% & 6 & 0.049\% & 5 & 0.015\% & 7 & 0.033\% \\
        \multicolumn{1}{ l }{\textbf{[W]R7}} & 7 & 0.021\% & 7 & 0.006\% & 6 & 0.049\% & 5 & 0.015\% & 7 & 0.033\% \\
        \multicolumn{1}{ l }{\textbf{[D]R8}} & 9 & 0.027\% & 7 & 0.006\% & 6 & 0.049\% & 5 & 0.015\% & 6 & 0.028\% \\
        \multicolumn{1}{ l }{\textbf{[D]R9}} & 8 & 0.024\% & 6 & 0.005\% & 6 & 0.049\% & 5 & 0.015\% & 6 & 0.028\% \\
        \multicolumn{1}{ l }{\textbf{[W]R10}} & 8 & 0.024\% & 6 & 0.005\% & 6 & 0.049\% & 5 & 0.015\% & 7 & 0.033\% \\
        \multicolumn{1}{ l }{\textbf{[D]R11}} & 9 & 0.027\% & 7 & 0.006\% & 6 & 0.049\% & 5 & 0.015\% & 7 & 0.033\% \\
        \multicolumn{1}{ l }{\textbf{[D]R12}} & 5 & 0.015\% & 6 & 0.005\% & 5 & 0.041\% & 5 & 0.015\% & 6 & 0.028\% \\
        \multicolumn{1}{ l }{\textbf{[W]R13}} & 4 & 0.012\% & 5 & 0.004\% & 5 & 0.041\% & 5 & 0.015\% & 6 & 0.028\% \\
        \multicolumn{1}{ l }{\textbf{[M]R14}} & 5 & 0.015\% & 5 & 0.004\% & 5 & 0.041\% & 5 & 0.015\% & 6 & 0.028\% \\
    \bottomrule
    
    \end{tabular}
    
 \label{tab-rq2results-agg}
\end{table}

\begin{table} [H]
    \noindent
    \caption{RQ2 results: Description of all potential faults, each associated with a specific fault ID and the HTTP status code, identified across all releases and all runs.}
    \centering
    \begin{tabular}{p{.03\textwidth}p{.11\textwidth}p{.78\textwidth}}\toprule
    \textbf{ID} & \textbf{Status code} & \textbf{Description}  \\
	\midrule
        F1 &502 \& 413& Server failed to handle the request/response, potentially large data or invalid data/response from another server.\\
        F2 &200& Invalid login details lead to success with the empty response.\\
        F3 &500& Server failed to understand the request. \\
        F4 &400& Error in parsing request headers. \\
        F5 &429& Multiple invalid login attempts continuously increase login time for the next attempt.\\
        F6 &200& Incorrect user ID results in success.  \\
        F7 &200& Empty/invalid JSON object for a device results in the success response with complete device information.\\
        F8 &409& Error while saving record based on ID which is not present in the database. \\
        F9 &200& Incorrect property values in JSON fetch all records for general tasks/surveys. \\
        F10 &500& Iterating over String type object results in server error. \\
        F11 &500& Database error when inserting records with missing fields. \\
        F12 &500& Server error when trying to schedule device's update\\
        F13 &400& Type mismatch when performing arithmetic operation.\\
        F14 &500& Error when converting Python Int type large value to Long in C.\\
        F15 &400& Performing array operation in Boolean type variable.\\
        F16 &500& Accessing Dict items from String/None type object.\\
        F17 &409& Violation of foreign key constraint in the database.\\
        F18 &500& Failure in handling long value results in a complete database query as part of the response. \\
	\bottomrule
	\end{tabular}
	\label{tab-rq2faultsinfo}
\end{table}

\subsubsection{RQ3 Results -- REST APIs Coverage}\label{RQ3Results}

For release R1, the results (\Cref{tab-rq3results}) show that all testing tools achieved an API coverage close to 77\%. 
From R2 to R14, the API coverage for all tools increased to 84\%. 
In the case of \RESTestCBT{} and \RESTestRT{}, API coverage reached approximately 81\%, while \RESTestART{} and \RESTestFT{} achieved slightly higher coverage of around 82\% from the R2 release onward.
The remaining tools \EvoMasterBB{}, \SchemathesisPT{}, \RESTlerF{}, \RESTlerFL{}, and \RestTestGenNET{} achieved a nearly consistent coverage of approximately 84\% across all releases from R2 through R14.
From the overall coverage results in \Cref{tab-rq3results-agg}, it can be observed that \EvoMaster{}, \Schemathesis{}, \RESTler{}, and \RestTestGen{} achieved comparable coverage, with only \RESTest{} showing slightly lower coverage.

Upon a detailed analysis of the coverage results generated by the Restats tool, we observed that all REST API testing tools easily achieve nearly 100\% path coverage, operation coverage, request and response content type coverage, parameter coverage, and parameter value coverage\footnote{A detailed description of each type of coverage is provided by Corradini et al.~\cite{corradini2021restats}.}. 
However, these tools could not achieve 100\% status code coverage and status code class coverage. 
As a result, their overall coverage reaches a maximum of only 84\%.  
Status code coverage is related to comparing returned status codes with those defined in the API schema~\cite{corradini2021restats}. 
This implies that the tests generated by these REST API testing tools could not cover all the documented status codes. 
Similarly, status code class coverage refers to the ratio of success codes (2XX) and error codes (4XX and 5XX) covered by tests~\cite{corradini2021restats}. 
This suggests that the tests generated by these REST API testing tools struggled to maintain a balance among different types of status code classes. 
It is also evident from the RQ1 results~\Cref{RQ1Results} that tests generated by all tools resulted in a higher number of 4XX errors than 5XX errors. 
The primary reason for the lower coverage of success codes (2XX) is that these tools often generated invalid inputs, resulting in more 4XX errors than successes.

The statistical comparison of the API test coverage, achieved by all the REST API testing tools, is presented in \Cref{tab-stcomp-rq1}.
It is noticeable that in most comparisons, no statistically significant differences were observed, which indicates that the tools' performance in terms of API coverage is comparable. 
In the few cases where significant differences were found, the effect sizes were small. 
Among the \RESTest{} variants, \RESTestART{} performed slightly better than \RESTestCBT{} and \RESTestRT{}. 
Similarly, \EvoMasterBB{}, \SchemathesisPT{}, \RESTlerF{}, and \RESTlerFL{} outperformed all four variants of \RESTest{} and \RestTestGenNET{}, while exhibiting comparable performance among themselves. 
Lastly, \RestTestGenNET{} only demonstrated better coverage compared to \RESTestCBT{}, \RESTestRT{}, and \RESTestFT{}.

\vspace{5pt}
\begin{rqres}{rq3res}
In terms of achieving higher API test coverage, \EvoMasterBB{}, \SchemathesisPT{}, \RESTlerF{}, and \RESTlerFL{} performed relatively well, reaching a maximum of 84\%. 
The main factor limiting full API test coverage is that none of the testing tools managed to reach 100\% in both status code coverage and status code class coverage. 
\end{rqres}

\begin{table} [H]
    \centering
    \footnotesize
    \noindent
    \caption{RQ3 results: REST API coverage achieved by each testing tool for each release from R1 to R14, reported as average values across 10 runs. Releases tagged with D, W, and M, denote daily, weekly, and monthly releases, respectively. }
    \begin{tabular}{@{} cl*{9}c @{}}\toprule 
        & & \RTCBT{} & \RTART{} & \RTRT{} & \RTFT{} 
        & \EMBB{} & \STPT{} & \RLFz{} 
        & \RLFL{} & \RTGNET{} \\
		\cmidrule{2-11}
        &\textbf{[W]R1} &76.73\%&78.7\%&76.73\%&77.62\%&77.16\%&77.16\%&77.16\%&77.16\%&76.67\%\\
        &\textbf{[D]R2} &81.18\%&82.57\%&81.21\%&82.01\%&83.81\%&84.07\%&84.04\%&84.04\%&83.93\%\\
        &\textbf{[D]R3} &81.29\%&82.67\%&81.29\%&82.06\%&84.12\%&84.13\%&84.13\%&84.12\%&84.0\%\\
        &\textbf{[W]R4} &81.3\%&82.67\%&81.3\%&82.06\%&84.14\%&84.14\%&84.14\%&84.14\%&84.01\%\\
        &\textbf{[D]R5} &81.32\%&82.67\%&81.32\%&82.06\%&84.16\%&84.16\%&84.16\%&84.16\%&84.02\%\\
        &\textbf{[D]R6} &81.32\%&82.67\%&81.32\%&82.06\%&84.16\%&84.16\%&84.16\%&84.16\%&84.02\%\\
        &\textbf{[W]R7} &81.33\%&82.67\%&81.35\%&82.06\%&84.17\%&84.18\%&84.17\%&84.17\%&84.05\%\\
        &\textbf{[D]R8} &81.35\%&82.67\%&81.35\%&82.06\%&84.18\%&84.18\%&84.18\%&84.18\%&84.05\%\\
        \rotnd{\rlap{~\textbf{Releases}}}
        &\textbf{[D]R9} &81.35\%&82.67\%&81.35\%&82.06\%&84.18\%&84.18\%&84.18\%&84.18\%&84.05\%\\
        &\textbf{[W]R10} &81.35\%&82.67\%&81.35\%&82.06\%&84.18\%&84.18\%&84.18\%&84.18\%&84.05\%\\
        &\textbf{[D]R11} &81.35\%&82.67\%&81.35\%&82.06\%&84.18\%&84.18\%&84.18\%&84.18\%&84.05\%\\
        &\textbf{[D]R12} &81.35\%&82.67\%&81.35\%&82.06\%&84.18\%&84.18\%&84.18\%&84.18\%&84.05\%\\
        &\textbf{[W]R13} &81.36\%&82.67\%&81.36\%&82.06\%&84.19\%&84.19\%&84.19\%&84.19\%&84.07\%\\
        &\textbf{[M]R14} &81.36\%&82.67\%&81.36\%&82.06\%&84.19\%&84.19\%&84.19\%&84.19\%&84.07\%\\
        
		\bottomrule
	\end{tabular}
	\label{tab-rq3results}
\end{table}

\begin{table} [H]
    \centering
    \footnotesize
    \noindent
    \caption{RQ3 results: Overall comparison of tools in terms of REST API coverage achieved by each testing tool for each release from R1 to R14, reported as average values across 10 runs. Releases tagged with D, W, and M, denote daily, weekly, and monthly releases, respectively. }
    
    \begin{tabular}{@{} cl*{5}c @{}}\toprule
        & & \textbf{\RESTest{}} & \textbf{\EvoMaster{}} & \textbf{\Schemathesis{}} & \textbf{\RESTler{}} & \textbf{\RestTestGen{}} \\
        \cmidrule{2-7}
        &\textbf{[W]R1} &77.45\% &77.16\% &77.16\% &77.16\% &76.67\% \\
        &\textbf{[D]R2} &81.99\% &83.81\% &84.07\% &84.04\% &83.93\% \\
        &\textbf{[D]R3} &82.08\% &84.12\% &84.13\% &84.13\% &84.00\% \\
        &\textbf{[W]R4} &82.08\% &84.14\% &84.14\% &84.14\% &84.01\% \\
        &\textbf{[D]R5} &82.09\% &84.16\% &84.16\% &84.16\% &84.02\% \\
        &\textbf{[D]R6} &82.09\% &84.16\% &84.16\% &84.16\% &84.02\% \\
        &\textbf{[W]R7} &82.10\% &84.17\% &84.18\% &84.17\% &84.05\% \\
        &\textbf{[D]R8} &82.10\% &84.18\% &84.18\% &84.18\% &84.05\% \\
        \rotnd{\rlap{~\textbf{Releases}}}
        &\textbf{[D]R9} &82.10\% &84.18\% &84.18\% &84.18\% &84.05\% \\
        &\textbf{[W]R10} &82.10\% &84.18\% &84.18\% &84.18\% &84.05\% \\
        &\textbf{[D]R11} &82.10\% &84.18\% &84.18\% &84.18\% &84.05\% \\
        &\textbf{[D]R12} &82.10\% &84.18\% &84.18\% &84.18\% &84.05\% \\
        &\textbf{[W]R13} &82.11\% &84.19\% &84.19\% &84.19\% &84.07\% \\
        &\textbf{[M]R14} &82.11\% &84.19\% &84.19\% &84.19\% &84.07\% \\
	\bottomrule
	\end{tabular}
    
    \label{tab-rq3results-agg}
\end{table}

\begin{table}[H]
    \noindent
    \small
    \rotatebox{90}{
    \begin{minipage}{1\textheight}
    \centering
    \caption{Statistical comparison of testing tools based on REST API coverage across releases R1--R14 and all runs. It includes the \textit{p} values calculated using the Wilcoxon test and the $\hat{A}_{12}$ values derived from the Vargha-Delaney effect size measure. }
    
	\begin{tabular}{@{}l l l l l l l l l l l l l l l l l l l@{}}\toprule 
        \multicolumn{1}{l }{\textbf{}} & \multicolumn{2}{c }{\textbf{\RTCBT{}}} & \multicolumn{2}{c }{\textbf{\RTART{}}}& \multicolumn{2}{c }{\textbf{\RTRT{}}} & \multicolumn{2}{c }{\textbf{\RTFT{}}} & \multicolumn{2}{c }{\textbf{\EMBB{}}} & \multicolumn{2}{c }{\textbf{\STPT{}}}& \multicolumn{2}{c }{\textbf{\RLFz{}{}}} & \multicolumn{2}{c }{\textbf{\RLFL{}}} & \multicolumn{2}{c }{\textbf{\RTGNET{}}} \\ 
	\cmidrule(lr){2-3}\cmidrule(ll){4-5} \cmidrule(ll){6-7}\cmidrule(ll){8-9}\cmidrule(ll){10-11}\cmidrule(ll){12-13}\cmidrule(ll){14-15}\cmidrule(ll){16-17}\cmidrule(ll){18-19}
	\multicolumn{1}{ l }{\textbf{}}  & \textit{p}-val& $\hat{A}_{12}$& \textit{p}-val& $\hat{A}_{12}$& \textit{p}-val& $\hat{A}_{12}$& \textit{p}-val& $\hat{A}_{12}$& \textit{p}-val& $\hat{A}_{12}$& \textit{p}-val& $\hat{A}_{12}$& \textit{p}-val& $\hat{A}_{12}$& \textit{p}-val& $\hat{A}_{12}$& \textit{p}-val& $\hat{A}_{12}$\\ 
	\cmidrule(lr){1-1}\cmidrule(lr){2-3}\cmidrule(ll){4-5} \cmidrule(ll){6-7}\cmidrule(ll){8-9}\cmidrule(ll){10-11}\cmidrule(ll){12-13}\cmidrule(ll){14-15}\cmidrule(ll){16-17}\cmidrule(ll){18-19} 
        
    \multicolumn{1}{ l }{\RTCBT{}} &-&-&\textcolor{gray}{1.0}&0.419&\textcolor{gray}{1.0}&0.5&\textcolor{gray}{1.0}&\textbf{0.527}&\textcolor{gray}{1.0}&0.394&\textcolor{gray}{1.0}&0.394&\textcolor{gray}{1.0}&0.394&\textcolor{gray}{1.0}&0.394&\textcolor{gray}{1.0}&0.445\\
    \multicolumn{1}{ l }{\RTART{}} &<0.05&\textbf{0.581}&-&-&<0.05&\textbf{0.581}&\textcolor{gray}{0.133}&\textbf{0.56}&\textcolor{gray}{1.0}&0.476&\textcolor{gray}{1.0}&0.476&\textcolor{gray}{1.0}&0.476&\textcolor{gray}{1.0}&0.476&\textcolor{gray}{1.0}&0.48\\
    \multicolumn{1}{ l }{\RTRT{}} &\textcolor{gray}{1.0}&\textbf{0.5}&\textcolor{gray}{1.0}&0.419&-&-&\textcolor{gray}{1.0}&\textbf{0.527}&\textcolor{gray}{1.0}&0.394&\textcolor{gray}{1.0}&0.394&\textcolor{gray}{1.0}&0.394&\textcolor{gray}{1.0}&0.394&\textcolor{gray}{1.0}&0.445\\
    \multicolumn{1}{ l }{\RTFT{}} &\textcolor{gray}{1.0}&0.473&\textcolor{gray}{1.0}&0.44&\textcolor{gray}{1.0}&0.473&-&-&\textcolor{gray}{1.0}&0.393&\textcolor{gray}{1.0}&0.393&\textcolor{gray}{1.0}&0.393&\textcolor{gray}{1.0}&0.393&\textcolor{gray}{1.0}&0.397\\
    \multicolumn{1}{ l }{\EMBB{}} &<0.05&\textbf{0.606}&<0.05&\textbf{0.524}&<0.05&\textbf{0.606}&<0.05&\textbf{0.607}&-&-&\textcolor{gray}{1.0}&0.498&\textcolor{gray}{1.0}&0.499&\textcolor{gray}{1.0}&0.499&0.003&\textbf{0.527}\\
    \multicolumn{1}{ l }{\STPT{}} &<0.05&\textbf{0.606}&<0.05&\textbf{0.524}&<0.05&\textbf{0.606}&<0.05&\textbf{0.607}&\textcolor{gray}{1.0}&\textbf{0.502}&-&-&\textcolor{gray}{1.0}&\textbf{0.501}&\textcolor{gray}{1.0}&\textbf{0.501}&0.001&\textbf{0.527}\\
    \multicolumn{1}{ l }{\RLFz{}} &<0.05&\textbf{0.606}&<0.05&\textbf{0.524}&<0.05&\textbf{0.606}&<0.05&\textbf{0.607}&\textcolor{gray}{1.0}&\textbf{0.501}&\textcolor{gray}{1.0}&0.499&-&-&\textcolor{gray}{1.0}&\textbf{0.5}&0.001&\textbf{0.527}\\
    \multicolumn{1}{ l }{\RLFL{}} &<0.05&\textbf{0.606}&<0.05&\textbf{0.524}&<0.05&\textbf{0.606}&<0.05&\textbf{0.607}&\textcolor{gray}{1.0}&\textbf{0.501}&\textcolor{gray}{1.0}&0.499&\textcolor{gray}{1.0}&0.5&-&-&0.002&\textbf{0.527}\\
    \multicolumn{1}{ l }{\RTGNET{}} &<0.05&\textbf{0.555}&\textcolor{gray}{0.994}&\textbf{0.52}&<0.05&\textbf{0.555}&<0.05&\textbf{0.603}&\textcolor{gray}{1.0}&0.473&\textcolor{gray}{1.0}&0.473&\textcolor{gray}{1.0}&0.473&\textcolor{gray}{1.0}&0.473&-&-\\
        
	\bottomrule
        \multicolumn{19}{l}{* The values shown in gray color indicate that $p-value>\alpha$, i.e., there is no statistically significant difference. The values presented in bold indicate that the}\\
        \multicolumn{19}{l}{tool listed on the left side outperformed the tool on the right side. Instances of self-comparison of tools are left blank in the table.}
	\end{tabular}
    
 \label{tab-stcomp-rq3}
 \end{minipage}}
\end{table}

\subsubsection{RQ4 Results -- Regressions}\label{RQ4Results}

The results for potential regressions identified between two subsequent releases, using the tests generated by all testing tools, are presented in \Cref{tab-rq4results}. 
It is important to note that we used the initial API specifications, which included all possible response types and remained unchanged throughout the experiment. 
Therefore, the regressions reported in our results were not documented in the specifications. 
It is evident that the tests from \EvoMasterBB{} and \SchemathesisPT{} led to finding a higher number of regressions in all release changelogs. 
The tests generated by \RESTlerF{} resulted in locating regressions in nearly all release changelogs, with the exceptions of R4-R5 and R7-R8. 
In the case of \RESTest{}, the tests from \RESTestART{} led to finding more regressions compared to \RESTestCBT{}, \RESTestRT{}, and \RESTestFT{}. 
Moreover, \RESTlerFL{} only performed better than all variants of \RESTest{}.
Both \RESTestRT{} and \RestTestGenNET{} showed equivalent performance in identifying only three regressions in the same set of release changelogs.
The overall tools comparison results in \Cref{tab-rq4results-agg} show that test cases generated by \EvoMaster{} led to finding the highest number of regressions in nearly all releases. 
However, the fewest regressions were identified from \RestTestGen{}-generated tests, while the remaining tools perform comparably.

\begin{figure}[H]
\centerline{\includegraphics[width=\linewidth, keepaspectratio]{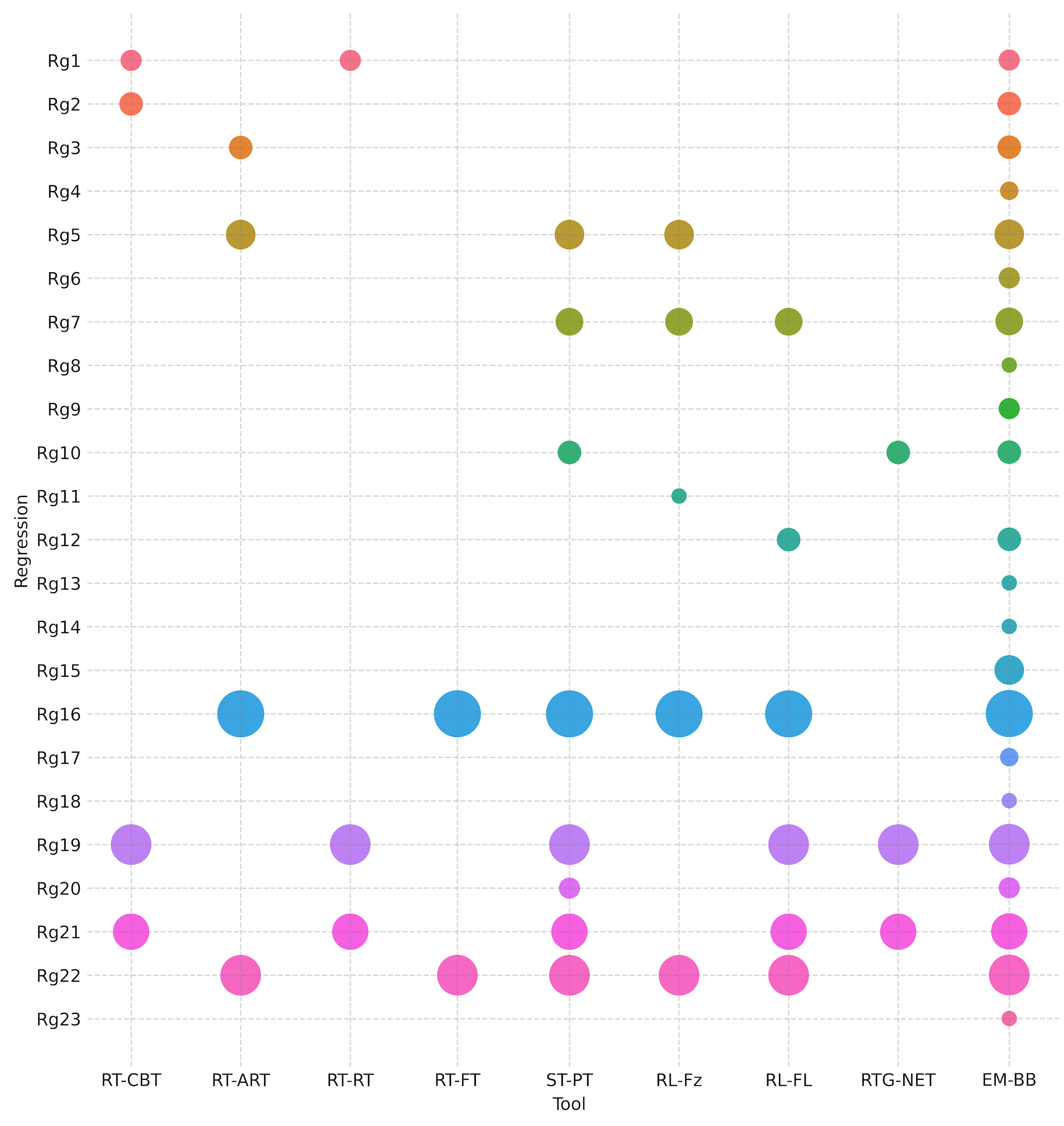}}
\caption{Regressions and their frequency of occurrence corresponding to each testing tool across releases R1--R14 and all runs. 
In this plot, each color-coded bubble represents a specific regression (shown on the y-axis) identified from the test cases generated by all tools (listed on the x-axis). The size of the bubble shows the frequency of the regression. 
Key findings indicate that most regressions were detected from the tests generated by \EvoMasterBB{}. 
Among all regressions, Rg16, Rg19, Rg21, and Rg22 occurred most frequently. 
Moreover, regression Rg11 was only found from tests generated by \RESTlerF{}, and Rg23 was solely detected from \EvoMasterBB{}-generated tests. 
Detailed descriptions of all regressions (Rg1--Rg23) are presented in~\Cref{tab-rq4regsinfo}. 
}
\label{fig:rq4regressions}
\end{figure}

In total, 23 unique potential regressions (Rg1-Rg23) were identified using the tests generated by all tools. 
It was observed that certain regressions appeared in multiple release changelogs while others were unique to particular tools. 
\Cref{tab-rq4regsinfo} outlines all identified regressions corresponding to different release changes. 
In addition, it was noted that certain regressions reappeared several times, from R1 to R14. 
The frequency of each regression's occurrence, corresponding to the tools whose tests revealed these regressions, is depicted in \Cref{fig:rq4regressions}.

Since the identified regressions in our study could be due to unintended faults or behavioral differences, some appear as fixes, others as enhancements, or failure propagation. 
Therefore, we classify all 23 regressions into four categories, as outlined below. 
\begin{enumerate}
    \item \textbf{Success-to-Failure:} APIs that initially returned a successful 200 status code, but in the next release, resulted in a 4XX/5XX error response. This implies that a functionality working in one release has failed in the next. The regressions classified under this category include Rg1, Rg3, Rg13, Rg14, and Rg18.
    \item \textbf{Failure-to-Success:} APIs that initially returned a 4XX/5XX error response but led to a successful 200 status code in the subsequent release. This suggests that functionality not working (potentially due to a bug) in one release has been fixed in the next release. The regressions Rg4, Rg6, Rg8, Rg10, Rg12, and Rg15 belong to this category.
    \item \textbf{Success-Enhancement/Reduction:} APIs that returned a successful 200 status code in both the initial and subsequent releases, but the response in the next release exhibited new enhancements or reductions. The only regression that falls into this category is Rg2. 
    \item \textbf{Failure-Shift:} APIs that initially returned one type of client/server-side error but later returned a different type of error. This indicates a shift in the type of failures. The regressions Rg5, Rg7, Rg9, Rg11, Rg16, Rg17, Rg19, Rg20, Rg21, Rg22, and Rg23 belong to this category. 
\end{enumerate}

In the \emph{Success-to-Failure} category, Rg1 was found in the \emph{User Tasks} API, where a 500 server error occurred while iterating the String object for the subsequent release. 
This regression could potentially be related to the fault F10, as discussed in the RQ2 results (\Cref{RQ2Results}). 
Rg1 was observed in the release changelogs of R1-R2 and R9-R10 through the tests generated by \RESTestCBT{}, \RESTestRT{}, and \EvoMasterBB{}. 
The regression Rg3 was identified due to 400 validation errors in \emph{Assignments} and \emph{Survey} endpoints of \emph{User Tasks} API. 
Rg3 was noticed in the changelogs of R1-R2, R3-R4, R8-R9, and R9-R10 using the tests generated by the \RESTestART{} and \EvoMasterBB{}. 
The regression Rg13 was discovered in the \emph{User Tasks} API, where a 500 server error was caused due to a database issue. 
The regression Rg14 was identified in \emph{User Tasks} API due to 404 errors (URL not found), signifying a shift in the endpoints. 
The regression Rg18 was noted when 500 server errors occurred while accessing data from the None type object. 
This regression could potentially be associated with the F16 fault, as discussed in the RQ2 results (\Cref{RQ2Results}). 
Both Rg13 and Rg14 were identified in the R2-R3 release changelog, while Rg18 was observed in the R3-R4 release changelog. 
Rg13, Rg14, and Rg18 were found in tests generated by \EvoMasterBB{}.

In the category \emph{Failure-to-Success}, several regressions were identified in various APIs. 
Rg4 was discovered in the endpoints of general tasks of the \emph{User Tasks} API, where errors `URL not found' were fixed, leading to 200 Ok responses. 
Rg6 was found in assignments and survey endpoints of \emph{User Tasks} API, where validation errors were fixed, resulting in 200 Ok responses. 
Rg8 was located in the \emph{Measurements} API, where additional details were provided in the 200 Ok responses, indicating both a fix and an enhancement. 
Rg10 was identified in logs, assignments, and survey endpoints of \emph{Users} and \emph{User Tasks} APIs due to "Not Authenticated" errors were resolved that led to 200 Ok responses with additional records. 
Rg12 was detected in assignments-related endpoints of \emph{User Tasks} API, in which parsing headers errors were fixed. 
Rg15 was identified in logs and general tasks endpoints of \emph{User} and \emph{User Tasks} APIs, in which 500 server errors were fixed and resulted in 200 Ok responses containing refined and additional information. 
Furthermore, the regressions Rg4 and Rg8 were observed in the R1-R2 release changelogs. 
Rg6 appeared in the changelogs for the releases R1-R2 and R8-R9, while Rg10 was noted in the releases R1-R2 and R2-R3. 
Rg12 was found in the release changelogs of R1-R2, R2-R3, R3-R4, and R9-R10. 
Rg15 was located in the release changelogs of R2-R3, R3-R4, R8-R9, and R9-R10.
The regressions Rg4, Rg6, Rg8, Rg12, and Rg15 were identified from tests generated by \EvoMasterBB{}. 
Furthermore, Rg8 was discovered through tests produced by both \EvoMasterBB{} and \RESTlerFL{}.

In the \emph{Success-Enhancement/Reduction} category, Rg2 was observed in the \emph{Users} API related to the message logs endpoint. 
This regression was identified in the changelogs for releases R1-R2, R3-R4, R8-R9, and R9-R10, demonstrating multiple changes in users' message logs during evolution. 
Moreover, this regression was found through the tests generated by \RESTestCBT{} and \EvoMasterBB{}.

Many regressions, classified as \emph{Failure-Shift}, signify unstable changes that occur during API evolution and lead to different failures. 
Among these regressions, Rg17, Rg19, Rg20, and Rg23 are related to \emph{Development} API, indicating continuous modifications in the API. 
The regressions Rg5, Rg11, Rg16, Rg20, and Rg22 are associated with search methods (e.g., search with name or ID) of \emph{Users} API. 
In addition, Rg16 and Rg22 were identified in the \emph{Reports} API during the generation or retrieval of report data.
The regression Rg9 was discovered in the subscriptions-related endpoints of the \emph{Invoicing} API. 
Regressions Rg7 and Rg21 were discovered in APIs related to \emph{Device} and \emph{Users}, indicating server errors when iterating over String-type objects. 
This regression could potentially be related to the F10 fault (RQ2 results \Cref{RQ2Results}).
Furthermore, it can be observed (from \Cref{fig:rq4regressions}) that the regressions Rg16, Rg19, Rg21, and R22 were the most frequently occurring across multiple release changelogs.

Interestingly, regressions in this category were detected from tests obtained from nearly all testing tools.
For instance, regression Rg5 was detected through tests generated by \RESTestART{}, \EvoMasterBB{}, \SchemathesisPT{}, and \RESTlerF{}. 
Similarly, regression Rg7 was identified by tests generated by tools \EvoMasterBB{}, \SchemathesisPT{}, \RESTlerF{}, and \RESTlerFL{}.
On the other hand, regressions Rg9, Rg17, and Rg23 were only found using tests obtained from \EvoMasterBB{}, while Rg11 was specifically identified from the tests produced by \RESTlerF{}.
The regression Rg16 was detected by a variety of testing tools including \RESTestART{}, \RESTestFT{}, \EvoMasterBB{}, \SchemathesisPT{}, \RESTlerF{}, and \RESTlerFL{}.
Both Rg19 and Rg21 were identified by tests from tools \RESTestCBT{}, \RESTestRT{}, \EvoMasterBB{}, \SchemathesisPT{}, \RESTlerFL{}, and \RestTestGenNET{}.
The regression Rg20 was found using tests obtained from \EvoMasterBB{} and \SchemathesisPT{}.
Lastly, regression Rg22 was located using tests generated by \RESTestART{}, \RESTestFT{}, \EvoMasterBB{}, \SchemathesisPT{}, \RESTlerF{}, and \RESTlerFL{}.

\vspace{5pt}
\begin{rqres}{rq4res}
In total, 23 unique potential regressions were detected across 14 releases, using tests generated by all tools. 
Among these, tests produced by \EvoMasterBB{} contributed to identifying 22 regressions. 
Moreover, the results showed various trends across release changes: some failures were fixed, new features were added, and certain failures recurred or propagated due to the ongoing development of the APIs under test. 
\end{rqres}

\begin{table} [H]
    \centering
    \footnotesize
    \noindent
    \caption{RQ4 results: Number of potential regressions found with various testing tools across evolving releases (R1--R14) and all runs.}
    \begin{tabular}{@{} cl*{9}c @{}}\toprule 
        & & \textbf{\RTCBT{}} & \textbf{\RTART{}} & \textbf{\RTRT{}} & \textbf{\RTFT{}} 
        & \textbf{\EMBB{}} & \textbf{\STPT{}} & \textbf{\RLFz{}} 
        & \textbf{\RLFL{}} & \textbf{\RTGNET{}} \\
	\cmidrule{2-11}
        &\textbf{R1-R2} &2&1&1&0&11&2&2&2&1\\
        &\textbf{R2-R3} &0&0&0&0&6&1&1&0&0\\
        &\textbf{R3-R4} &0&1&0&0&7&2&1&0&0\\
        &\textbf{R4-R5} &0&1&0&0&1&2&0&0&0\\
        &\textbf{R5-R6} &0&0&1&1&3&1&1&0&1\\
        &\textbf{R6-R7} &0&0&1&1&4&1&1&0&1\\
        &\textbf{R7-R8} &0&0&0&1&3&1&0&0&0\\
        &\textbf{R8-R9} &0&0&0&1&6&1&1&0&0\\ 
        &\textbf{R9-R10} &0&1&0&0&6&1&1&0&0\\
        \rotnd{\rlap{~\textbf{Release changelog}}}
        &\textbf{R10-R11} &0&1&0&0&2&2&1&1&0\\
        &\textbf{R11-R12} &1&0&0&0&4&2&1&2&0\\
        &\textbf{R12-R13} &1&0&0&0&3&2&1&1&0\\ 
        &\textbf{R13-R14} &0&0&0&0&2&2&1&1&0\\ 
		\bottomrule
	\end{tabular}
	\label{tab-rq4results}
\end{table}

\begin{table} [H]
    \centering
    \footnotesize
    \noindent
    \caption{RQ4 results: Overall comparison of tools in terms of number of potential regressions found with various testing tools across evolving releases (R1--R14) and all runs.}
    
    \begin{tabular}{@{} cl*{5}c @{}}\toprule
        & & \textbf{\RESTest{}} & \textbf{\EvoMaster{}} & \textbf{\Schemathesis{}} & \textbf{\RESTler{}} & \textbf{\RestTestGen{}} \\
        \cmidrule{2-7}
        &\textbf{R1-R2} &2 & 11 & 2 & 2 & 1 \\
        &\textbf{R2-R3} &0 & 6 & 1 & 1 & 0 \\
        &\textbf{R3-R4} &1 & 7 & 2 & 1 & 0 \\
        &\textbf{R4-R5} &1 & 1 & 2 & 0 & 0 \\
        &\textbf{R5-R6} &1 & 3 & 1 & 1 & 1 \\
        &\textbf{R6-R7} &2 & 4 & 1 & 1 & 1 \\
        &\textbf{R7-R8} &1 & 3 & 1 & 0 & 0 \\
        &\textbf{R8-R9} &1 & 6 & 1 & 1 & 0 \\ 
        &\textbf{R9-R10} &1 & 6 & 1 & 1 & 0 \\
        \rotnd{\rlap{~\textbf{Release changelog}}}
        &\textbf{R10-R11} &1 & 2 & 2 & 1 & 0 \\
        &\textbf{R11-R12} &1 & 4 & 2 & 2 & 0 \\
        &\textbf{R12-R13} &1 & 3 & 2 & 1 & 0 \\ 
        &\textbf{R13-R14} &0 & 2 & 2 & 1 & 0 \\
	\bottomrule
	\end{tabular}
    
    \label{tab-rq4results-agg}
\end{table}

\begin{table} [H]
    \noindent
    \caption{RQ4 results: Description of all potential regressions corresponding to each regression ID and release changelog, derived from release R1 to R14 and across all runs.}
    \centering
    \begin{tabular}{p{.05\textwidth}p{.3\textwidth}p{.6\textwidth}}\toprule
    \textbf{ID} & \textbf{Release changelog} & \textbf{Description}  \\
	\midrule
        \textbf{Rg1}&R1-R2, R9-R10&200 Ok to 500 Server error while iterating over String object\\ 
        \textbf{Rg2}&R1-R2, R3-R4, R8-R9, R9-R10&200 Ok to 200 Ok with additional info or different responses\\
        \textbf{Rg3}&R1-R2, R3-R4, R8-R9, R9-R10&200 Ok to 400 Validation error\\
        \textbf{Rg4}&R1-R2&404 Not Found to 200 Ok\\
        \textbf{Rg5}&R1-R2, R3-R4, R4-R5, R5-R6, R6-R7&502 Unsatisfied request to 400 Validation error\\
        \textbf{Rg6}&R1-R2, R8-R9&400 Validation error to 200 Ok with additional details \\
        \textbf{Rg7}&R1-R2, R2-R3&502 Unsatisfied request to 500 Server error\\
        \textbf{Rg8}&R1-R2&502 Unsatisfied request to 200 Ok with additional details \\
        \textbf{Rg9}&R1-R2, R6-R7, R12-R13&502 Unsatisfied request to 500 Server error caused by a database issue\\
        \textbf{Rg10}&R1-R2, R2-R3&404 Not Authenticated to 200 Ok (with additional details/record)\\
        \textbf{Rg11}&R1-R2&400 Error parsing headers to 400 Input Validation error\\
        \textbf{Rg12}&R1-R2, R2-R3, R3-R4, R9-R10&400 Error parsing headers to 200 Ok with additional details \\
        \textbf{Rg13}&R2-R3&200 Ok to 500 Server error caused by a database issue\\
        \textbf{Rg14}&R2-R3&200 Ok to 404 Request URL not found\\
        \textbf{Rg15}&R2-R3, R2-R3, R3-R4, R8-R9, R9-R10&500 Server error to 200 Ok with additional details\\
        \textbf{Rg16}&R2-R3, R3-R4, R4-R5, R5-R6, R6-R7, R7-R8, R8-R9, R9-R10, R10-R11, R11-R12, R13-R14&400 Validation error (403 No access) to 502 Unsatisfied request\\
        \textbf{Rg17}&R3-R4, R9-R10&500 Server error to 404 Request URL not found\\
        \textbf{Rg18}&R3-R4&200 Ok to 500 Server error while accessing data from None type object\\ 
        \textbf{Rg19}&R3-R4, R5-R6, R6-R7, R7-R8, R10-R11, R11-R12, R12-R13, R13-R14&500 Server error to 502 Unsatisfied request\\
        \textbf{Rg20}&R3-R4, R10-R11&404 Not Found to 502 Unsatisfied request\\
        \textbf{Rg21}&R4-R5, R6-R7, R7-R8, R8-R9, R11-R12, R12-R13, R13-R14&502 Unsatisfied request to 500 Server error\\
        \textbf{Rg22}&R4-R5, R6-R7, R8-R9, R9-R10, R10-R11, R11-R12, R12-R13, R13-R14&502 Unsatisfied request to 404 Request URL not found (403 No access, 400 Input Validation error)\\
        \textbf{Rg23}&R8-R9&404 Not Found to 500 Server error caused by a database issue\\
	\bottomrule
	\end{tabular}
	\label{tab-rq4regsinfo}
\end{table}

\subsubsection{RQ5 Results -- Cost Overhead}\label{RQ5Results}
\Cref{tab-rq5cost2} presents the testing cost overhead for each tool, based on their failure and fault detection ratios across all releases. 
In terms of failure detection, all tools exhibited high cost overhead for R1, primarily due to the lowest corresponding failure detection ratios outlined in \Cref{tab-rq1results}. 
From releases R2 to R14, the cost overhead decreased for all tools, with \SchemathesisPT{} showing the lowest overhead, and \RESTestART{} and \RESTestFT{} showing the highest. 
Both \RESTlerF{} and \RESTlerFL{}, as well as \RESTestART{} and \RESTestFT{}, and \RESTestCBT{} and \RESTestRT{} exhibit nearly similar cost overheads. 
Regarding fault detection, all tools demonstrated a high cost overhead (approximately 99\%) across all releases, mainly due to the lowest corresponding fault detection ratios presented in \Cref{tab-rq1results}. 
According to the overall comparison results in \Cref{tab-rq5cost3}, \Schemathesis{} has the lowest failure detection cost overhead, followed by \RestTestGen{} with the second lowest overhead. 
Both \EvoMaster{} and \RESTler{} showed comparable overhead, while \RESTest{} incurred the highest cost overhead. 
In terms of fault detection, all tools exhibited comparable cost overhead.

It is important to note that although all tools generated numerous test cases, only a few led to the detection of failures, and even fewer revealed faults. 
This resulted in reduced failure and fault detection rates and increased cost overheads. 
However, as discussed in~\Cref{RQ3Results}, test coverage consistently improved for all tools, especially from R2 to R14. 
Some tools also achieved this increased coverage with comparatively lower overhead by detecting more failures and faults. 
For example, from R2 to R14, \RestTestGenNET{} demonstrated high failure detection ratios (\Cref{RQ1Results}) along with improved coverage, while \RESTestCBT{} and \RESTestRT{} showed high fault detection (\Cref{RQ2Results}) with increased coverage (\Cref{RQ3Results}). 
Furthermore, in the case of \EvoMasterBB{}, despite demonstrating a high-cost overhead, the results from \Cref{RQ4Results} indicate that tests generated by \EvoMasterBB{} lead to the detection of a higher number of potential regressions.

The overall analysis indicates that a high cost overhead does not necessarily imply reduced testing quality. 
Instead, these findings highlight a trade-off among tools, where the effectiveness varies depending on the specific testing objective, whether it is failure detection, fault identification, coverage improvement, or regression finding. 
Although a lower cost overhead is generally desirable, it may be acceptable when the priority is ensuring release quality through high coverage, finding more faults/failures, or detecting regressions. 
This trade-off is context-dependent and influenced by both the application domain and the type of release. 
In healthcare applications like the one studied in our work, the emphasis is on quality; therefore, detecting failures and faults is typically prioritized over cost considerations. 
This is especially true for production releases, where stability is critical for end users, including patients, doctors, and public health authorities. 
In contrast, during rapid releases (e.g., daily or weekly) for an IoT-based healthcare application, controlling cost overhead becomes critical to prevent damaging medical devices due to excessive test execution. 
Therefore, practitioners should begin by clearly defining their testing objectives, release priorities, and acceptable cost overhead, and then select the most appropriate tool accordingly.

\vspace{5pt}
\begin{rqres}{rq5res}
In most cases, \Schemathesis{} demonstrated the lowest failure detection cost overhead, around 72\%.
All REST API testing tools exhibited a high fault detection cost overhead (99\%). 
The primary cause of this high cost is that the tests generated by each tool often led to numerous duplicate failures, with fewer instances of identifying potential faults and regressions. 
Nevertheless, despite this substantial cost, tests generated by certain tools led to the finding of several potential faults and regressions. 
\end{rqres}

\begin{table}[H]
    \noindent
    \tiny
    \rotatebox{90}{
    \begin{minipage}{1\textheight}
    \centering
    \caption{RQ5 results: Testing cost overhead for each tool, calculated based on the percentage of failures detected and faults identified throughout all releases (R1--R14), reported as average values across 10 runs. Releases tagged with D, W, and M, denote daily, weekly, and monthly releases, respectively. }
    
	\begin{tabular}{@{}l l l l l l l l l l l l l l l l l l l@{}}\toprule
        \multicolumn{1}{l }{\textbf{}} & \multicolumn{2}{c }{\textbf{\RTCBT{}}} & \multicolumn{2}{c }{\textbf{\RTART{}}}& \multicolumn{2}{c }{\textbf{\RTRT{}}} & \multicolumn{2}{c }{\textbf{\RTFT{}}} & \multicolumn{2}{c }{\textbf{\EMBB{}}} & \multicolumn{2}{c }{\textbf{\STPT{}}}& \multicolumn{2}{c }{\textbf{\RLFz{}{}}} & \multicolumn{2}{c }{\textbf{\RLFL{}}} & \multicolumn{2}{c }{\textbf{\RTGNET{}}} \\ 
		\cmidrule(lr){2-3}
		\cmidrule(ll){4-5} \cmidrule(ll){6-7}\cmidrule(ll){8-9}\cmidrule(ll){10-11}\cmidrule(ll){12-13}\cmidrule(ll){14-15}\cmidrule(ll){16-17}\cmidrule(ll){18-19}
		\multicolumn{1}{ l }{\textbf{}}  &Failure&Fault&Failure&Fault&Failure&Fault&Failure&Fault&Failure&Fault&Failure&Fault&Failure&Fault&Failure&Fault&Failure&Fault\\ 
		\cmidrule(lr){1-1}\cmidrule(lr){2-3}
		\cmidrule(ll){4-5} \cmidrule(ll){6-7}\cmidrule(ll){8-9}\cmidrule(ll){10-11}\cmidrule(ll){12-13}\cmidrule(ll){14-15}\cmidrule(ll){16-17}\cmidrule(ll){18-19} 

        \multicolumn{1}{ l }{\textbf{R1}}&97.61\%&99.91\%&98.78\%&99.93\%&97.4\%&99.9\%&98.66\%&99.88\%&99.87\%&100.0\%&99.75\%&99.98\%&99.87\%&99.97\%&99.82\%&99.97\%&99.85\%&99.99\%\\
        \multicolumn{1}{ l }{\textbf{R2}}&93.81\%&99.98\%&97.56\%&99.97\%&93.83\%&99.98\%&98.47\%&99.95\%&91.95\%&100.0\%&72.47\%&99.97\%&90.54\%&99.97\%&90.17\%&99.97\%&85.33\%&99.98\%\\
        \multicolumn{1}{ l }{\textbf{R3}}&93.76\%&99.98\%&97.55\%&99.97\%&93.79\%&99.98\%&98.51\%&99.93\%&91.95\%&100.0\%&72.52\%&99.97\%&90.54\%&99.97\%&90.22\%&99.97\%&85.34\%&99.97\%\\
        \multicolumn{1}{ l }{\textbf{R4}}&92.65\%&99.96\%&96.35\%&99.95\%&92.69\%&99.96\%&96.74\%&99.92\%&91.6\%&99.99\%&72.46\%&99.95\%&90.54\%&99.97\%&90.22\%&99.97\%&83.49\%&99.97\%\\
        \multicolumn{1}{ l }{\textbf{R5}}&92.61\%&99.96\%&96.36\%&99.95\%&92.63\%&99.96\%&96.69\%&99.92\%&91.6\%&99.99\%&64.51\%&99.95\%&90.53\%&99.97\%&90.16\%&99.97\%&83.45\%&99.97\%\\
        \multicolumn{1}{ l }{\textbf{R6}}&92.63\%&99.96\%&96.34\%&99.95\%&92.64\%&99.96\%&96.72\%&99.93\%&90.9\%&99.99\%&72.52\%&99.95\%&90.53\%&99.97\%&90.2\%&99.97\%&83.47\%&99.97\%\\
        \multicolumn{1}{ l }{\textbf{R7}}&92.63\%&99.96\%&96.36\%&99.95\%&92.68\%&99.96\%&96.76\%&99.93\%&91.6\%&99.99\%&72.52\%&99.95\%&90.54\%&99.97\%&90.17\%&99.97\%&83.46\%&99.97\%\\
        \multicolumn{1}{ l }{\textbf{R8}}&92.69\%&99.96\%&96.35\%&99.95\%&92.64\%&99.96\%&96.67\%&99.92\%&91.59\%&99.99\%&72.48\%&99.95\%&90.54\%&99.97\%&90.21\%&99.97\%&83.47\%&99.97\%\\
        \multicolumn{1}{ l }{\textbf{R9}}&92.64\%&99.96\%&96.31\%&99.95\%&92.63\%&99.96\%&96.67\%&99.92\%&89.96\%&100.0\%&72.53\%&99.95\%&90.53\%&99.97\%&90.18\%&99.97\%&83.46\%&99.97\%\\
        \multicolumn{1}{ l }{\textbf{R10}}&92.64\%&99.96\%&96.35\%&99.95\%&92.68\%&99.96\%&96.74\%&99.92\%&91.27\%&100.0\%&72.5\%&99.95\%&90.53\%&99.97\%&90.18\%&99.97\%&83.45\%&99.97\%\\
        \multicolumn{1}{ l }{\textbf{R11}}&92.63\%&99.96\%&96.32\%&99.95\%&92.65\%&99.96\%&96.66\%&99.92\%&90.53\%&99.99\%&72.44\%&99.95\%&90.54\%&99.97\%&90.19\%&99.97\%&83.44\%&99.97\%\\
        \multicolumn{1}{ l }{\textbf{R12}}&93.83\%&99.97\%&97.56\%&99.97\%&93.85\%&99.98\%&98.47\%&99.93\%&89.33\%&100.0\%&72.52\%&99.96\%&90.5\%&99.97\%&90.16\%&99.97\%&85.35\%&99.97\%\\
        \multicolumn{1}{ l }{\textbf{R13}}&93.85\%&99.98\%&97.58\%&99.98\%&93.81\%&99.97\%&98.52\%&99.95\%&91.94\%&100.0\%&72.48\%&99.96\%&90.55\%&99.97\%&90.16\%&99.97\%&85.34\%&99.97\%\\
        \multicolumn{1}{ l }{\textbf{R14}}&93.82\%&99.97\%&97.56\%&99.97\%&93.84\%&99.98\%&98.45\%&99.93\%&91.0\%&100.0\%&72.5\%&99.96\%&90.53\%&99.97\%&90.17\%&99.97\%&85.35\%&99.97\%\\
        
	\bottomrule
	\end{tabular}
    
 \label{tab-rq5cost2}
 \end{minipage}}
\end{table}

\begin{table}[H]
    \noindent
    \footnotesize
    \centering
    \caption{RQ5 results: Overall comparison of tools in terms of testing cost overhead for each tool, calculated based on the percentage of failures detected and faults identified throughout all releases (R1--R14), reported as average values across 10 runs. Releases tagged with D, W, and M, denote daily, weekly, and monthly releases, respectively. }
    
	\begin{tabular}{@{}l l l l l l l l l l l l@{}}\toprule
        \multicolumn{1}{l }{\textbf{}} & \multicolumn{2}{c }{\textbf{\RESTest{}}} & \multicolumn{2}{c }{\textbf{\EvoMaster{}}} & \multicolumn{2}{c }{\textbf{\Schemathesis{}}} & \multicolumn{2}{c }{\textbf{\RESTler{}}} & \multicolumn{2}{c }{\textbf{\RestTestGen{}}} \\ 
        \cmidrule(lr){2-3}\cmidrule(ll){4-5}\cmidrule(ll){6-7}\cmidrule(ll){8-9}\cmidrule(ll){10-11}
        \multicolumn{1}{ l }{\textbf{}} & Failure & Fault & Failure & Fault & Failure & Fault & Failure & Fault & Failure & Fault \\ 
        \cmidrule(lr){1-1}\cmidrule(lr){2-3}\cmidrule(ll){4-5}\cmidrule(ll){6-7}\cmidrule(ll){8-9}\cmidrule(ll){10-11} 
    
        \multicolumn{1}{ l }{\textbf{R1}} & 98.61\% & 99.91\% & 99.87\% & 100.0\% & 99.75\% & 99.98\% & 99.85\% & 99.97\% & 99.85\% & 99.99\% \\
        \multicolumn{1}{ l }{\textbf{R2}} & 95.92\% & 99.97\% & 91.95\% & 100.0\% & 72.47\% & 99.97\% & 90.35\% & 99.97\% & 85.33\% & 99.98\% \\
        \multicolumn{1}{ l }{\textbf{R3}} & 95.90\% & 99.97\% & 91.95\% & 100.0\% & 72.52\% & 99.97\% & 90.38\% & 99.97\% & 85.34\% & 99.97\% \\
        \multicolumn{1}{ l }{\textbf{R4}} & 94.86\% & 99.94\% & 91.60\% & 99.99\% & 72.46\% & 99.95\% & 90.38\% & 99.97\% & 83.49\% & 99.97\% \\
        \multicolumn{1}{ l }{\textbf{R5}} & 94.82\% & 99.94\% & 91.60\% & 99.99\% & 64.51\% & 99.95\% & 90.35\% & 99.97\% & 83.45\% & 99.97\% \\
        \multicolumn{1}{ l }{\textbf{R6}} & 94.83\% & 99.95\% & 90.90\% & 99.99\% & 72.52\% & 99.95\% & 90.37\% & 99.97\% & 83.47\% & 99.97\% \\
        \multicolumn{1}{ l }{\textbf{R7}} & 94.86\% & 99.95\% & 91.60\% & 99.99\% & 72.52\% & 99.95\% & 90.36\% & 99.97\% & 83.46\% & 99.97\% \\
        \multicolumn{1}{ l }{\textbf{R8}} & 94.84\% & 99.95\% & 91.59\% & 99.99\% & 72.48\% & 99.95\% & 90.38\% & 99.97\% & 83.47\% & 99.97\% \\
        \multicolumn{1}{ l }{\textbf{R9}} & 94.82\% & 99.96\% & 89.96\% & 100.0\% & 72.53\% & 99.95\% & 90.36\% & 99.97\% & 83.46\% & 99.97\% \\
        \multicolumn{1}{ l }{\textbf{R10}} & 94.85\% & 99.95\% & 91.27\% & 100.0\% & 72.50\% & 99.95\% & 90.36\% & 99.97\% & 83.45\% & 99.97\% \\
        \multicolumn{1}{ l }{\textbf{R11}} & 94.82\% & 99.95\% & 90.53\% & 99.99\% & 72.44\% & 99.95\% & 90.37\% & 99.97\% & 83.44\% & 99.97\% \\
        \multicolumn{1}{ l }{\textbf{R12}} & 95.80\% & 99.97\% & 89.33\% & 100.0\% & 72.52\% & 99.96\% & 90.33\% & 99.97\% & 85.35\% & 99.97\% \\
        \multicolumn{1}{ l }{\textbf{R13}} & 95.94\% & 99.98\% & 91.94\% & 100.0\% & 72.48\% & 99.96\% & 90.35\% & 99.97\% & 85.34\% & 99.97\% \\
        \multicolumn{1}{ l }{\textbf{R14}} & 95.80\% & 99.97\% & 91.00\% & 100.0\% & 72.50\% & 99.96\% & 90.35\% & 99.97\% & 85.35\% & 99.97\% \\
        
	\bottomrule
	\end{tabular}
    
 \label{tab-rq5cost3}
\end{table}

\subsection{Threats to Validity}
As with any experiment, potential threats to validity may arise in our experiments. In the following, we outline each potential threat that could impact our experimental results and the corresponding mitigation strategies.

\subsubsection{External validity threat}
The threat to external validity is related to the generalizability of experimental findings. 
A potential threat to the external validity of our experiments could arise from the REST APIs and tools selected for testing. 
Another potential threat could be due to APIs accessed through a staging environment created for testing purposes. 
In our experiments, we used 17 REST APIs with 120 endpoints from a real-world healthcare IoT application, along with its 14 evolving releases during DevOps. 
These APIs, tested in the staging environment, eventually transition to the production environment, aligning with the current practice of Oslo City’s industry partner. 
We utilized this experimental apparatus, provided by Oslo City’s healthcare department, in our experiments. 
It is worth noting that this application is operational across multiple municipalities in Norway, including Oslo, and is also being used in several other countries. 
Given the large scope of our application under test, the APIs utilized in our experiments constitute a representative sample. 
Moreover, we carefully selected REST API testing tools that align with the dynamics of our APIs under test. 
These tools, being state-of-the-art and well-established, make a good representative set. 
While our findings may not generalize to all healthcare applications, this limitation is common in empirical studies~\cite{sartaj2023testing,khan2019aspectocl}.

\subsubsection{Internal validity threat}
The threat to interval validity can occur due to uncontrolled factors. 
A possible threat to the internal validity of our experiments may occur due to any instrumentation performed on the REST API testing tools or the REST APIs under test. 
To mitigate this threat, we selected REST API testing tools in their executable form, without modifying their code. 
In the case of REST APIs under test, we only have black-box access to these APIs, with all implementations carried out by the industry partner.
Another possibility of internal validity threat can be due to tool inputs, including the API schema and test configuration, particularly when using \RESTest{}. 
To reduce this threat, we relied on the documentation provided for each REST API testing tool to ensure proper setup and usage.
Furthermore, we conducted multiple sessions with the technical team from Oslo City and their industry partner. 
The primary purpose of these sessions was to demonstrate the setup process and acquire their feedback. 
An additional potential threat could arise from the parallel execution of tools, which may dominate CPU usage and result in unequal resource distribution. 
To handle this, we adopted a parallel execution setup inspired by existing work of a similar nature~\cite{zhang2023open}. 
Following that setup, we configured multiple threads to execute parallel jobs while efficiently utilizing the available logical processors and CPU cores.

\subsubsection{Construct validity threat}
Construct validity threat is associated with the interpretation of results. 
We presented our evaluation findings using metrics commonly employed in similar experimental contexts~\cite{martin2022online,kim2022automated,zhang2023open}. 
Another potential threat to construct validity could arise from the manual analysis of faults and regressions. 
To mitigate this threat, we shared our experimental data, which includes generated test cases, failure reports, test coverage, regressions, and fault catalogs, with Oslo City's healthcare department for their review and feedback. 
Due to the sensitive nature of the data involved in this study, relating to API information and patient-associated data, we are bound by non-disclosure agreements that prevent us from publicly releasing a replication package or data set.

\subsubsection{Conclusion validity threat}
The threat to conclusion validity arises from the selection of analysis metrics, measures, or statistical tests that may influence the accurate drawing of conclusions. 
A potential threat to the validity of the conclusions of experiments may arise due to the randomness factor of REST API testing tools. 
We ran each tool 10 times with a fixed one-hour duration to address the randomness factor. 
This experimental setting is aligned with existing empirical studies~\cite{kim2022automated,zhang2023open}. 
To analyze and report the results of our experiments, we employed metrics such as API failures, faults, test coverage, and regressions in evolving releases, which are commonly used in similar studies~\cite{martin2022online,godefroid2020differential}. 
Moreover, we evaluated the results using non-parametric statistical tests, specifically the Wilcoxon signed-rank test, Vargha-Delaney \^{A}${}_{12}$ effect size measure, and Holm's correction to adjust p-values, following recommended guidelines~\cite{arcuri2011practical}.

\begin{figure}[htbp]
\centerline{\includegraphics[width=\linewidth, keepaspectratio]{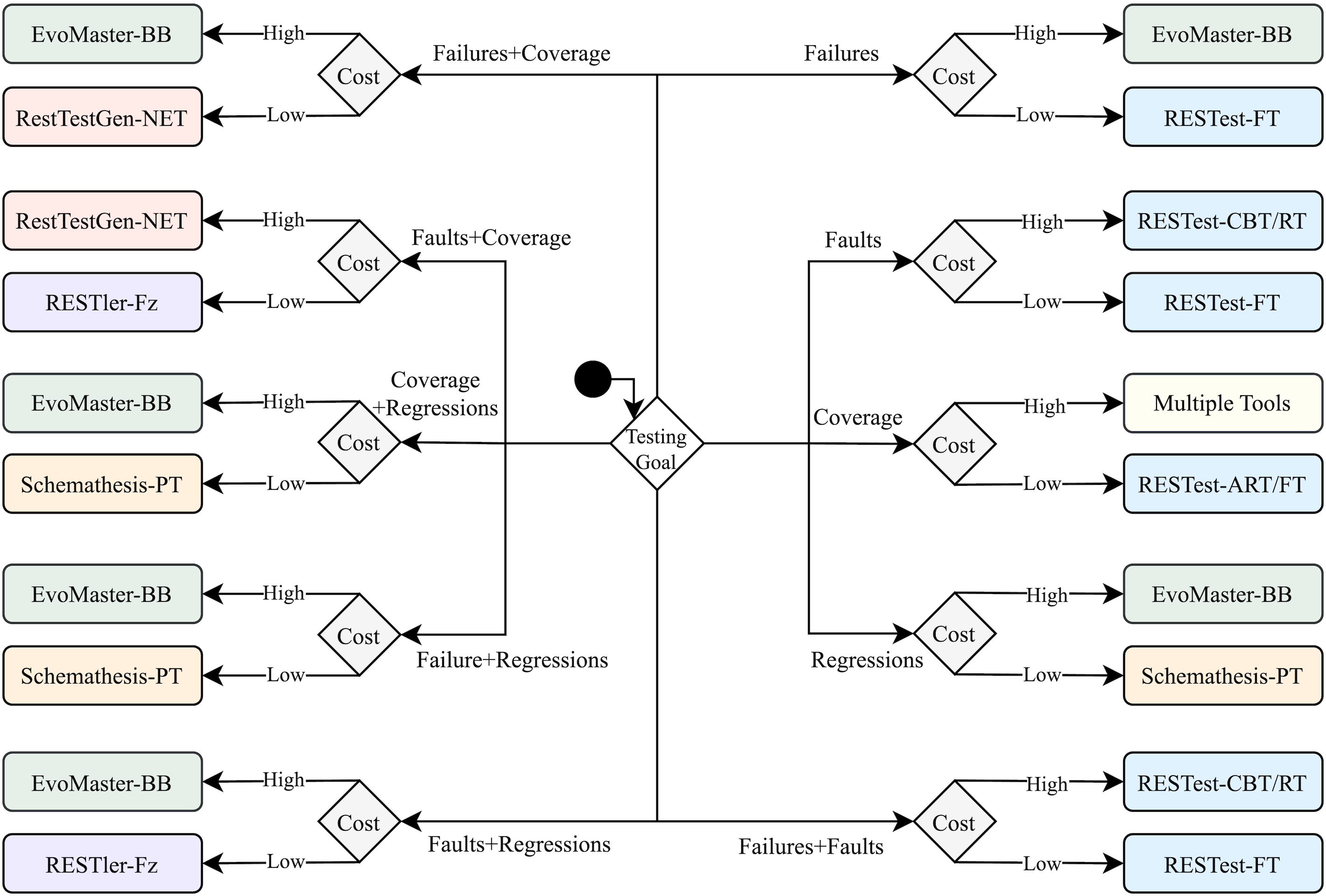}}
\caption{Workflow illustrating recommendations for practitioners in selecting REST API testing tools for healthcare applications.}
\label{fig:guide}
\end{figure}

\section{Tool Selection Recommendations}\label{sec:recommendations}
Based on our experimental findings with testing the REST API of an industrial healthcare application, we provide recommendations to guide practitioners in selecting suitable REST API testing tools for similar healthcare systems. 
\Cref{fig:guide} presents a workflow serving as a visual guide to illustrate the selection process. 
As observed in the experimental results, there is no one-size-fits-all tool that is universally effective for all testing purposes. 
The performance and effectiveness of each tool can vary depending on the specific testing context. 
Therefore, testers need to define the testing goal, i.e., whether they are interested in analyzing failures, faults, coverage, regressions, or a combination of these aspects. 
In the following, we discuss recommendations considering the testing goal and cost overhead. 

\begin{itemize}
    \item If the testing goal is to find the maximum number of \emph{failures} and a high-cost overhead is acceptable, \EvoMasterBB{} can be a suitable choice. However, if the goal is to identify as many failures as possible with minimum cost overhead, \RESTestFT{} seems to be a reasonable alternative. 
    \item For detecting maximum \emph{unique potential faults}, variants of \RESTest{} such as \RESTestCBT{}, \RESTestRT{}, and \RESTestFT{} can be employed interchangeably. Nevertheless, if the priority is to keep the cost overhead as low as possible, \RESTestFT{} is an appropriate choice. 
    \item If the testing goal is to achieve \emph{maximum coverage} and high-cost overhead is bearable, one can choose from a range of tools such as \EvoMasterBB{}, \SchemathesisPT{}, \RESTlerF{}, \RESTlerFL{}, and \RestTestGenNET{}. However, for reasonable coverage with minimum cost overhead, either \RESTestART{} or \RESTestFT{} would be suitable.  
    \item When the testing goal is to discover a higher number of \emph{regressions} across evolving releases and a high-cost overhead is tolerable, \EvoMasterBB{} can be an appropriate selection. On the other hand, if the requirement is for a low-cost overhead, \SchemathesisPT{} can be considered as an alternative.
    \item If the testing goal is to identify the maximum number of \emph{failures} and \emph{faults}, and high-cost overhead is not a concern, one can opt for \RESTestCBT{} or \RESTestRT{}. However, if the requirement is to minimize cost overhead while detecting a higher number of failures and faults, \RESTestFT{} can be an appropriate choice. 

    \item When the testing goal is to identify the maximum number of \emph{failures} while achieving \emph{high coverage}, and if high-cost overhead is acceptable, \EvoMasterBB{} can be a suitable option. Conversely, if there is a need to reduce cost overhead, \RestTestGenNET{} can be viewed as a viable alternative. 
    \item If the testing goal is to detect the highest number of \emph{unique faults} while ensuring \emph{high coverage} with a sizable high-cost overhead, \RestTestGenNET{} could be an appropriate choice. However, if the same testing goal needs to be achieved with lower cost overhead, \RESTlerF{} appears to be a suitable alternative. 
    \item When testing aims to achieve \emph{high coverage} and detect the highest number of \emph{regressions} across multiple releases, \EvoMasterBB{} can be utilized if a high-cost overhead is acceptable. On the other hand, if there is a need to reduce cost overhead while achieving the same testing goal, \SchemathesisPT{} appears to be a reasonable alternative. 
    \item If testing aims to uncover a maximum number of \emph{failures} and \emph{regressions} without high-cost overhead constraints, \EvoMasterBB{} can be an appropriate selection. Conversely, if the intent is to reduce cost overhead while maintaining the same testing goal, \SchemathesisPT{} can be considered a suitable option. 
    \item If the testing goal is to identify the maximum number of \emph{unique faults} and \emph{regressions}, and if a high-cost overhead is tolerable, \EvoMasterBB{} could be an appropriate choice. However, if the goal is to maintain low costs while still uncovering maximum unique faults and regressions, \RESTlerF{} can be selected as an alternative option. 
\end{itemize}

All tool recommendations outlined above are formulated by carefully considering key testing-related aspects, such as cost and coverage. 
In the Oslo City context, we shared all results, findings, and recommendations with their team and industry partner, both during in-person meetings and through this paper. 
While they found these insightful, the decision to select a specific tool remained entirely theirs. 
Our role was to provide tool configurations, test results, and a selection guide to support their decision-making based on their testing needs and resource constraints. 
It is also important to note that our objective was not to conduct an industrial evaluation of testing tools with practitioners, but to analyze their effectiveness for the given REST API of the healthcare IoT application.

\section{Experiences, Insights, and Challenges}\label{sec:insights}
Based on our experiments with REST API testing tools, we discuss experiences, insights, and challenges in the following sections. 

\subsection{Language-specific Test Inputs}
The healthcare application used in our experiments operates in several countries and provides multilingual support. 
This trait is common among web services of multinational companies, which typically support multiple languages. 
An important aspect of testing REST APIs of such applications is using test inputs in all supported languages, necessitating the generation of \textit{language-specific test inputs}. 
For instance, if an application supports Norwegian, Swedish, and English, generating test inputs in these languages is needed to test the application's APIs from a language perspective. 
Moreover, if an application is region-specific with an API supporting a particular language, it would require testing tools to generate language-specific inputs. 
For instance, in our experiments with the Karie and Medido device APIs, we found that when the language setting was Norwegian, the APIs expected inputs in Norwegian. 
When REST API testing tools sent inputs in English, the API returned a 400 Bad Request error. 
A similar behavior was observed in another healthcare application, the Cancer Registry of Norway, which also processes cancer case data in Norwegian~\cite{isaku2023cost}.

The REST API testing tools used in our experiments mainly generate test inputs in English. 
Therefore, augmenting these tools to support the generation of test inputs specific to various languages emerges as a significant factor. 
Another potential avenue to explore is the automatic recognition of supported languages by these tools, followed by the generation of test inputs tailored to each identified language. 
Given the wide range of applications and rising popularity of large language models (LLMs), generating multilingual test inputs with LLMs could be a valuable enhancement to REST API testing tools. 
This approach would broaden the applicability of these tools in a multilingual context.

\subsection{Automation Level}
Each REST API testing tool supports different levels of automation. 
Our observations are based on running these tools locally using their provided Java or Python executables, rather than containerized setups like Docker, which may result in different user experiences. 
A common input for all tools is an OpenAPI specification in Swagger or JSON format, which can be easily converted between formats; for instance, \RestTestGen{} provides automated support for this conversion. 
We found \EvoMaster{} and \Schemathesis{} straightforward to use in black-box mode, requiring only the OpenAPI specification and basic parameter settings. 
In contrast, tools like \RestTestGen{}, \RESTler{}, and \RESTest{} require additional configurations. 
\RestTestGen{} requires defining configuration files for each API and testing technique, including tool-specific and API-specific settings. 
\RESTler{} necessitates the installation of Microsoft .NET and running a Compile phase for each API individually, which initially resulted in errors that required manual debugging to resolve configuration issues. 
Similarly, \RESTest{} requires manual setup of API properties and test configuration files; although it can generate a configuration skeleton, it must be tailored manually. 
While these tools necessitate manual effort for initial configuration and setup, the subsequent testing process is automated. 
However, in a DevOps context, the manual effort required for API-related configurations can increase considerably due to the continuous evolution of APIs.  
A potential solution to alleviate manual effort involves utilizing automated tools that can infer REST API specifications as their implementation undergoes evolution. 
For example, a tool like RESTSpecIT~\cite{decrop2024you} can be used for this purpose.

\subsection{Regression Test Selection, Minimization, and Prioritization}
Numerous techniques are presented in the literature for prioritization, minimization, and selection of the test suite in the context of regression testing~\cite{yoo2012regression,greca2023state}. 
However, to the best of our knowledge, their applicability to prioritize, minimize, and select regression test cases for REST APIs remains undetermined. 
Proposing new techniques for regression test suite prioritization, minimization, or selection falls outside the scope of this paper.
Therefore, in our experiments, we used \emph{retest-all} strategy, which involves running all the tests generated for the initial release on all subsequent releases. 
Given the daily or weekly release routine within the DevOps process, executing all test cases becomes impractical due to the significant overhead of API calls.
For IoT applications' REST APIs dependent on various third-party applications, the frequent execution of numerous tests is infeasible due to the API call limit imposed by these external applications.
Furthermore, in DevOps environments, API specifications typically evolve along with the implementation. 
As a result, some test cases become obsolete, which requires strategies such as test case minimization to systematically execute tests on subsequent releases. 
A potential area for future research could focus on devising strategies for prioritization, minimization, or selection of test cases specifically for the regression testing of REST APIs of IoT-based applications. 
The regression testing outcomes presented in this work could serve as a foundation for further research. 
These findings could also be used to refine or customize existing methods for test selection, minimization, and prioritization within the REST API testing context.

\subsection{REST API Regression Testing Tool}%

Regression testing of continuously evolving REST APIs necessitates a specialized tool that supports automation. Until now, approaches and tools targeting REST API testing have primarily focused on OpenAPI specifications, REST API test generation and execution, or regression testing. 
Using separate tools for each task in the regression testing of evolving REST APIs can lead to significant manual effort and compatibility challenges. 
Therefore, one potential direction worth exploring is the creation of a toolchain that leverages existing tools designed for individual issues. 
This approach could streamline the process and enhance efficiency in the regression testing of evolving REST APIs.

\subsection{Continuous Regression Testing in DevOps}
Typically, unintended faults or behavioral changes identified through regression testing are investigated and fixed before moving to the next release. 
However, our experiment was not designed in this way, as it would have required continuous involvement from industry practitioners to debug and fix issues for each release. 
This level of involvement was not feasible within the scope of our project due to the significant time and resources required, including specialized personnel (e.g., software engineers and testers), additional medical devices, and dedicated computing infrastructure. 
In our study, we shared all experimental results with the Oslo City team and their industry partner upon completion of the experiment and analysis. 
Due to organizational policies and confidentiality rules, we received general positive feedback indicating that the results were insightful, without confirmation of whether specific regressions or faults were present or fixed. 
Conducting a study with continuous regression testing in a DevOps context would require active practitioners' engagement throughout the experiment. 
It would also require applying techniques such as regression test selection, minimization, or prioritization to systematically run tests. 
Therefore, such a study represents a promising direction for future research.

\subsection{Findings Comparison with Existing Studies}
As our study focuses on black-box REST API testing, we compare our experimental findings and conclusions with those reported in previous black-box testing studies. 
The work by Corradini et al.~\cite{corradini2021empirical} presented a comparison of four tools (\RestTestGen{}, \RESTler{}, bBOXRT, and \RESTest{}) based on robustness and coverage. 
They found \RESTler{} to be the most robust and \RESTest{} the least robust. 
Although our analysis did not focus on robustness, we encountered no execution failures beyond initial configurations. 
Regarding coverage, their study reported that \RestTestGen{} achieved high coverage, whereas our results showed comparable coverage across \RestTestGen{}, \RESTler{}, and \RESTest{}, with \RESTest{} slightly lower. 
Their general conclusions were that (i) \RESTler{} is a mature but time-consuming tool, (ii) \RestTestGen{} is better suited for time-constrained testing, (iii) combining multiple input generation techniques could be useful, and (iv) achieving high input coverage is easier than output coverage. 
Our study only supports their conclusion regarding high input coverage. 
However, we did not observe other conclusions, as we did not assess the time consumption of each tool or explore combinations of input generation techniques.

In the work by Kim et al.~\cite{kim2022automated}, 10 tools were compared in black-box and white-box modes based on code coverage, error responses, and practical implications. 
Although their findings are based on code coverage, ours focuses on API coverage. 
A common finding with our work is that erroneous requests are sometimes accepted, which, in our case, enabled identifying faults. 
Their conclusions highlighted that (i) sample input parameter values can improve test generation, (ii) natural language processing (NLP) can be used to interpret parameter descriptions, (iii) black-box tools could not achieve high coverage, and (iv) \SchemathesisPT{} and \EvoMasterBB{} outperformed other tools. 
Our study does not report the first two conclusions. 
For the latter two, our findings differ: all black-box tools achieved over 82\% coverage, and the effectiveness of each tool varied by context, such as \RESTest{} outperformed in fault detection and \EvoMasterBB{} outperformed in regression identification. 
A possible reason for these differences is that we used more recent versions of the tools than those evaluated in~\cite{kim2022automated}, and the newer or upgraded features may have contributed to improved coverage and fault detection.

In the work by Zhang et al.~\cite{zhang2023open}, seven tools were evaluated in black-box and white-box modes based on line coverage, time budget, faults, coverage-fault correlation for white-box testing, and current challenges. 
Their black-box findings indicated that \EvoMasterBB{} and \SchemathesisPT{} achieved the highest line coverage, longer time budgets yielded better results, and each tool detected only a few fault types. 
Our study did not analyze line coverage or various time budgets, but found \RESTest{} tests effective in identifying most types of faults. 
Their overall conclusions were that (i) many tools lack robustness and crash frequently, (ii) API schemas may contain faults, (iii) a high number of interactions with databases, (iv) the need for mocking external live services, and (v) NLP can be used to infer constraints on string input parameters. 
In comparison, none of these conclusions are reported in our study, as our analysis focused specifically on black-box testing. 
For instance, we did not study aspects such as robustness, API specification faults, or mocking of external services. 
Moreover, some of their conclusions were derived from white-box testing contexts, such as database interactions.

\section{Conclusion}\label{sec:conclusion}
Given the numerous available REST API testing tools, a key concern of the Oslo City Healthcare Department is selecting a suitable tool to test real-world IoT applications in healthcare in a DevOps environment. 
Taking this into account, we conducted an empirical study involving five well-established and state-of-the-art REST API testing tools: \RESTest{}, \EvoMaster{}, \Schemathesis{}, \RESTler{}, and \RestTestGen{}. 
These tools were employed to test an industrial healthcare IoT application comprising 17 REST APIs with 120 endpoints. 
The experiments were carried out using 14 releases (including daily, weekly, and monthly) that experienced continuous development and evolution during the DevOps process.
Given the black-box access to the APIs of our application under test, we set up black-box testing techniques supported by REST API testing tools. 
These configurations resulted in a total of 10 tools, namely \RESTestCBT{}, \RESTestART{}, \RESTestFT{}, \RESTestRT{}, \EvoMasterBB{}, \SchemathesisPT{}, \RESTlerF{}, \RESTlerFL{}, \RestTestGenNET{}, and \RestTestGenMAST{}. 
Our experiments focused on analyzing failures, faults, test coverage, regressions in REST APIs under test, and cost overhead.

The results revealed that \EvoMasterBB{} generated the highest number of test cases compared to all other tools. 
On the other hand, \RestTestGenMAST{} did not generate any tests, indicating that none of the APIs have security vulnerabilities. 
In terms of API failures, all tools generated a higher number of client-side errors than server-side errors. 
Among all, \EvoMasterBB{} generated tests led to the highest number of API failures. 
The tests produced by all tools led to the identification of 18 unique potential faults across 14 releases, with \RESTestCBT{} and \RESTestRT{} being the primary contributors to these faults. 
As for API test coverage, the results indicated that all tools attained a coverage level of 82-84\%, with the lowest coverage observed in status code and status code class coverages. 
The regression results showed that tests generated by \EvoMasterBB{} led to finding the highest number of regressions. 
In total, 23 unique potential regressions were identified in all 14 releases. 
Finally, the cost analysis demonstrated that each tool generates a significant number of tests, often leading to numerous duplicate failures, thereby resulting in a cost overhead exceeding 70\%.

\section*{Acknowledgements}
This research work is a part of the WTT4Oslo project (No. 309175) funded by the Research Council of Norway. All the experiments reported in this paper are conducted in a laboratory setting of Simula Research Laboratory; therefore, they do not by any means reflect the quality of services Oslo City provides to its citizens. Moreover, these experiments do not reflect the quality of services various vendors provide to Oslo City.

\bibliographystyle{unsrt}
\bibliography{refs}

\begin{thebibliography}{10}

\bibitem{sartaj2023hita}
Hassan Sartaj, Shaukat Ali, Tao Yue, and Julie~Marie Gjøby.
\newblock {HITA: An Architecture for System-level Testing of Healthcare IoT Applications}.
\newblock In {\em European Conference on Software Architecture}, pages 451--468, Cham, 2023. Springer.

\bibitem{oslocity}
Oslo Kommune.
\newblock Norwegian health authority.
\newblock \url{https://www.oslo.kommune.no/etater-foretak-og-ombud/helseetaten/}, 2024.
\newblock [Online; accessed 05-May-2024].

\bibitem{martin2021restest}
Alberto Martin-Lopez, Sergio Segura, and Antonio Ruiz-Cort{\'e}s.
\newblock {RESTest: automated black-box testing of RESTful web APIs}.
\newblock In {\em Proceedings of the 30th ACM SIGSOFT International Symposium on Software Testing and Analysis}, pages 682--685, 2021.

\bibitem{atlidakis2019restler}
Vaggelis Atlidakis, Patrice Godefroid, and Marina Polishchuk.
\newblock {RESTler: Stateful REST API Fuzzing}.
\newblock In {\em 2019 IEEE/ACM 41st International Conference on Software Engineering (ICSE)}, pages 748--758. IEEE, 2019.

\bibitem{hatfield2022deriving}
Zac Hatfield-Dodds and Dmitry Dygalo.
\newblock Deriving semantics-aware fuzzers from web {API} schemas.
\newblock In {\em Proceedings of the ACM/IEEE 44th International Conference on Software Engineering: Companion Proceedings}, pages 345--346, 2022.

\bibitem{wu2022combinatorial}
Huayao Wu, Lixin Xu, Xintao Niu, and Changhai Nie.
\newblock {Combinatorial Testing of RESTful APIs}.
\newblock In {\em Proceedings of the 44th International Conference on Software Engineering}, pages 426--437, 2022.

\bibitem{corradini2022resttestgen}
Davide Corradini, Amedeo Zampieri, Michele Pasqua, and Mariano Ceccato.
\newblock {RestTestGen: An Extensible Framework for Automated Black-box Testing of RESTful APIs}.
\newblock In {\em 2022 IEEE International Conference on Software Maintenance and Evolution (ICSME)}, pages 504--508. IEEE, 2022.

\bibitem{arcuri2018evomaster}
Andrea Arcuri.
\newblock {EvoMaster}: Evolutionary multi-context automated system test generation.
\newblock In {\em 2018 IEEE 11th International Conference on Software Testing, Verification and Validation (ICST)}, pages 394--397. IEEE, 2018.

\bibitem{kim2022automated}
Myeongsoo Kim, Qi~Xin, Saurabh Sinha, and Alessandro Orso.
\newblock {Automated test generation for REST APIs: no time to rest yet}.
\newblock In {\em Proceedings of the 31st ACM SIGSOFT International Symposium on Software Testing and Analysis}, pages 289--301, 2022.

\bibitem{martin2021black}
Alberto Martin-Lopez, Andrea Arcuri, Sergio Segura, and Antonio Ruiz-Cort{\'e}s.
\newblock {Black-box and white-box test case generation for RESTful APIs: Enemies or allies?}
\newblock In {\em 2021 IEEE 32nd International Symposium on Software Reliability Engineering (ISSRE)}, pages 231--241. IEEE, 2021.

\bibitem{arcuri2023building}
Andrea Arcuri, Man Zhang, Asma Belhadi, Bogdan Marculescu, Amid Golmohammadi, Juan~Pablo Galeotti, and Susruthan Seran.
\newblock Building an open-source system test generation tool: lessons learned and empirical analyses with {EvoMaster}.
\newblock {\em Software Quality Journal}, 31(3):947--990, 2023.

\bibitem{godefroid2020differential}
Patrice Godefroid, Daniel Lehmann, and Marina Polishchuk.
\newblock {Differential regression testing for REST APIs}.
\newblock In {\em Proceedings of the 29th ACM SIGSOFT International Symposium on Software Testing and Analysis}, pages 312--323, 2020.

\bibitem{sartaj2023testing}
Hassan Sartaj, Shaukat Ali, Tao Yue, and Kjetil Moberg.
\newblock {Testing Real-World Healthcare IoT Application: Experiences and Lessons Learned}.
\newblock In {\em Proceedings of the 31st ACM Joint European Software Engineering Conference and Symposium on the Foundations of Software Engineering}, ESEC/FSE 2023, page 2044–2049. Association for Computing Machinery, 2023.

\bibitem{beizer2003software}
Boris Beizer.
\newblock {\em Software testing techniques (2nd Ed.)}.
\newblock Dreamtech Press, 2003.

\bibitem{yoo2012regression}
Shin Yoo and Mark Harman.
\newblock Regression testing minimization, selection and prioritization: a survey.
\newblock {\em Software testing, verification and reliability}, 22(2):67--120, 2012.

\bibitem{greca2023state}
Renan Greca, Breno Miranda, and Antonia Bertolino.
\newblock State of practical applicability of regression testing research: A live systematic literature review.
\newblock {\em ACM Computing Surveys}, 55(13s):1--36, 2023.

\bibitem{wtsoslo}
Helsedirektoratet.
\newblock National welfare technology program.
\newblock \url{https://www.helsedirektoratet.no/tema/velferdsteknologi}, 2024.
\newblock [Online; accessed 05-May-2024].

\bibitem{sartaj2024modelbased}
Hassan Sartaj, Shaukat Ali, Tao Yue, and Kjetil Moberg.
\newblock Model-based digital twins of medicine dispensers for healthcare {IoT} applications.
\newblock {\em Software: Practice and Experience}, 54(6):1172--1192, 2024.

\bibitem{sartaj2024uncertainty}
Hassan Sartaj, Shaukat Ali, and Julie~Marie Gjøby.
\newblock Uncertainty-aware environment simulation of medical devices digital twins.
\newblock {\em Software and Systems Modeling}, 24:651–677, 2025.

\bibitem{sartaj2024medet}
Hassan Sartaj, Shaukat Ali, and Julie~Marie Gjøby.
\newblock {MeDeT: Medical Device Digital Twins Creation with Few-shot Meta-learning}.
\newblock {\em ACM Transactions on Software Engineering and Methodology}, 34(6):1--36, 2025.

\bibitem{dias2018brief}
Jo{\~a}o~Pedro Dias, Fl{\'a}vio Couto, Ana~CR Paiva, and Hugo~Sereno Ferreira.
\newblock A brief overview of existing tools for testing the internet-of-things.
\newblock In {\em 2018 IEEE International Conference on Software Testing, Verification and Validation Workshops (ICSTW)}, pages 104--109. IEEE, 2018.

\bibitem{ahmad2016model}
Abbas Ahmad, Fabrice Bouquet, Elizabeta Fourneret, Franck Le~Gall, and Bruno Legeard.
\newblock Model-based testing as a service for {IoT} platforms.
\newblock In {\em Leveraging Applications of Formal Methods, Verification and Validation: Discussion, Dissemination, Applications: 7th International Symposium, ISoLA 2016, Imperial, Corfu, Greece, October 10-14, 2016, Proceedings, Part II 7}, pages 727--742, Cham, 2016. Springer.

\bibitem{leotta2018acceptance}
Maurizio Leotta, Diego Clerissi, Dario Olianas, Filippo Ricca, Davide Ancona, Giorgio Delzanno, Luca Franceschini, and Marina Ribaudo.
\newblock {An acceptance testing approach for Internet of Things systems}.
\newblock {\em IET Software}, 12(5):430--436, 2018.

\bibitem{amalfitano2017towards}
Domenico Amalfitano, Nicola Amatucci, Vincenzo De~Simone, Vincenzo Riccio, and Fasolino~Anna Rita.
\newblock Towards a thing-in-the-loop approach for the verification and validation of {IoT} systems.
\newblock In {\em Proceedings of the 1st ACM Workshop on the Internet of Safe Things}, pages 57--63, 2017.

\bibitem{wang2022understanding}
Tao Wang, Kangkang Zhang, Wei Chen, Wensheng Dou, Jiaxin Zhu, Jun Wei, and Tao Huang.
\newblock Understanding device integration bugs in smart home system.
\newblock In {\em Proceedings of the 31st ACM SIGSOFT International Symposium on Software Testing and Analysis}, pages 429--441, 2022.

\bibitem{hu2022ct}
Linghuan Hu, W~Eric Wong, D~Richard Kuhn, Raghu~N Kacker, and Shuo Li.
\newblock {CT-IoT}: a combinatorial testing-based path selection framework for effective {IoT} testing.
\newblock {\em Empirical Software Engineering}, 27:1--38, 2022.

\bibitem{garn2022combinatorial}
Bernhard Garn, Dominik-Philip Schreiber, Dimitris~E Simos, Rick Kuhn, Jeff Voas, and Raghu Kacker.
\newblock Combinatorial methods for testing internet of things smart home systems.
\newblock {\em Software Testing, Verification and Reliability}, 32(2):e1805, 2022.

\bibitem{sotiriadis2014towards}
Stelios Sotiriadis, Nik Bessis, Eleana Asimakopoulou, and Navonil Mustafee.
\newblock Towards simulating the internet of things.
\newblock In {\em 2014 28th International Conference on Advanced Information Networking and Applications Workshops}, pages 444--448. IEEE, 2014.

\bibitem{golmohammadi2022testing}
Amid Golmohammadi, Man Zhang, and Andrea Arcuri.
\newblock {Testing RESTful APIs: A Survey}.
\newblock 33(1), nov 2023.

\bibitem{martin2022online}
Alberto Martin-Lopez, Sergio Segura, and Antonio Ruiz-Cort{\'e}s.
\newblock {Online testing of RESTful APIs: Promises and challenges}.
\newblock In {\em Proceedings of the 30th ACM Joint European Software Engineering Conference and Symposium on the Foundations of Software Engineering}, pages 408--420, 2022.

\bibitem{alonso2022arte}
Juan~C Alonso, Alberto Martin-Lopez, Sergio Segura, Jose~Maria Garcia, and Antonio Ruiz-Cortes.
\newblock {ARTE: Automated Generation of Realistic Test Inputs for Web APIs}.
\newblock {\em IEEE Transactions on Software Engineering}, 49(1):348--363, 2022.

\bibitem{godefroid2020intelligent}
Patrice Godefroid, Bo-Yuan Huang, and Marina Polishchuk.
\newblock {Intelligent REST API data fuzzing}.
\newblock In {\em Proceedings of the 28th ACM Joint Meeting on European Software Engineering Conference and Symposium on the Foundations of Software Engineering}, pages 725--736, 2020.

\bibitem{corradini2022automated}
Davide Corradini, Amedeo Zampieri, Michele Pasqua, Emanuele Viglianisi, Michael Dallago, and Mariano Ceccato.
\newblock {Automated black-box testing of nominal and error scenarios in RESTful APIs}.
\newblock {\em Software Testing, Verification and Reliability}, 32(5):e1808, 2022.

\bibitem{arcuri2019restful}
Andrea Arcuri.
\newblock {RESTful API automated test case generation with EvoMaster}.
\newblock {\em ACM Transactions on Software Engineering and Methodology (TOSEM)}, 28(1):1--37, 2019.

\bibitem{felicio2023rapitest}
Duarte Fel{\'\i}cio, Jos{\'e} Sim{\~a}o, and Nuno Datia.
\newblock {RapiTest: Continuous Black-Box Testing of RESTful Web APIs}.
\newblock {\em Procedia Computer Science}, 219:537--545, 2023.

\bibitem{karlsson2020quickrest}
Stefan Karlsson, Adnan {\v{C}}au{\v{s}}evi{\'c}, and Daniel Sundmark.
\newblock {QuickREST: Property-based test generation of OpenAPI-described RESTful APIs}.
\newblock In {\em 2020 IEEE 13th International Conference on Software Testing, Validation and Verification (ICST)}, pages 131--141. IEEE, 2020.

\bibitem{kim2023enhancing}
Myeongsoo Kim, Davide Corradini, Saurabh Sinha, Alessandro Orso, Michele Pasqua, Rachel Tzoref-Brill, and Mariano Ceccato.
\newblock {Enhancing REST API Testing with NLP Techniques}.
\newblock In {\em Proceedings of the 32nd ACM SIGSOFT International Symposium on Software Testing and Analysis}, pages 1232--1243, 2023.

\bibitem{kim2023adaptive}
Myeongsoo Kim, Saurabh Sinha, and Alessandro Orso.
\newblock {Adaptive REST API Testing with Reinforcement Learning}.
\newblock In {\em 2023 38th IEEE/ACM International Conference on Automated Software Engineering (ASE)}, pages 446--458. IEEE, 2023.

\bibitem{foley2025apirl}
Myles Foley and Sergio Maffeis.
\newblock {APIRL: Deep Reinforcement Learning for REST API Fuzzing}.
\newblock In {\em Proceedings of the AAAI Conference on Artificial Intelligence}, volume~39, pages 191--199, 2025.

\bibitem{kim2025llamaresttest}
Myeongsoo Kim, Saurabh Sinha, and Alessandro Orso.
\newblock {LlamaRestTest: Effective REST API Testing with Small Language Models}.
\newblock {\em arXiv preprint arXiv:2501.08598}, 2025.

\bibitem{apifuzzer}
KissPeter.
\newblock {APIFuzzer — HTTP API Testing Framework}.
\newblock \url{https://github.com/KissPeter/APIFuzzer}, 2024.
\newblock [Online; accessed 20-April-2024].

\bibitem{tcases}
Cornutum.
\newblock {Tcases: A Model-Based Test Case Generator}.
\newblock \url{https://github.com/Cornutum/tcases}, 2024.
\newblock [Online; accessed 20-April-2024].

\bibitem{dredd}
Apiary.
\newblock {Dredd — HTTP API Testing Framework}.
\newblock \url{https://github.com/apiaryio/dredd}, 2024.
\newblock [Online; accessed 20-April-2024].

\bibitem{liu2022morest}
Yi~Liu, Yuekang Li, Gelei Deng, Yang Liu, Ruiyuan Wan, Runchao Wu, Dandan Ji, Shiheng Xu, and Minli Bao.
\newblock {Morest: model-based RESTful API testing with execution feedback}.
\newblock In {\em Proceedings of the 44th International Conference on Software Engineering}, pages 1406--1417, 2022.

\bibitem{ed2018automatic}
Hamza Ed-Douibi, Javier Luis~C{\'a}novas Izquierdo, and Jordi Cabot.
\newblock {Automatic generation of test cases for REST APIs: A specification-based approach}.
\newblock In {\em 2018 IEEE 22nd International Enterprise Distributed Object Computing Conference (EDOC)}, pages 181--190. IEEE, 2018.

\bibitem{laranjeiro2021black}
Nuno Laranjeiro, Jo{\~a}o Agnelo, and Jorge Bernardino.
\newblock {A black box tool for robustness testing of REST services}.
\newblock {\em IEEE Access}, 9:24738--24754, 2021.

\bibitem{segura2018metamorphic}
Sergio Segura, Jos{\'e}~A Parejo, Javier Troya, and Antonio Ruiz-Cort{\'e}s.
\newblock {Metamorphic testing of RESTful web APIs}.
\newblock In {\em Proceedings of the 40th International Conference on Software Engineering}, pages 882--882, 2018.

\bibitem{stallenberg2021improving}
Dimitri Stallenberg, Mitchell Olsthoorn, and Annibale Panichella.
\newblock {Improving test case generation for REST APIs through hierarchical clustering}.
\newblock In {\em 2021 36th IEEE/ACM International Conference on Automated Software Engineering (ASE)}, pages 117--128. IEEE, 2021.

\bibitem{atlidakis2020checking}
Vaggelis Atlidakis, Patrice Godefroid, and Marina Polishchuk.
\newblock {Checking security properties of cloud service REST APIs}.
\newblock In {\em 2020 IEEE 13th International Conference on Software Testing, Validation and Verification (ICST)}, pages 387--397. IEEE, 2020.

\bibitem{mirabella2021deep}
A~Giuliano Mirabella, Alberto Martin-Lopez, Sergio Segura, Luis Valencia-Cabrera, and Antonio Ruiz-Cort{\'e}s.
\newblock {Deep learning-based prediction of test input validity for RESTful APIs}.
\newblock In {\em 2021 IEEE/ACM Third International Workshop on Deep Learning for Testing and Testing for Deep Learning (DeepTest)}, pages 9--16. IEEE, 2021.

\bibitem{corradini2023automated}
Davide Corradini, Michele Pasqua, and Mariano Ceccato.
\newblock {Automated black-box testing of mass assignment vulnerabilities in RESTful APIs}.
\newblock In {\em 2023 IEEE/ACM 45th International Conference on Software Engineering (ICSE)}, pages 2553--2564. IEEE, 2023.

\bibitem{pando2022software}
B~Pando and Abraham D{\'a}vila.
\newblock Software testing in the {DevOps} context: A systematic mapping study.
\newblock {\em Programming and Computer Software}, 48(8):658--684, 2022.

\bibitem{poth2022integration}
Alexander Poth, Olsi Rrjolli, and Andreas Riel.
\newblock Integration-and system-testing aligned with cloud-native approaches for {DevOps}.
\newblock In {\em 2022 IEEE 22nd International Conference on Software Quality, Reliability, and Security Companion (QRS-C)}, pages 201--208. IEEE, 2022.

\bibitem{rangnau2020continuous}
Thorsten Rangnau, Remco~v Buijtenen, Frank Fransen, and Fatih Turkmen.
\newblock Continuous security testing: A case study on integrating dynamic security testing tools in ci/cd pipelines.
\newblock In {\em 2020 IEEE 24th International Enterprise Distributed Object Computing Conference (EDOC)}, pages 145--154. IEEE, 2020.

\bibitem{chen2020performance}
Jinfu Chen.
\newblock Performance regression detection in {DevOps}.
\newblock In {\em Proceedings of the ACM/IEEE 42nd International Conference on Software Engineering: Companion Proceedings}, pages 206--209, New York, NY, USA, 2020. Association for Computing Machinery.

\bibitem{bertolino2023devopret}
Antonia Bertolino, Guglielmo~De Angelis, Antonio Guerriero, Breno Miranda, Roberto Pietrantuono, and Stefano Russo.
\newblock {DevOpRET}: Continuous reliability testing in devops.
\newblock {\em Journal of Software: Evolution and Process}, 35(3):e2298, 2023.

\bibitem{abdulla2023automated}
Hasan Hameed Hasan~Ahmed Abdulla and Fawzi~Abdulaziz Albalooshi.
\newblock Automated testing for {DevOps} in {GitHub} environment: A comprehensive analysis.
\newblock In {\em 2023 International Conference on Advanced Mechatronics, Intelligent Manufacture and Industrial Automation (ICAMIMIA)}, pages 833--838. IEEE, 2023.

\bibitem{patel2022state}
Akshit~Raj Patel and Sulabh Tyagi.
\newblock The state of test automation in {DevOps}: a systematic literature review.
\newblock In {\em Proceedings of the 2022 Fourteenth International Conference on Contemporary Computing}, pages 689--695, New York, NY, USA, 2022. Association for Computing Machinery.

\bibitem{corradini2021empirical}
Davide Corradini, Amedeo Zampieri, Michele Pasqua, and Mariano Ceccato.
\newblock {Empirical comparison of black-box test case generation tools for RESTful APIs}.
\newblock In {\em 2021 IEEE 21st International Working Conference on Source Code Analysis and Manipulation (SCAM)}, pages 226--236. IEEE, 2021.

\bibitem{zhang2023open}
Man Zhang and Andrea Arcuri.
\newblock {Open Problems in Fuzzing RESTful APIs: A Comparison of Tools}.
\newblock {\em ACM Transactions on Software Engineering and Methodology}, 32(6):1--45, 2023.

\bibitem{karie}
AceAge.
\newblock {Karie Medicine Dispenser}.
\newblock \url{https://kariehealth.com/}, 2025.
\newblock [Online; accessed 18-April-2025].

\bibitem{medido}
Medido.
\newblock {Automatic Medicine Dispenser Medido}.
\newblock \url{https://medido.com/en/}, 2025.
\newblock [Online; accessed 18-April-2025].

\bibitem{pilly}
Responssenteret.
\newblock {Pilly SMS medicine dispenser}.
\newblock \url{https://responssenteret.no/responsskolen/brukere/manualer-videoer/Pilly.php}, 2025.
\newblock [Online; accessed 18-April-2025].

\bibitem{lin2023forest}
Jiaxian Lin, Tianyu Li, Yang Chen, Guangsheng Wei, Jiadong Lin, Sen Zhang, and Hui Xu.
\newblock {foREST: A Tree-based Black-box Fuzzing Approach for RESTful APIs}.
\newblock In {\em 2023 IEEE 34th International Symposium on Software Reliability Engineering (ISSRE)}, pages 695--705. IEEE, 2023.

\bibitem{martin2020restest}
Alberto Martin-Lopez, Sergio Segura, and Antonio Ruiz-Cort{\'e}s.
\newblock {RESTest: Black-box constraint-based testing of RESTful web APIs}.
\newblock In {\em Service-Oriented Computing: 18th International Conference, ICSOC 2020, Dubai, United Arab Emirates, December 14--17, 2020, Proceedings 18}, pages 459--475. Springer, 2020.

\bibitem{martin2021specification}
Alberto Martin-Lopez, Sergio Segura, Carlos M{\"u}ller, and Antonio Ruiz-Cort{\'e}s.
\newblock Specification and automated analysis of inter-parameter dependencies in web {APIs}.
\newblock {\em IEEE Transactions on Services Computing}, 15(4):2342--2355, 2021.

\bibitem{sartaj2021testing}
Hassan Sartaj, Muhammad~Zohaib Iqbal, and Muhammad~Uzair Khan.
\newblock Testing cockpit display systems of aircraft using a model-based approach.
\newblock {\em Software and Systems Modeling}, 20(6):1977--2002, 2021.

\bibitem{ratcliff1988pattern}
John~W Ratcliff, David~E Metzener, et~al.
\newblock Pattern matching: The gestalt approach.
\newblock {\em Dr. Dobb’s Journal}, 13(7):46, 1988.

\bibitem{python3difflib}
Python~Software Foundation.
\newblock {\em difflib — Helpers for computing deltas}, 2024.
\newblock [Online; accessed 10-July-2024].

\bibitem{tsikerdekis2018persistent}
Michail Tsikerdekis.
\newblock Persistent code contribution: a ranking algorithm for code contribution in crowdsourced software.
\newblock {\em Empirical Software Engineering}, 23(4):1871--1894, 2018.

\bibitem{gabel2010scalable}
Mark Gabel, Junfeng Yang, Yuan Yu, Moises Goldszmidt, and Zhendong Su.
\newblock Scalable and systematic detection of buggy inconsistencies in source code.
\newblock In {\em Proceedings of the ACM international conference on Object oriented programming systems languages and applications}, OOPSLA '10, pages 175--190, New York, NY, USA, 2010. Association for Computing Machinery.

\bibitem{elbaum2002test}
Sebastian Elbaum, Alexey~G Malishevsky, and Gregg Rothermel.
\newblock Test case prioritization: A family of empirical studies.
\newblock {\em IEEE transactions on software engineering}, 28(2):159--182, 2002.

\bibitem{corradini2021restats}
Davide Corradini, Amedeo Zampieri, Michele Pasqua, and Mariano Ceccato.
\newblock {Restats: A Test Coverage Tool for RESTful APIs}.
\newblock In {\em 2021 IEEE International Conference on Software Maintenance and Evolution (ICSME)}, pages 594--598, 2021.

\bibitem{martin2019test}
Alberto Martin-Lopez, Sergio Segura, and Antonio Ruiz-Cort{\'e}s.
\newblock {Test coverage criteria for RESTful web APIs}.
\newblock In {\em Proceedings of the 10th ACM SIGSOFT International Workshop on Automating TEST Case Design, Selection, and Evaluation}, pages 15--21, 2019.

\bibitem{shapiro1965analysis}
Samuel~Sanford Shapiro and Martin~B Wilk.
\newblock An analysis of variance test for normality (complete samples).
\newblock {\em Biometrika}, 52(3-4):591--611, 1965.

\bibitem{arcuri2011practical}
Andrea Arcuri and Lionel Briand.
\newblock A practical guide for using statistical tests to assess randomized algorithms in software engineering.
\newblock In {\em Proceedings of the 33rd international conference on software engineering}, pages 1--10, New York, NY, USA, 2011. Association for Computing Machinery.

\bibitem{holm1979simple}
Sture Holm.
\newblock A simple sequentially rejective multiple test procedure.
\newblock {\em Scandinavian journal of statistics}, pages 65--70, 1979.

\bibitem{wagner2021code}
Stefan Wagner and Marvin Wyrich.
\newblock Code comprehension confounders: A study of intelligence and personality.
\newblock {\em IEEE Transactions on Software Engineering}, 48(12):4789--4801, 2021.

\bibitem{marculescu2022faults}
Bogdan Marculescu, Man Zhang, and Andrea Arcuri.
\newblock {On the faults found in REST APIs by automated test generation}.
\newblock {\em ACM Transactions on Software Engineering and Methodology (TOSEM)}, 31(3):1--43, 2022.

\bibitem{khan2019aspectocl}
Muhammad~Uzair Khan, Hassan Sartaj, Muhammad~Zohaib Iqbal, Muhammad Usman, and Numra Arshad.
\newblock {AspectOCL}: using aspects to ease maintenance of evolving constraint specification.
\newblock {\em Empirical Software Engineering}, 24(4):2674--2724, 2019.

\bibitem{isaku2023cost}
Erblin Isaku, Hassan Sartaj, Christoph Laaber, Tao Yue, Shaukat Ali, Thomas Schwitalla, and Jan~F Nyg{\aa}rd.
\newblock {Cost Reduction on Testing Evolving Cancer Registry System}.
\newblock In {\em 2023 IEEE International Conference on Software Maintenance and Evolution (ICSME)}, pages 508--518. IEEE, 2023.

\bibitem{decrop2024you}
Alix Decrop, Gilles Perrouin, Mike Papadakis, Xavier Devroey, and Pierre-Yves Schobbens.
\newblock {You Can REST Now: Automated Specification Inference and Black-Box Testing of RESTful APIs with Large Language Models}.
\newblock {\em arXiv preprint arXiv:2402.05102}, 2024.

\end{thebibliography}

\end{document}